\documentclass[preprint]{aastex}
\usepackage{graphicx}
\usepackage{epsfig}
\usepackage{epstopdf}
\usepackage{rotating}
\usepackage{float}
\usepackage{natbib}
 
\usepackage[font=scriptsize]{caption}
\usepackage[font=scriptsize]{subcaption}
\newcommand{\civ}{\ion{C}{4}}

\newcommand{\aox}{$\alpha_{\rm{ox}} \,$}

\newcommand{\aro}{$\alpha_{\rm{ro}} \,$}
\newcommand{\daro}{$\Delta \alpha_{\rm{ro}} \,$}

\newcommand{\ltwofive}{$L_{2500\,\rm{\AA}}$}
\newcommand{\lledd}{$L/L_{\rm{Edd}}$}

\shorttitle{Mean and Extreme Radio Properties of Quasars}
\shortauthors{Kratzer et al.}

\begin{document}

\title{Mean and Extreme Radio Properties of Quasars and the Origin of Radio Emission}
\author{Rachael M. Kratzer,\altaffilmark{1} Gordon T. Richards\altaffilmark{1}}
\altaffiltext{1}{Department of Physics, Drexel University, Philadelphia, PA.}

\begin{abstract}

  We investigate the evolution of both the radio-loud fraction (RLF)
  and (using stacking analysis) the mean radio-loudness of quasars.
  We consider how these properties evolve as a function of redshift
  and luminosity, black hole (BH) mass and accretion rate, and
  parameters related to the dominance of a wind in the broad emission
  line region.  We match the FIRST source catalog to samples of
  luminous quasars (both spectroscopic and photometric), primarily
  from the Sloan Digital Sky Survey.  After accounting for
  catastrophic errors in BH mass estimates at high-redshift, we find
  that both the RLF and the mean radio luminosity increase for
  increasing BH mass and decreasing accretion rate.  Similarly both
  the RLF and mean radio loudness increase for quasars which are
  argued to have weaker radiation line driven wind components of the
  broad emission line region.  In agreement with past work, we find
  that the RLF increases with increasing optical/UV luminosity and decreasing
  redshift while the mean radio-loudness evolves in the exact opposite
  manner.  This difference in behavior between the mean radio-loudness
  and the RLF in $L-z$ may indicate selection effects that bias our
  understanding of the evolution of the RLF; deeper surveys in the
  optical and radio are needed to resolve this discrepancy.  Finally,
  we argue that radio-loud (RL) and radio-quiet (RQ) quasars may be
  parallel sequences but where only RQ quasars at one extreme of the
  distribution are likely to become RL, possibly through slight
  differences in spin and/or merger history.

\end{abstract}


\section{Introduction}
\label{sec:Intro}

Quasars were first identified by the 3rd Cambridge Catalog of Radio
Sources \citep{Edge59}.  Although their extra-Galactic nature
\citep{Schmidt63} and viable energy source \citep{LB69} have been
determined, we still lack a complete understanding of why some active
galactic nuclei (AGN) are strong radio sources and others are not.  In
particular, it is generally not possible to use information outside of
the radio part of the spectrum to reliably predict whether an {\em
  individual} quasar will be radio-loud or not.

Radio-loudness has been defined both in the absolute sense
\citep[e.g.,][]{Peacock86} and in the relative sense
\citep[e.g.,][]{Kellermann89}; see Section~\ref{sec:aro}.  Regardless
of how the radio-loud (RL) boundary is imposed, many researchers have
argued that the distribution exhibits a bimodality
\citep[e.g.,][]{Strittmatter80, Kellermann89, Miller90, Visnovsky92,
  Goldschmidt99}, with fewer quasars lying between the objects with
powerful radio emission and the much larger population with very
little radio emission.  To set the stage for our work, it is worth
spending some time reviewing the literature regarding this argument.
In particular, while it would seem that the {\em literature} itself is
bimodal on the question of radio bimodality, we will argue that all of
these studies are actually in reasonable agreement.

While it had been known that some 10\% of AGN and, for that matter,
giant elliptical galaxies were strong radio sources, both
\citet{Strittmatter80} and \citet{Kellermann89} found the
radio-loudness distribution to be bimodal; they observed relatively
few quasars between the bulk of the quasar population and its
radio-loud tail. \citet{Kellermann89} specifically analyzed Very Large
Array (VLA) data of 114 Palomar-Green (PG) quasars \citep{SG83}, and,
while the PG quasars may not be representative of the average Sloan
Digital Sky Survey (SDSS; \citealt{York00}) quasar \citep{Jester05},
they are generally the best-studied quasars. \citet{Kellermann89}
further noted that the bimodality is not obviously due to beaming (see
also \citealt{Barvainis05} and \citealt{Ulvestad05}).

\citet{White00} showed that the depth of the FIRST (Faint Images of
the Radio Sky at Twenty Centimeters; \citealt{FIRST95}) data fills in
where prior radio samples of quasars were lacking and argued that the
historical (apparent) bimodality is not real. \citet{Ivezic} pointed
out that there are selection effects due to the limits in
the optical magnitude and radio flux that must be taken into
consideration in such analyses. In short, going deeper in the radio
without also going deeper in the optical does, indeed, yield more
radio-intermediate sources but without the commensurate ability to
find more RL sources. When this bias is taken into account,
\citet{Ivezic} demonstrated that the data are consistent with a formal
bimodality in the radio-loudness distribution.

In tallying papers for and against a radio bimodality, \citet{Cira03a}
and \citet{Cira03b} are two of the papers that always appear in the
against column. However, the analysis in both papers shows a
distribution with two modes. In \citet{Cira03b}, two components
are used in their fit of the distribution, and both a single Gaussian
and a flat distribution are rejected. Thus, these papers would be
better categorized as providing evidence in support of a bimodality
yet demonstrating that there is not a barren gap between the peaks as
some have characterized the ``bimodality''.

\citet{Xu99} and \citet{Sikora07} demonstrate the bimodality in a
different manner using heterogeneous combinations of multiple
subsamples of data. When radio luminosities are plotted as a function
of optical luminosity \citep{Sikora07} or [\ion{O}{3}] line luminosity
\citep{Xu99}, their samples split into two: a radio-quiet (RQ)
sequence and a RL sequence 3--4 order of magnitudes louder (see
\citealt{Sikora07}, Fig.~1 and \citealt{Xu99}, Fig.~1). The different
subsamples give the perception of the sort of gap that \citet{Cira03a,
  Cira03b} have argued against; however, this is largely due to the
sample selection. The important thing is that, for a given optical or
[\ion{O}{3}] line luminosity, the radio luminosity spans 5 or 7 orders
of magnitude, respectively. As described in Section \ref{sec:Data},
our sample is nicely complementary to these investigations.

\citet{Rafter09} also present arguments against a radio bimodality. In
an attempt to make an unbiased sample, they follow the prescription of
\citet{Ivezic}; however, we would argue that their cutting of $\log
R>2$ objects and removal of optical sources whose lack of FIRST radio
detections require them to be RQ actually biases their sample. Accounting for these issues, their distribution is consistent
with the arguments by \citet{Ivezic} for bimodality.

\citet{Mahony12} investigate the radio luminosity distribution of an
X-ray selected sample of low-redshift (broad-lined) quasars observed
at 20GHz.  While they argue that there is ``no clear evidence for a
bimodal distribution'', their distributions (both in radio-luminosity
and $\log R$) would be poorly fit by a single Gaussian component.  While
there is no {\em gap} in the population, the distributions are
consistent with other samples and generally exhibit two
modes. Moreover, the findings are consistent with the argument by
\citet{Xu99} that using the core radio properties minimizes the
differences between the RL and RQ distributions.

\citet{Singal13} notably take a different approach by looking
separately at the optical and radio quasar luminosity
functions. However, they use the SDSS DR7 quasars without
limiting to the ``uniform" sample which was designed for statistical
analysis; see \citet{Richards06} and Section~\ref{sec:DR7QC}.  As a
result, they include objects selected via the ``serendipity" branch of
the quasar target selection algorithm \citep{Stoughton02} which is
radio-biased both explicitly by radio-selection and implicitly through
X-ray selection \citep{Miller11}. That issue aside, \citet{Singal13}
find that radio-loudness increases with redshift, which they note as
being contrary to \citet{Jiang07}. However, \citet{Jiang07}
investigate the radio-loud {\em fraction} and not the mean radio
loudness, so there is no contradiction. Indeed, \citet{White07}
similarly found an increase in the mean radio-loudness with
redshift. We shall explore this point again in Section
\ref{sec:results}.

\citet{Singal13} further find no radio bimodality in the
radio-loudness distribution; however, their analysis is limited to
objects with radio-detections, which severely limits their ability to
probe the full RQ distribution. This restriction to
radio-detections would be appropriate where non-detections do not
distinguish between RL and RQ. However,
\citet{Jiang07} limit their analysis to $i<18.9$ where a non-detection
by FIRST is virtually equivalent (modulo incompleteness at the FIRST
detection limit, see Section~\ref{sec:firstlim}) to being
radio-quiet. We would therefore argue that their analysis is not able
to accurately test for a bimodality in the radio-loudness distribution.

Arguably, \citet{Balokovic} present the best summary of the question
of radio-loud bimodality in quasars. Using Monte Carlo simulations,
they show that the radio-loudness probability distribution function is
consistent with radio luminosity being dependent upon optical
luminosity and is inconsistent with a single distribution in the ratio
of radio-to-optical luminosity. While they were not able to confirm or
reject the hypothesis of the distribution being formally bimodal, the
important result is an empirical dichotomy. That is, two components
are needed to fit the distribution, even if there is not a clear
minimum between those distributions. Indeed, no recent analyses have
actually argued for a desert at intermediate radio-loudnesses; whether
the dichotomous distribution is additionally bimodal or not is a
matter of semantics.

Generally speaking, the data appear consistent with the argument by
\citet{White07} that the radio-loudness distribution is indeed
double-peaked but that the dip between the peaks is more modest than
the standard binary RL/RQ classification suggests. Arguments
contradicting the bimodal nature of the distribution generally are
either based on data that actually does show a bimodal distribution or
that is analyzed in a biased manner as emphasized by \citet[Section
4.2]{Ivezic}.  Nevertheless, the distinction is not very large and is
subject to a number of biases due to redshift, limiting magnitudes in
both the optical and radio, and inherent selection effects in the
quasar population.

We suggest that there is little utility for further discussion about
whether the population is bimodal or not without deeper data (over a
sufficient area) in {\em both} the optical and the radio.
Going deeper in the optical while maintaining the FIRST depth will
artificially enhance the bimodality as the only new sources will have
$\log R >2$.  Similarly going deeper in the radio while maintaining
the SDSS depth will necessarily fill in the radio-intermediate and
radio-quiet population, artificially reducing any true
bimodality. Only by going deeper at both wavelengths can more progress
be made; see Section~\ref{sec:diag}. As such, instead of further
analysis of the shape of the radio-loudness distribution, in this
paper we will instead focus on extending the demographics of the
investigation of the radio properties of quasars, providing new
constraints on the problem.

The issue of bimodality aside, it remains that there are quasars with
strong radio emission and those without.  Many have speculated that
these two classes of quasars must be governed by similar physical
processes \citep{Barthel89, UrryandP95, Shankar10} since they only differ in the amount of radio emission observed.  Still others
have suggested that there really are two different types of quasars
\citep{Moore84,Peacock86,Miller90}. More recently, high black hole
mass and/or low values of the mass-weighted accretion rate (the
Eddington ratio; \lledd) have been implicated as being the primary
drivers of the differences \citep{Laor00,Lacy01,Ho02}.


\citet{Wilson95} argue that the biggest difference between RL and RQ
quasars is the rate at which the central black hole spins. Since the thermal
emission from RL and RQ quasars are so similar
\citep{Neugebauer86,Sanders89, Steidel91}, their black hole masses and
accretion rates must also be comparable. \citet{Richards11} recovers
this same basic conclusion.  The remaining black hole property of
spin, which has been shown to be responsible for the collimation of
radio jets in the presence of an accretion disk
\citep{BZ77,BP82,Blandford90}, is arguably the most plausible
explanation for the difference between RL and RQ quasars (and may not
be independent of mass). Unfortunately, it is extremely difficult to
accurately determine the spin of a black hole.

While the exact reasons behind the difference in radio emission for RL
and RQ quasars has yet to be confidently explained, many have
uncovered valuable properties of these objects that aid in
understanding them.  Our work herein follows and builds upon two of
those investigations: one looking at the radio-loud tail of the
population \citep{Jiang07} and one looking at the mean radio
properties of quasars \citep{White07} (via stacking analyses).

A stacking analysis is an important part of the conversation about the
nature of radio emission in quasars as essentially all quasars are
radio emitters when probed to deep limits \citep[e.g.,][]{Wals05,
  White07,Kimball11}.  At the faintest radio luminosities the radio
emission in quasars is likely due to starburst emission
\citep[e.g.,][]{Condon02, Kimball11}. However, even if a starburst
could produce radio luminosities as high as $10^{31}$ ergs
$\rm{s}^{-1}$ (whereas Kimball finds the peak to be $10^{29}$), that
still leaves a considerable population of radio-quiet quasars that are
neither formally radio-loud nor consistent with a starburst origin
\citep{Blundell98,Jiang10,Zakamska14}.  Indeed, \citet{Ulvestad05},
using high resolution radio imaging, and \citet{Barvainis05}, using
variability studies, both conclude that RQ quasars are just weaker
version of RL quasars, while \citet{Blundell07} argue that disk winds
are responsible for radio emission in RQ quasars.  


By presenting a unique synthesis of these two perspectives (both the
mean and extreme radio properties of quasars) and by adding new
dimensions to these analyses, both by increasing the sample sizes and
considering new parameters, we hope to further constrain our
understanding of the nature of both quasars themselves and their
(occasional) radio exuberance.  Ultimately, the goal is to understand
the production of radio emission to the extent that the radio
properties of an {\em individual} quasar can be predicted by referencing the properties of that quasar in other parts of the electromagnetic
spectrum.

The structure of the paper is as follows.  Section \ref{sec:Data}
begins with a detailed account of the surveys from which our sources
are drawn.  Those familiar with the SDSS and FIRST data sets can skip
to Section~\ref{sec:methods} or even Section~\ref{sec:results}.
Section~\ref{sec:methods} describes the methods and metrics that we
use to conduct our analyses.  Section~\ref{sec:results} considers the
{\em mean} (using the stacking analysis) and {\em extreme} (using the
radio-loud fraction) radio properties of quasars, including
luminosity and redshift (Section \ref{sec:Lz}), Principal
Component and \ion{C}{4} parameters (Section \ref{sec:CIV}), black hole
properties (Section \ref{sec:BH}),
and optical/UV color (Section \ref{sec:StacksColor}). The implications
of our findings are discussed in Section \ref{sec:Discuss}, and we
summarize in Section \ref{sec:Conclusions}.

For the entirety of this paper we employ the accepted cosmology of a
flat universe with $H_0 = 70 \, \rm{km} \, \rm{s}^{-1} \,
\rm{Mpc}^{-1}$, $\Omega_{m} = 0.3$, and $\Omega_{\Lambda} = 0.7$
\citep{Spergel07}.  We will use the term ``quasar'' throughout to
describe luminous AGNs, regardless of their radio properties.

\section{Data}
\label{sec:Data}

\subsection{Sloan Digital Sky Survey}
\label{sec:SDSS}

The Sloan Digital Sky Survey (SDSS; \citealt{York00}) is an optical
survey that has mapped more than 10,000 square degrees of sky located
in the northern galactic hemisphere and partially along the Celestial
Equator. Photometry was performed with a wide-field 2.5-m telescope
\citep{Gunn06} in five magnitude bands between 3,000 and 10,000 \AA \,
($ugriz$; \citealt{Fuku96, Gunn98}). Spectra between 3,800 and 9,200 \AA
\, were also gathered with a pair of double spectrographs.

Reducing the SDSS data entails: correcting defective data (i.e.,
cosmic rays and dead pixels), ascertaining the contributions of noise
to the measured brightnesses from within the CCD as well as from the
earth's atmosphere, calculating the point-spread functions (PSFs) of
the CCD array as a function of time and location, pinpointing objects
of interest, combining the data from the five optical bands, fitting
simple models to each located object, separating the images of objects
that overlap, and determining positions, magnitudes, and shapes of
these objects \citep{York00, Hogg01, Lupton01}. Refinements were made
to the astrometry \citep{Pier03} and photometry \citep{Ivezic04} of
sources as the properties specific to this survey became apparent. The
photometric data are corrected for Galactic dust reddening using the
maps from \citet{Schlegel98} and are reported in terms of the AB
magnitude scale \citep{OkeGunn83}.

The spectroscopic data are automatically reduced by the spectroscopic
pipeline \citep{Stoughton02} which extracts, corrects, and calibrates
the spectra, determines the spectral types, and measures the
redshifts. The reduced spectra are then stored in the operational
database.

Our primary optical data set consists of objects from the SDSS Seventh
Data Release (DR7; \citealt{DR7}), but we will also make use of the
quasar data from the continuation project (SDSS-III;
\citealt{SDSSIII}) as discussed in Section \ref{sec:MC}.

\subsection{FIRST}
\label{sec:Match}

The Very Large Array (VLA) FIRST survey (Faint Images of the Radio Sky
at Twenty Centimeters; \citealt{FIRST95}) covers about the same sky
area as SDSS.  FIRST radio fluxes were obtained in the VLA's
B-configuration at 20 cm (1.4 GHz).  Images of the radio sky were
taken for 165 seconds each with an angular resolution of 5$\arcsec$, a
typical RMS sensitivity of 0.15 mJy $\rm{bm}^{-1}$, and an approximate
threshold flux density of 1.0 mJy $\rm{bm}^{-1}$. The 2012 February 16
catalog contains over 946,000 sources, but only a fraction of these
sources can be matched to known quasars and are processed as described
in Section \ref{sec:RLF}. Additionally, 99.9\% of the FIRST pointings
are blank sky \citep{White07}, and these measurements will be used to
perform stacking analyses of quasars described in
Section~\ref{sec:Stacks}.

We initially matched each optically confirmed quasar to the peak flux
of the closest radio source within $1\farcs5$, but this technique
would only be robust if all of our radio sources were
unresolved. Although only about 5\% of matched optical-radio sources
include lobes and less than 10\% of SDSS-FIRST quasars have radio
morphologies other than point sources\footnote{See \citet{Ivezic},
  Section~3.8 for a complete discussion of the demographics of complex
  radio sources from FIRST.}, these radio fluxes must be
underestimates due to the resolving out of the extended emission at
faint magnitudes and/or high redshift \citep{Hodge2011}; therefore, we will still
attempt to account for a wider variety of radio emission
configurations (core, lobe, etc.). In order to avoid systematically
underestimating the total luminosity of resolved objects that may have
a faint radio core but bright extended radio lobes, we choose to use
the total integrated flux of all radio components associated with each
optically confirmed quasar for our radio-loud fraction analysis.

We have followed the approach of \citet{Jiang07} to find the total
integrated radio flux associated with each optical source. Optically confirmed quasars with more than one FIRST source within a 30" matching radius
are assigned total integrated radio fluxes equivalent to the sum of
the individual integrated fluxes of their matched FIRST objects. If
only one FIRST detection lies within the 30" matching radius of an
optical source, the matching radius is further limited to 5" (in order
to limit spurious contamination by random single matches at
$>$5''). The total integrated radio flux for optical sources with only
one radio match within 5" is simply the integrated flux of that
matched radio source. Expanding the matching radius to 10" for single
radio sources would only increase the number of core radio sources by
$\sim 2.6$\%, so we opt for the 5" matching radius to reduce the
number of false matches included in our analyses. Finally, optical
sources that have only one FIRST match between 5" and 30" are
considered radio non-detections. See Section \ref{sec:diag} for a
discussion of possible complications associated with radio
measurements and how we plan to address them.

\subsection{DR7 Quasar Catalog}
\label{sec:DR7QC}

Our main quasar sample comes from the SDSS DR7 \citep{DR7} Quasar
Catalog \citep{DR7QC}.  It consists of 105,783 spectroscopically
confirmed quasars brighter than $M_i = -22.0$. The majority of these
objects were originally chosen according to the algorithm described by
\citet{Richards02a} for spectroscopic follow-up based on their
location in SDSS color space.  Low-redshift quasars ($z < 3$) were
limited to $i < 19.1$ in $ugri$ color space, while high-redshift
quasars ($z \geq 3.0$) were limited to $i < 20.1$ in $griz$ color
space. Additionally, irrespective of their location in color space,
\citet{Richards02a} included objects with FIRST point sources within
$2\farcs0$ and eliminated objects with unreliable photometric data.

The goal of the SDSS quasar survey was to construct the largest
possible quasar catalog to its given flux limits. As a number of
different algorithms were used to select quasars and some of these
algorithms changed in the early part of the survey
\citep{Stoughton02}, the quasar sample is not sufficiently uniform for
statistical analyses. Section~2 of \citet{Richards02a} discusses how
the sample can be limited to a more uniform selection for the sake of
statistical analysis. Approximately 60,000 quasars belong to the
uniform sample; these are the objects chosen by the final quasar
target selection algorithm of \citet{Richards02a}. This restriction ensures a more
self-consistent subsample and allows us to test whether the full
quasar catalog results are biased by selection effects.

However, we note that the so-called ``uniform'' sample was not meant
to be {\em radio} uniform. The fraction of quasars selected because of
their radio properties (as compared to the total number of quasars
selected) is non-uniform in situations where the completeness of the
optical selection is reduced. For our purposes, this is primarily over
redshifts $2.2<z<3.5$, where optical selection is rather incomplete
due to confusion with the stellar locus. In this redshift region, the
fraction selected because of radio properties is artificially
high. Thus, our analyses of the ``uniform'' sample will need to be
further restricted in redshift space in order to avoid biasing the
radio properties of the SDSS quasar sample.  Nevertheless, the uniform
sample is more radio-uniform than the full DR7 quasar sample.

\citet{Shen11} extend the DR7 Quasar Catalog by improving upon the
continuum and emission line measurements calculated by the SDSS
pipeline (specifically H$\alpha$, H$\beta$, \ion{Mg}{2}, and \civ);
these emission line measurements are implemented in our analyses
reported in Section~\ref{sec:CIV}.  By applying their refined spectral
fits, \citet{Shen11} also estimate the virial masses of the black
holes powering these quasars.  The 
BH masses derived
from \ion{Mg}{2} and \civ\ (and used in Section~\ref{sec:BH}) have
been updated
according the prescriptions described by \citet[Equation 3]{Rafiee11}
and \citet[Equation 3]{Park13}, respectively.  Additionally, we used
the improved redshifts of \citet{HewettWild} with these samples rather
than the redshifts cataloged in SDSS.

It is worth noting the differences between the DR7 quasar sample
(especially the ``uniform" subsample) and the samples analyzed by
\citet{Xu99} and \citet{Sikora07}. Those two papers built samples for
analysis that included a very wide range of AGN types: Seyferts,
broad-lined radio galaxies, and luminous quasars (including both
spiral and elliptical hosts). The SDSS quasars, while spanning the
largest redshift range of any monolithic quasar survey, are actually a
more homogenous sample of objects and nicely complement these broader
analyses. Specifically, the SDSS quasars generally only sample the
bright end of the luminosity function and are limited to luminosities
that distinguish them from lower-luminosity Seyfert AGNs.  For further
discussion, see Section~\ref{sec:Discuss}.

\subsection{Master Quasar Catalog}
\label{sec:MC}

Since restricting the DR7 quasar sample to ``uniform'' quasars reduces
the number of objects in our study considerably, we make an attempt to
extend our investigations to a larger sample of quasars. Thus, the
final dataset that we draw sources from is a ``Master'' Quasar Catalog
compiled by Richards et al.\ (2014, in prep.). It contains over 1.5
million sources and over 250,000 of those have confirming
spectroscopy. This dataset is a ``catalog of catalogs" consisting of
sources within the SDSS-I/II/III survey areas and draws objects from
the following sources:
\begin{itemize}
\item{SDSS I/II: \citet{DR7QC}}
\item{2QZ: \citet{2QZ}}
\item{2SLAQ: \citet{2SLAQ}}
\item{AUS: Croom et al. (in prep.)}
\item{AGES: \citet{AGES}}
\item{COSMOS: \citet{COSMOS07, COSMOS09}}
\item{SDSS-III: \citet{SDSSIII,SDSSIII13}}
\item{\citet{Richards09} Photometric Catalog}
\item{\citet{Bovy11} Photometric Catalog}
\item{\citet{Papovich06}}
\item{\citet{Glikman06}}
\item{\citet{Maddox12}}
\end{itemize}

This quasar sample is, of course, highly inhomogeneous but does
represent nearly every quasar known fainter than $i\sim16$ (including
candidate photometric quasars) at the time of Data Release 9 of
SDSS-III \citep{DR9} and extends the sample significantly in terms of
high-$z$ quasars, reddened quasars, and quasars over
$2.2<z<3.5$. Because of this sample's inhomogeneity, we will also
consider more homogeneous subsamples in our analyses.

\subsection{Our Samples}
\label{sec:OurSamples}

We initially define four subsamples of data (denoted A, B, C, and D)
for our analyses of the radio properties of quasars.  We will focus on
the results from Sample B (which is the most robust, see below),
supplemented with Sample D as needed.  All four sample definitions are
presented here for the sake of completeness.

Sample A is simply the entire DR7 Quasar Catalog, while Sample B is
comprised of objects from the DR7 Quasar Catalog that are flagged
``uniform'' as discussed above. Sample B is the most robust of our
four samples, suffering from the fewest selection effects (especially
when limited to $z<2.2$ and $i<18.9$); however, analysis of the other
samples is important to expand the total number of sources and the
redshift/luminosity ranges covered (at the expense of introducing
biases).

Sample C consists of those objects from the Master Quasar Catalog that
have spectroscopic redshifts.  This is our largest sample of confirmed
quasars; however, it has the strongest selection biases and does not
include sources as faint as those from Sample D.

Sample D is our attempt to create the largest possible sample while
minimizing selection biases.  To increase the size of the sample and
extend to fainter limits while maintaining a high level of uniformity,
Sample D includes quasar {\em candidates} that were identified by {\em
  both} the NBCKDE algorithm \citep{Richards09} and the XDQSO algorithm
\citep{Bovy11}. Thus, two independent algorithms agreed that these
objects are highly likely to be quasars. For the majority of these
objects we must rely on the photometric redshifts reported by these
two catalogs; however, if spectroscopic redshifts exist for the
objects in Sample D, the spectroscopic redshifts are utilized
instead. To make Sample D as robust as possible, we further limit it
to those objects identified by the XDQSO algorithm as having only one
significant peak (exceeding a probability of 80\%) in the photometric
redshift probability distribution function.

For our analyses we exclude quasars that show signs of dust
reddening/extinction. We do so by eliminating quasars with $\Delta
(g-i)>0.5$ which discards $\sim 6$\% of the objects in Sample B.
$\Delta (g-i)$ is defined to remove the dependence of color on
redshift (due to emission features), making it roughly equivalent to
$\alpha_{\rm opt}$, the underlying continuum in the optical-UV part
of the spectral energy distribution (SED); see \citet{Richards03}.

\begin{figure}[h!]
\plottwo{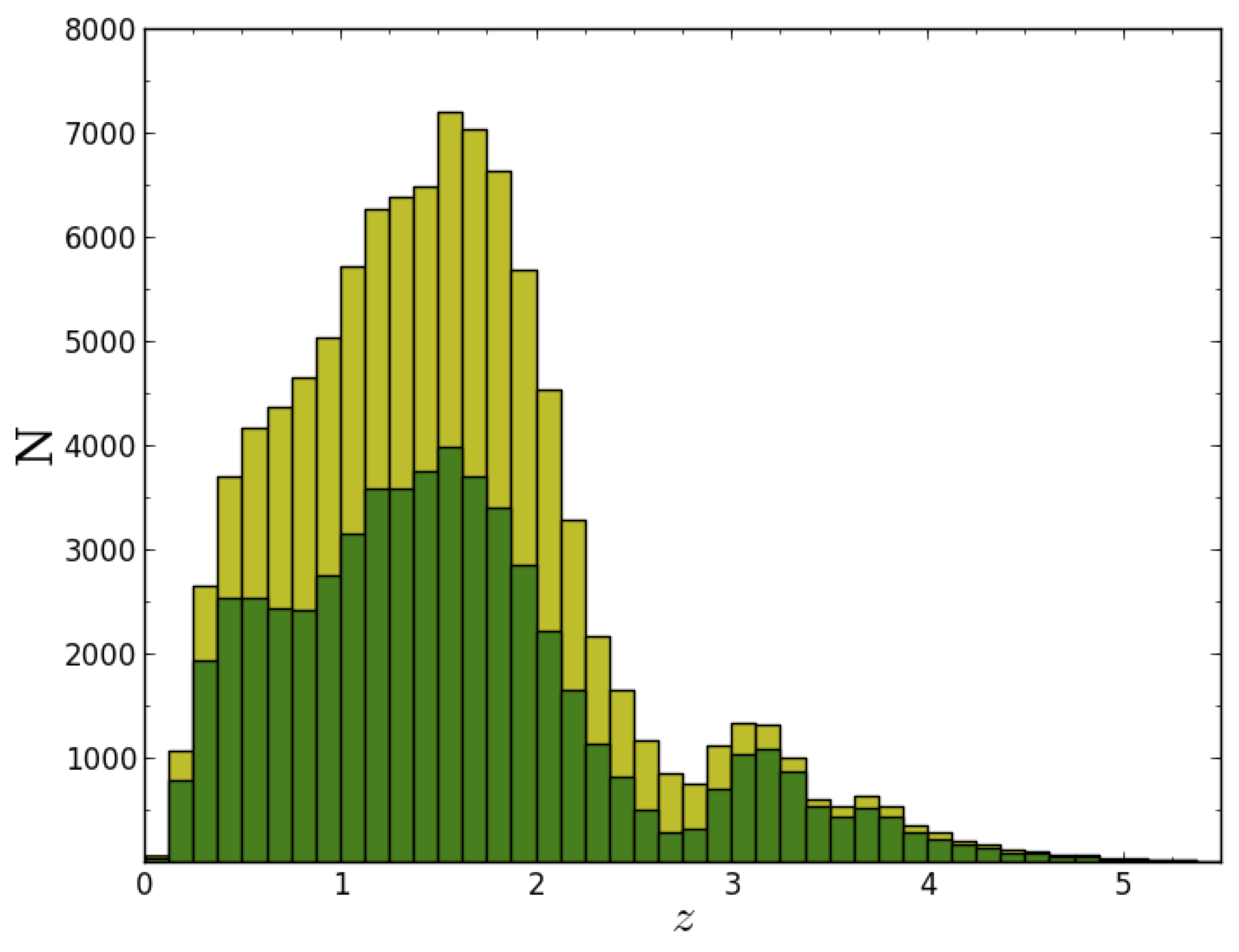}{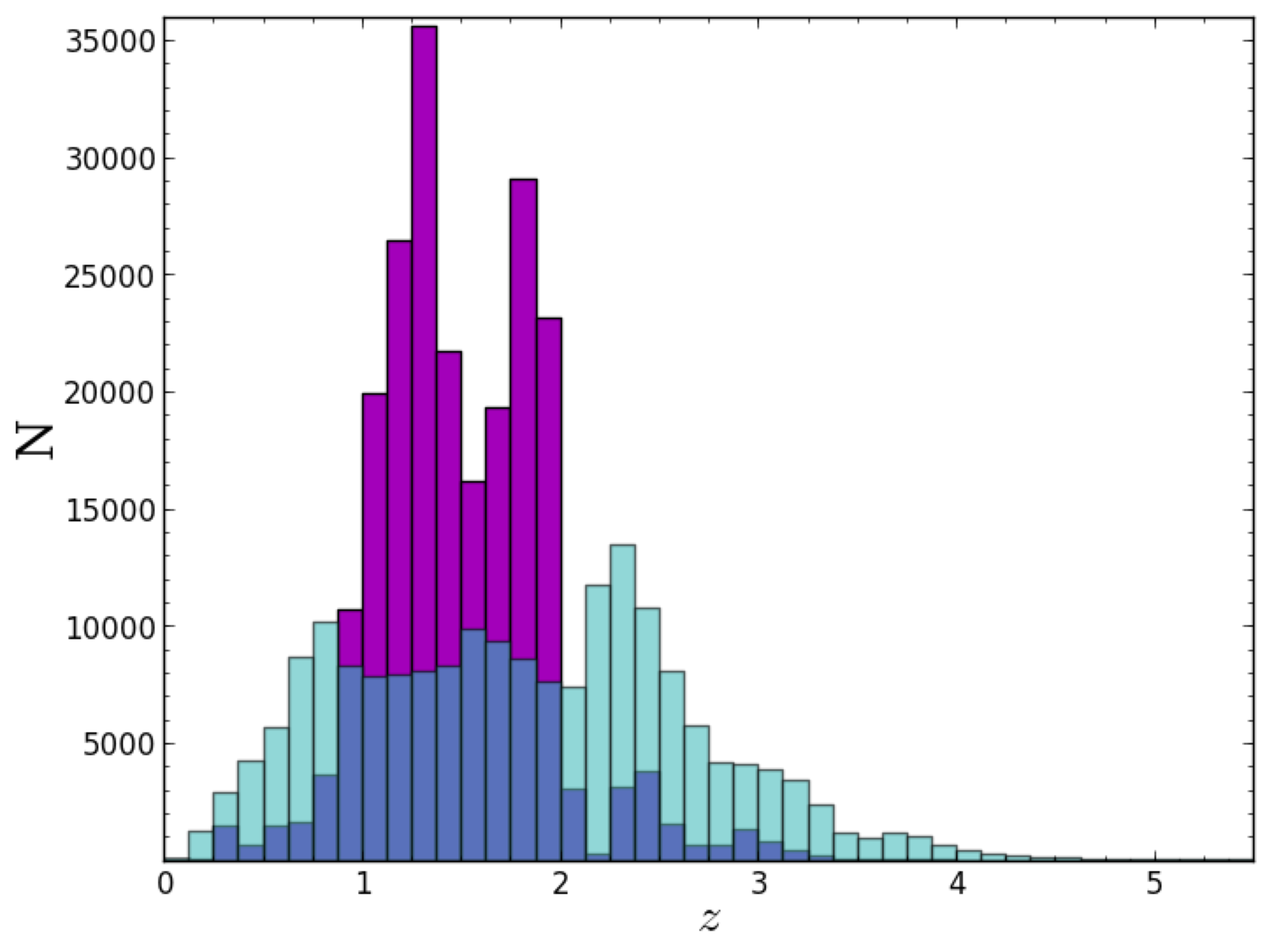}
\caption{Histograms of the redshift distribution for Sample A (left; yellow; 93,362 quasars), Sample B (left; green; 55,302 quasars), Sample C (right; blue; 181,720 quasars), and Sample D (right; purple; 210,825 quasars). Note that the two plots use different y-axis scalings. Sample C fills in the redshift distribution gap seen near $z\sim2.7$ in Samples A and B, while Sample D vastly increases the sample size by probing deeper than the spectroscopic samples (A,B,C) at the expense of the redshift distribution (and redshift accuracy).}\label{fig:f1}
\end{figure}

\begin{figure}[h!]
\plottwo{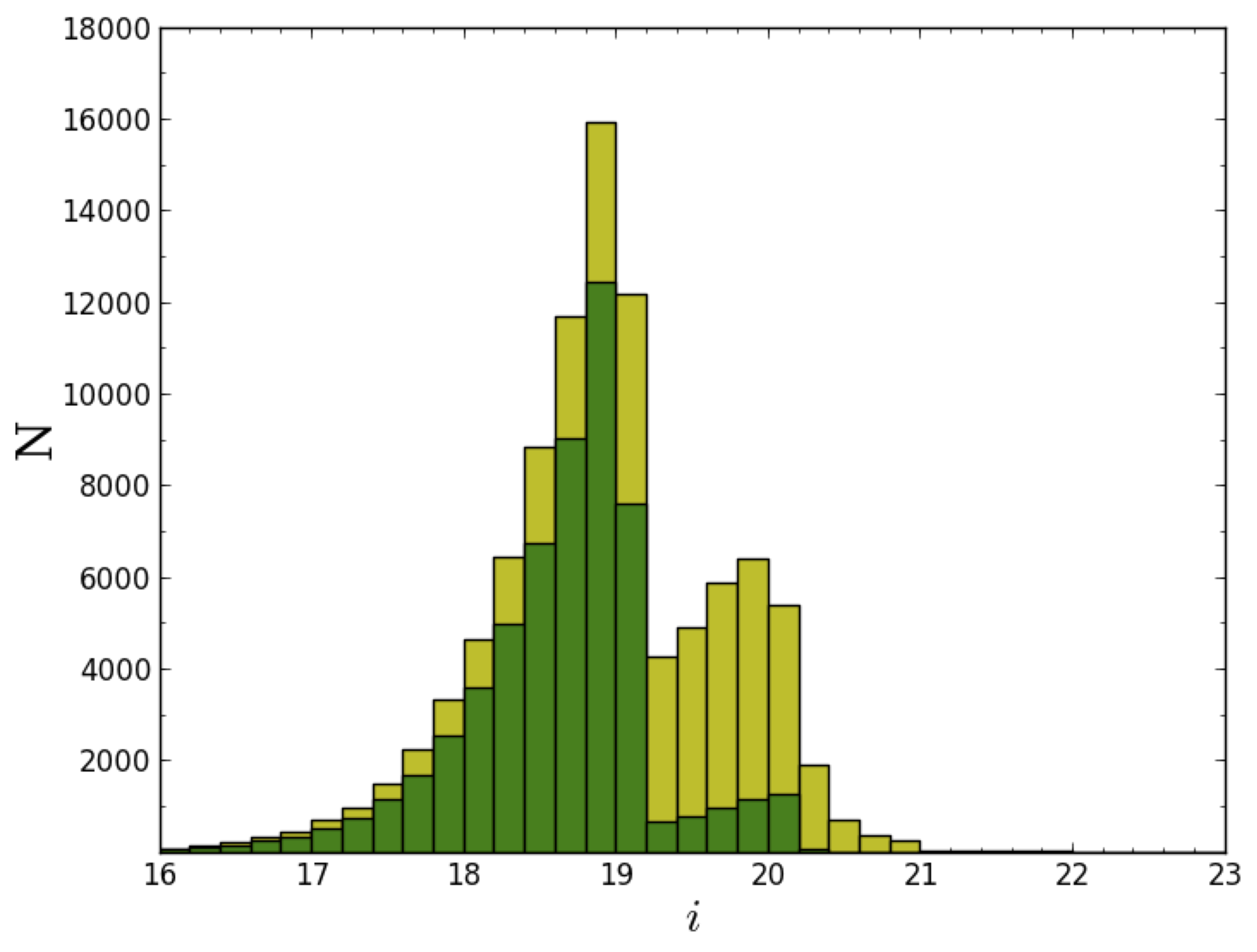}{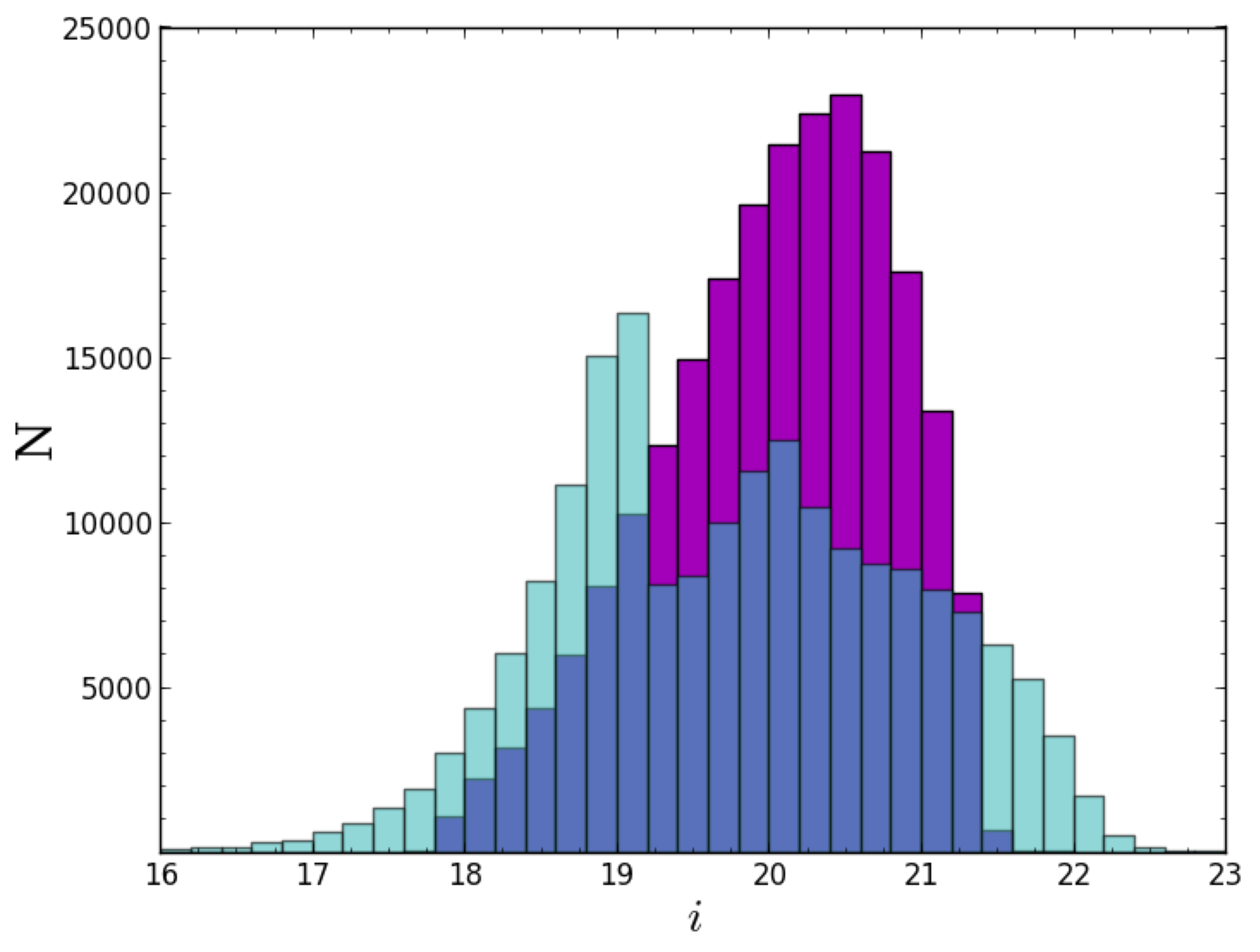}
\caption{Histograms of the $i$-band magnitude distribution for all four samples (as outlined in Fig.~\ref{fig:f1}). Again, note that the two plots use different y-axis scalings. The distribution of Sample A (left; yellow) reflects the different limiting magnitudes for $z<3$ and $z>3$ quasar selections in SDSS. Sample C (right; blue) demonstrates the extension to fainter magnitudes by the SDSS-III BOSS quasars, while Sample D (right; purple) probes more faint objects (while eliminating brighter objects where the efficiency of photometric quasar selection is lower due to higher stellar density).}\label{fig:f2}
\end{figure}

Figures \ref{fig:f1} and \ref{fig:f2} show histograms of the redshift
distribution and $i$-band magnitude distribution of Samples A, B, C,
and D. Sample B will be used for our primary analyses. The other three
samples (particularly Sample D) enable us to provide guidance on how
the radio properties change with redshift, luminosity, and apparent
magnitude beyond the limits of Sample B.

\subsection{Diagnostics}
\label{sec:diag}

We present diagnostics of the radio and optical properties of our
quasar samples as they relate to determining how any biases might
complicate our understanding of the physics of quasar radio emission
through better demographical analyses.  These analyses have been
performed on all four samples that we defined above, but results will
only be shown for Samples B and D.  Based on this analysis, we will
limit our discussion in Section~\ref{sec:results} to Sample B, noting
where the other samples provide additional information.

\subsubsection{Optical Luminosity and Redshift}

\begin{figure}[h!]
\epsscale{0.6}
\plotone{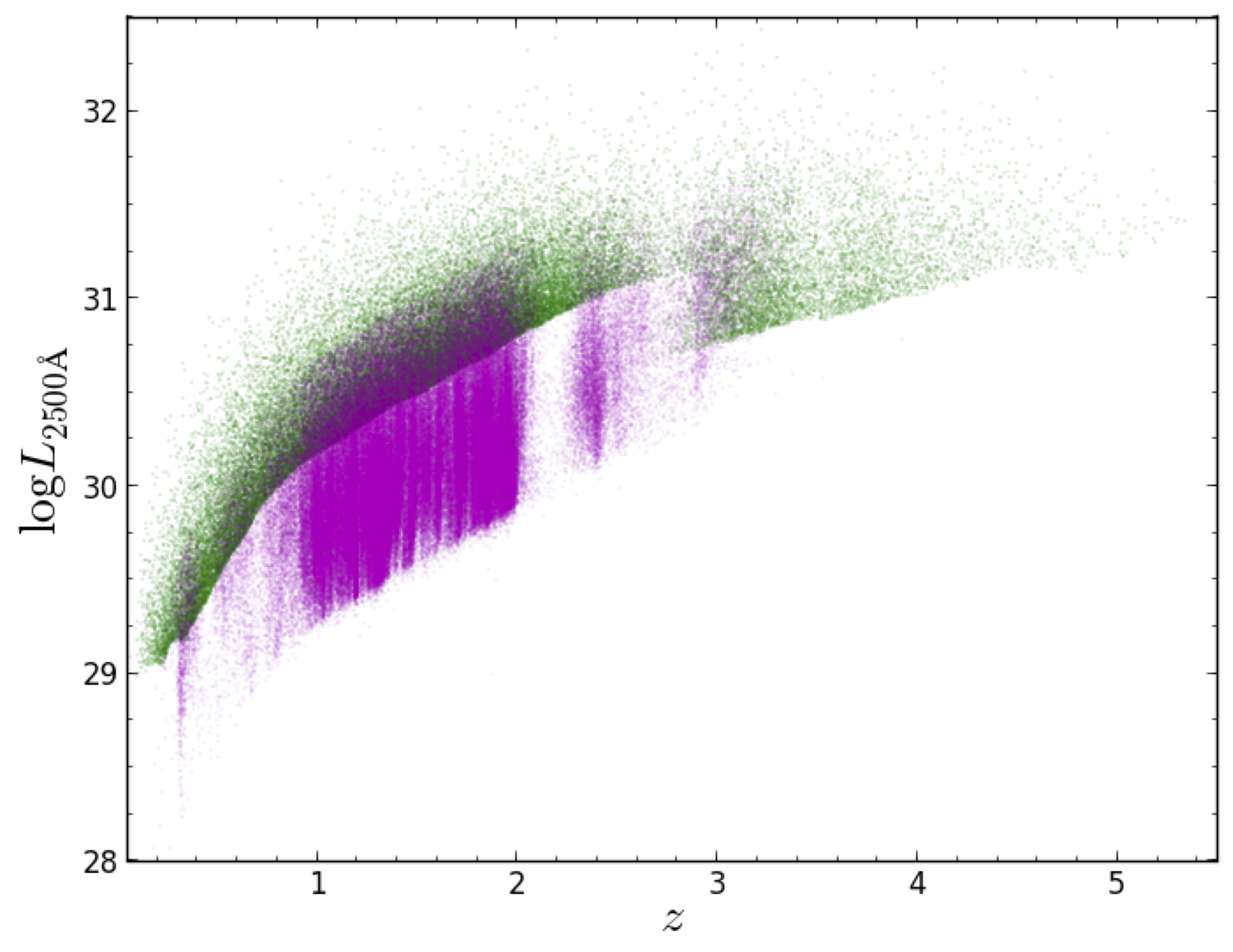}
\caption{\ltwofive\ corrected to $z=2$ as a function of redshift. The cuts made to Sample A to create Sample B (green; 55k quasars) has the effect of removing the faintest sources with $z<3$. After making cuts to Sample D (purple; 210k quasars) to create a relatively uniform subsample, photometric redshift degeneracies manifest themselves as bands of missing objects.}
\label{fig:f3}
\end{figure}

Figure \ref{fig:f3} shows the relationship between redshift and
$K$-corrected \ltwofive\ for Samples B and D.  Two
issues arise with regard to these parameters. First is the
flux-limited nature inherent to blind surveys: because of our samples'
fixed magnitude limits, there is an inherent degeneracy between the
redshifts and luminosities of the objects in our samples. Thus, some
caution is needed to ensure that, for example, an observed
characteristic of high-luminosity objects is not instead an inherent
characteristic of high-redshift objects. As such, in our analyses in
Section~\ref{sec:Lz}, we will consider the radio-loudness of quasars
as a function of both properties simultaneously.

The optical luminosity itself is subject to its own corrections.  Specifically,
$K$-corrections need to be applied to ensure that we are comparing
fluxes emitted in the same rest-frame wavelength range as opposed to
fluxes received in the same observed-frame wavelength range.
Traditionally, $K$-corrections are applied to extrapolate the power
emitted by the object in its rest frame within the filter's
bandpass. \citet{Richards06}, adopting \citet{Wisotzki2000} and
\citet{Blanton03}, argue against $K$-correcting to $z=0$ since most
quasars are not found in the local universe. To decrease the errors
associated with extrapolating from high redshift objects to $z=0$, we
follow \citet{Richards06} and $K$-correct closer to the median
redshift of our samples, $z=2$.

Lastly, we consider the redshift differences between the subsamples
(as seen in both Figures~\ref{fig:f1} and \ref{fig:f3}).  Sample B
extends to $z \approx 5.5$ with $29 < \log{L_{2500\, \rm \AA}}< 33$,
where the uniform restriction means that faint sources with $z < 2.7$
have been removed. Sample D fills in at $z\sim2.7$ and represents our
efforts to create a larger, relatively uniform sample. However, we do
so at the expense of certain redshift regions: the bands of missing
objects in Sample D indicate where photometric redshift degeneracies
exist.

\subsubsection{Radio-Loudness}

\begin{figure}[h!]
\epsscale{0.6}
\plotone{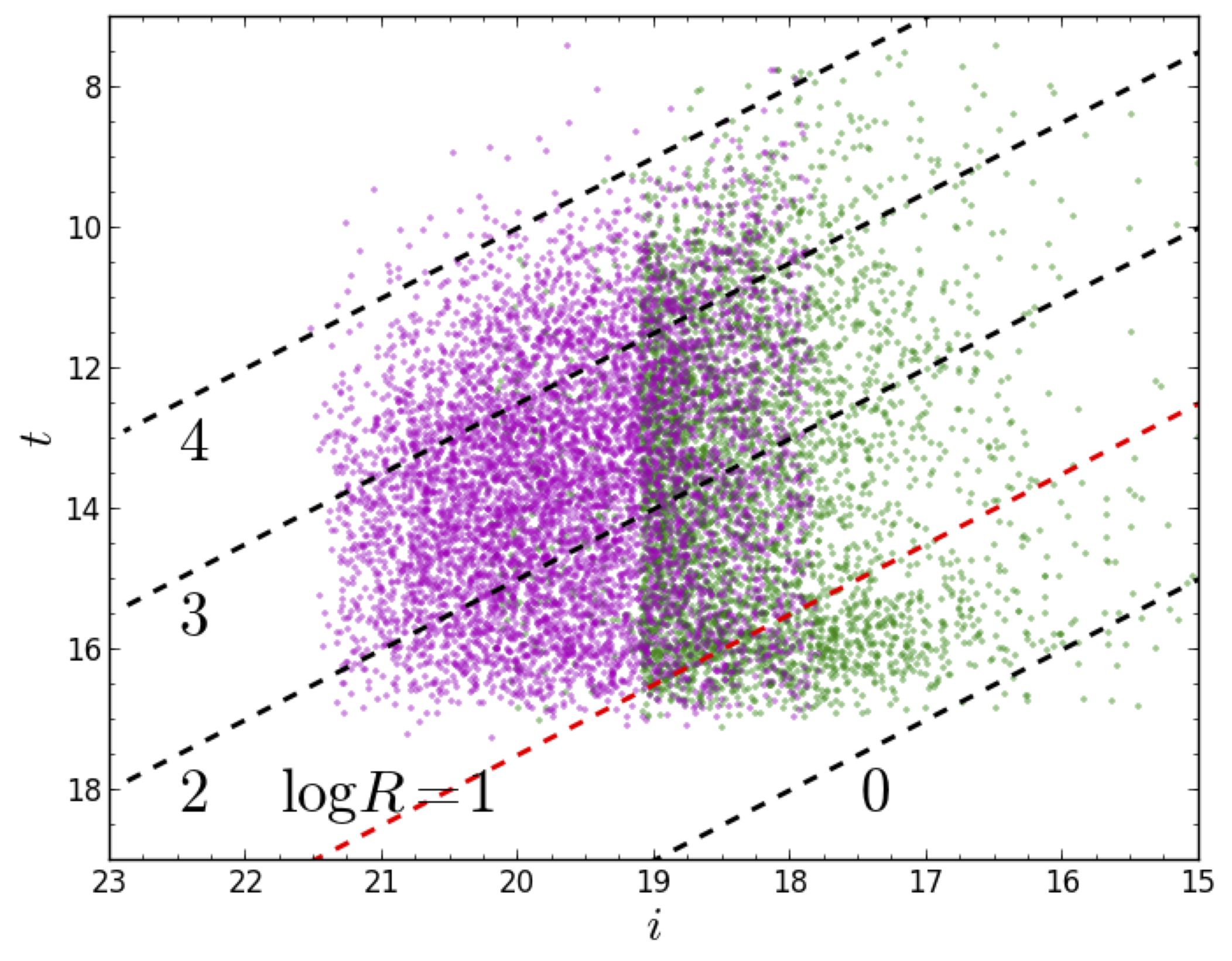}
\caption{Radio magnitude ($t$) of detected FIRST sources as a function of $i$ (Sample B: green, 4590 quasars; Sample D: purple, 6580 quasars). The dashed diagonal lines represent different values of the radio-loudness parameter, $\log{R}$, with the boundary between RL and RQ, $\log{R} = 1$, highlighted in red. It is apparent that the radio flux limit significantly restricts the distribution of $\log{R}$ to predominately RL values (for radio {\em detected} sources).}\label{fig:f4}
\end{figure}

The standard definition of radio-loud is based on the ratio of radio to optical fluxes according to
\begin{equation}
R = \frac{f_{6 \, \rm{cm}}}{f_{4400 \, \rm{\AA}}},
\label{eq:R}
\end{equation}
where $f_{6 \, \rm{cm}}$ is the 6 cm (5 GHz) measured radio flux,
$f_{4400 \, \rm{\AA}}$ is the $4400 \, \rm{\AA}$ measured optical
flux, and sources are considered radio-loud if $\log{R} > 1$
\citep{Schmidt70,Kellermann89}.  As emphasized by \citet[Section
4.2]{Ivezic}, the distribution of measured $\log{R}$ values can be
significantly affected by both the optical and radio flux limits of a
survey.  As such, we examine these properties from our samples in
Figure~\ref{fig:f4}. In this plot, the total integrated radio flux
(see Section \ref{sec:Match}) of each object is converted to a radio
magnitude, $t$, in terms of the AB magnitude scale
\citep{OkeGunn83}. From \citet{Ivezic},
\begin{equation}
t = -2.5 \log \left( \frac{f_{\rm int}}{3631 \, \rm Jy} \right).
\label{eq:t}
\end{equation}

The maximum radio magnitude of $t=16.4$ corresponds to the 1 mJy
detection limit of FIRST. The SDSS survey limits of $i=19.1$ (for low
redshift) and $i=20.2$ (for high redshift) \citep{Richards02a} are
readily seen in Sample B. The lines of constant $\log{R}$
show that these magnitude limits have a direct effect on the possible
values of $\log{R}$ that can be measured for a given data set. For
this reason, \citet{Ivezic} suggested exploring the histogram of
$\log{R}$ values in a parameter space that runs perpendicular to the
lines of $\log{R}$ within regions that are not bounded by the apparent
magnitude limits (see Figure 19 of \citealt{Ivezic}).  Note that only
Sample B can be considered reasonably complete to radio-loud sources
at the boundary of its optical flux limit ($i<19.1$).

\begin{figure}[h!]
\epsscale{0.6}
\plotone{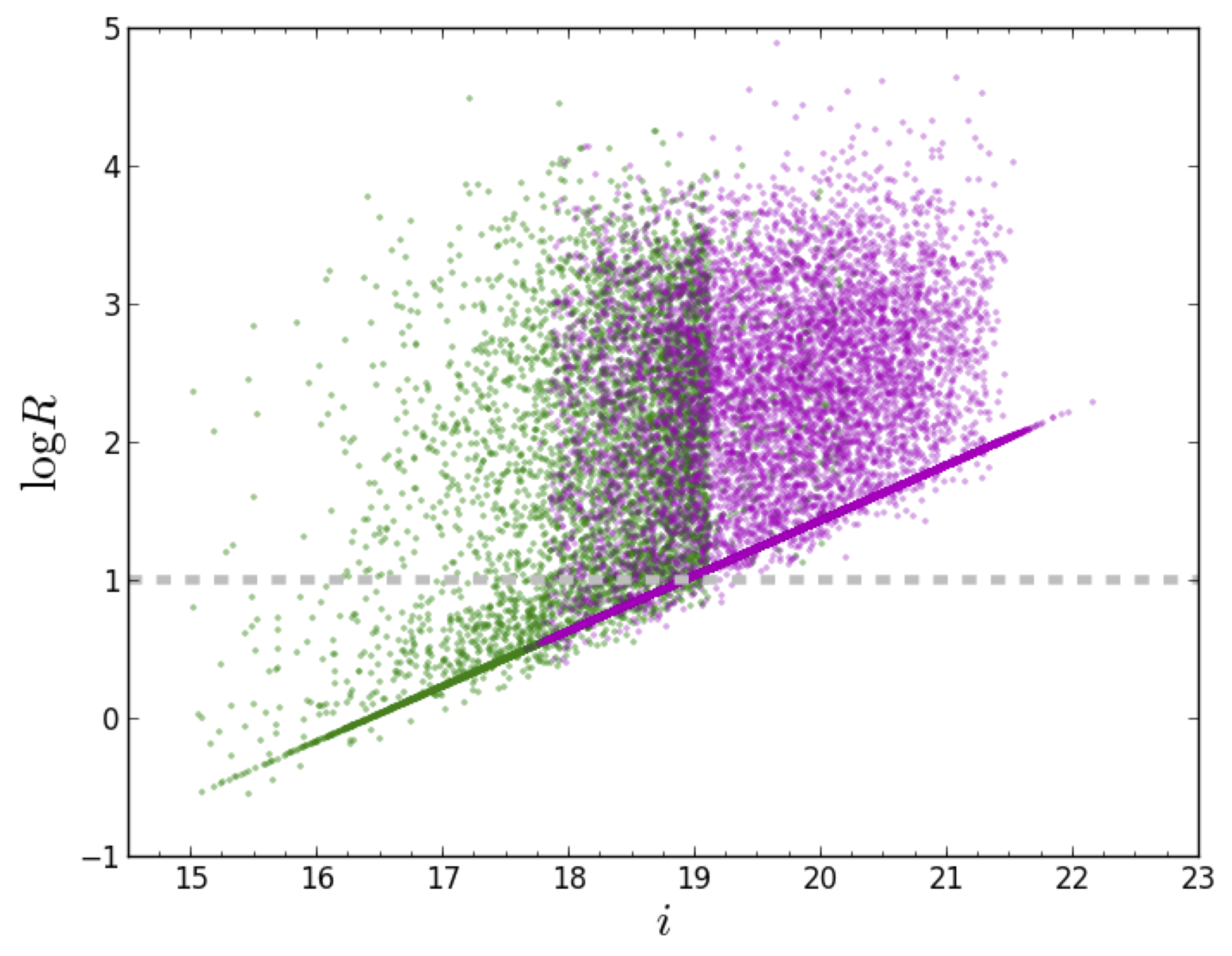}
\caption{Radio-loudness, $\log R$, as a function of $i$ for optically confirmed quasars within the FIRST observing area (Sample B: green; Sample D: purple).  Quasars undetected by FIRST are assigned a flux of 1 mJy to calculate $\log R$. Quasars undetected by FIRST at $i<18.9$ can be considered as RQ, but (optically) fainter objects can still be RL despite being undetected by FIRST.}
\label{fig:f5}
\end{figure}

Figure~\ref{fig:f5} illustrates a further limitation of the radio
loudness parameter, $\log{R}$, by showing how it depends upon optical
$i$-band magnitude. We calculated $\log R$ using
Equation~\ref{eq:R}; if a particular object is within the FIRST
observing area but has not been detected, $\log R$ is computed using
the FIRST detection threshold flux of 1 mJy; this results in the
artificial diagonal lines in the plots. The conventional division
between RL and RQ ($\log R = 1$; \citealt{Kellermann89}) is plotted as
a horizontal dashed gray line. Figure \ref{fig:f5} illustrates that
all of our samples exhibit RL incompleteness for objects fainter than
$i \approx 18.9$. At fainter magnitudes, it is quite possible for an
object to be intrinsically radio-loud but remain undetected by FIRST.
On the other hand, objects brighter than $i=18.9$ that are not
detected by FIRST should be classified as RQ even if they are
eventually detected in the radio at a lower flux limit (modulo the
incompleteness near the FIRST flux limits as discussed in Section
\ref{sec:firstlim}).  Our analysis of the RLF will concentrate on
Sample B as both Figures~\ref{fig:f4} and \ref{fig:f5} show that
non-detections in the radio for Sample D could still be formally radio
loud.  A survey to 10$\times$ the depth of FIRST (or $\sim15\,\mu$Jy
at 20\,cm) would be needed to detect all radio-loud quasars at the
depth of SDSS photometry.

\subsubsection{Radio Luminosity}
\label{sec:RadLum}

\begin{figure}[h!]
\epsscale{0.6}
\plotone{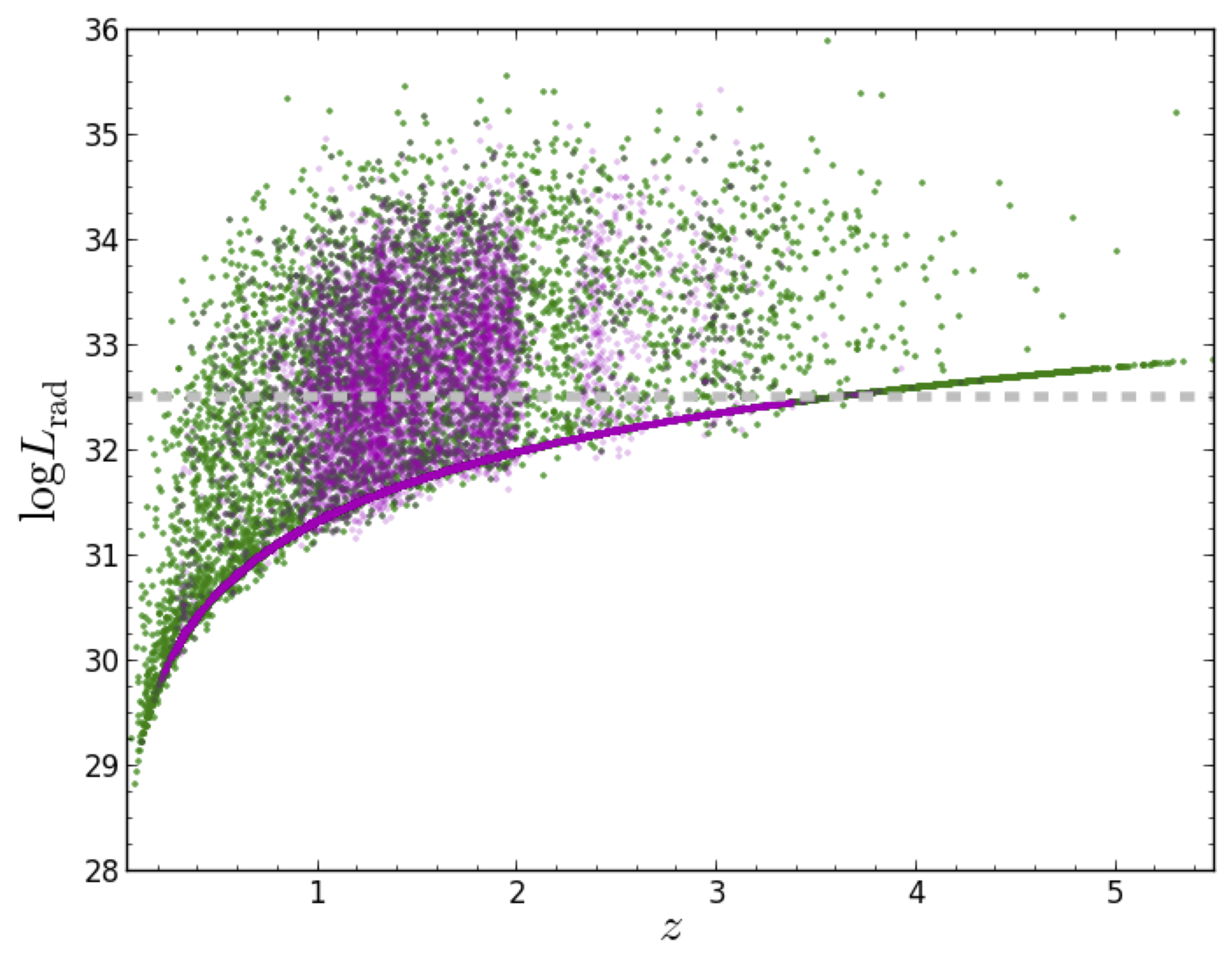} 
\caption{Radio luminosity as a function of redshift for optically confirmed quasars within the FIRST observing area (Sample B: green; Sample D: purple).  Quasars undetected by FIRST are assigned a flux of 1 mJy to calculate $L_{\rm rad}$. The horizontal dashed gray line denotes a division between RL and RQ quasars employed by \citet{Jiang07}. It is evident that high-redshift quasars ($z \ga 3.5$) can simultaneously be intrinsically radio-loud {\em and} FIRST non-detections. Alternatively, FIRST non-detections for lower redshifts ($z \la 3.5$) strongly indicate that an object is RQ.}
\label{fig:f6}
\end{figure}

Using the ratio of radio and optical fluxes as a measure of radio
loudness is preferred if those parameters are correlated.  If, on the
other hand, the radio and optical fluxes do not depend on one another,
an absolute radio flux or power is a more significant boundary between
RL and RQ quasars \citep{Peacock86, Miller90, Ivezic};
\citet{Goldschmidt99} used $P_{5 \rm GHz} = 10^{24}$ W $\rm{Hz}^{-1}
\, \rm{sr}^{-1}$ or $L_{20 \rm cm} = 10^{31}$ erg $\rm{s}^{-1} \,
\rm{Hz}^{-1}$ as the limit between RL and RQ.  As such, examination of
the radio luminosity distributions leads to additional biases in our
samples that must be considered. Figure \ref{fig:f6} illustrates how
radio luminosity depends on redshift for Samples B and D.

As with Figure~\ref{fig:f5}, the FIRST flux limit is
obvious in this plot and demarcates what redshifts (as opposed to
fluxes) beyond which our sample is incomplete to radio-loud
quasars. An alternate boundary (instead of $\log R$) between RL and RQ
quasars utilized by \citet{Jiang07}, $L_{20 \rm cm} = 10^{32.5}$ erg
$\rm{s}^{-1} \, \rm{cm}^{-2} \, \rm{Hz}^{-1}$, depends only on radio
luminosity and is denoted with a horizontal dashed gray line. In all
four samples, RL incompleteness exists above $z \ga 3.5$. That is, as
in Figure \ref{fig:f5}, it is possible for a high-redshift quasar to
be intrinsically radio-loud, but still not be detected in FIRST. Thus,
we should limit our most robust radio-loud fraction analysis to
redshifts lower than this.  On the other hand, a FIRST non-detection
for lower redshifts is a strong indication that the object is
radio-quiet.  In Section \ref{sec:RLF} we will take advantage of this
fact by treating FIRST non-detections (brighter than $i=18.9$ and with
$z<3$) as confirmed RQ objects.

Furthermore, as with the optical, the spectral indices of quasars in
the radio have a fairly large range, spanning at least
$-1<\alpha_\nu<0$.  Since our samples cover a large range of redshift,
they must also span a large range of the rest-frame radio spectrum. As
such, $K$-corrections to the rest-frame wavelength are important to
consider. In a manner similar to \ltwofive\ (see Figure \ref{fig:f3}), we follow
\citet{Richards06} and define an equivalently $z=2$ $K$-corrected
radio luminosity as
\begin{equation}
\frac{L_{\rm{rad}}}{4 \pi \, LD^2} = f_{\rm{int}} \,10^{-23} \, \frac{(1+2)^{\alpha_\nu}}{(1+z)^{(1+\alpha_\nu)}}
\label{eq:RadLum}
\end{equation}
where $L_{\rm{rad}}$ is measured in $\rm{erg} \, \rm{s}^{-1} \,
\rm{Hz}^{-1}$, luminosity distance (LD) is measured in cm, integrated
radio flux is measured in Jy, and the redshifts were taken from the
optically detected objects. Here $\alpha_\nu$ is the radio spectral
index, and we use $\alpha_\nu=-0.5$ for the entirety of this analysis
as we have a combination of flat-spectrum ($\alpha_\nu\sim 0$) and
steep-spectrum ($\alpha_\nu\sim -1$) sources in our samples.
\citet{Kimball08} provide spectral indices for individual sources;
however, non-simultaneity means that variability can skew the values.
If simultaneous radio flux measurements in two bandpasses were
available, it would be preferable to use radio spectral indices
measured for each individual object.  Figure~14 in \citet{Richards06}
illustrates how much error is induced by the wrong choice of spectral
index, shows how the K-correction to $z=2$ serves to minimize that (for a
population that peaks closer to $z=2$ than $z=0$), and suggests that an incorrect choice of spectral index should not have a large impact on our analyses.  Note that
this choice of $K$-correction means that any sample that uses the
radio luminosity to define RL quasars will be biased towards including
flatter spectrum (larger $\alpha$) sources at $z>2$ and steeper
spectrum (more negative $\alpha$) sources at $z<2$.

\subsubsection{Extended Flux Underestimation}
\label{sec:RadExt}

A serious issue to consider when using integrated fluxes is that these
measurements are underestimated for resolved FIRST sources ($>$ 10")
\citep{FIRST95}. The analysis by \citet{Jiang07} ignores this possible
complication, asserting that these highly extended radio sources are rare and so bright that, despite the underestimation of integrated flux, they will undoubtedly be considered RL.

\begin{figure}[h!]
\epsscale{0.6}
\plotone{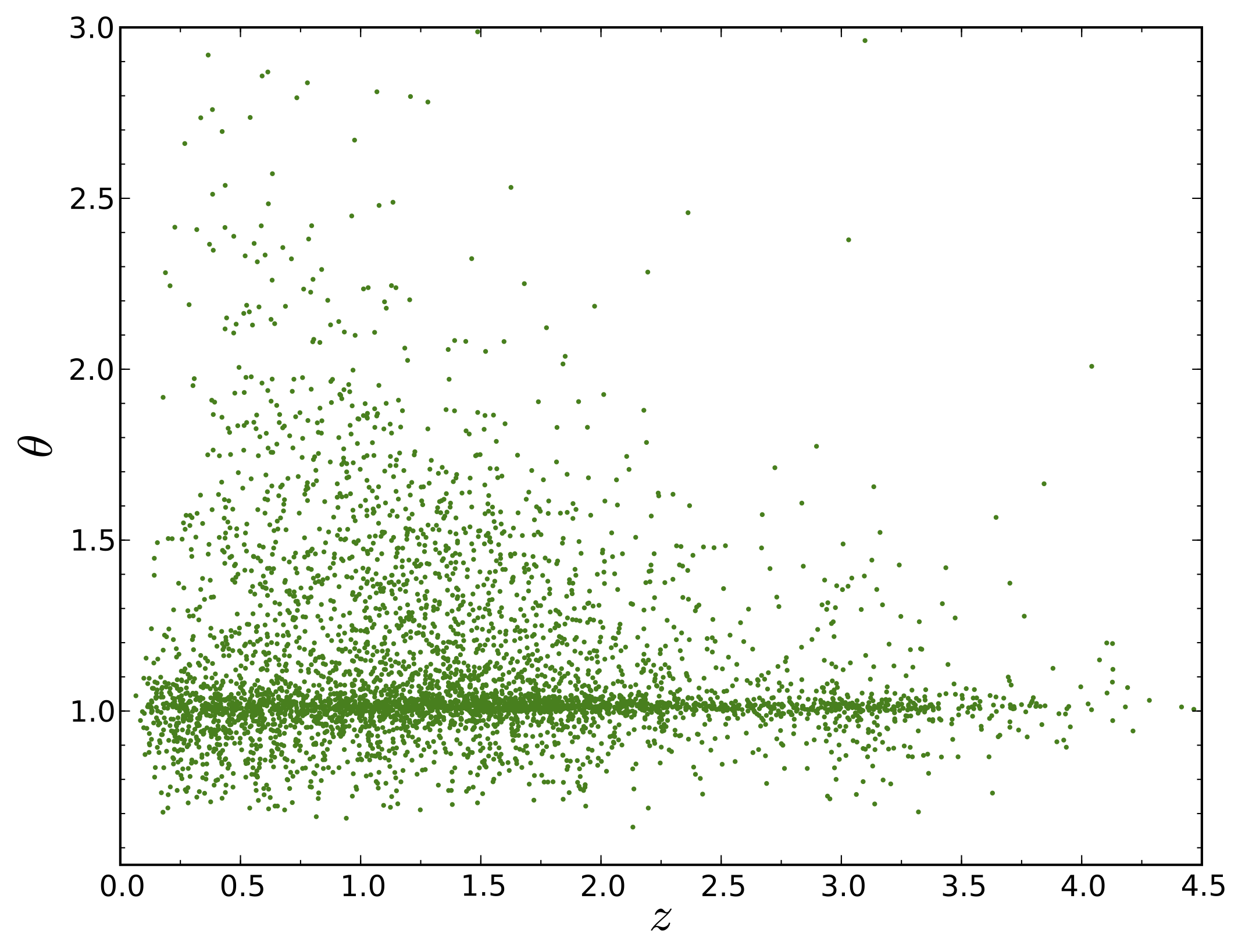}
        \caption{Source size, indicated by the ratio of the integrated to peak flux ($\theta$), as a function of redshift for FIRST-detected quasars in Sample B.  Beyond $z=1.5$, the relative fraction of extended sources falls.  However, this is not a large effect for our analysis as the majority of our sources are not resolved.}
\label{fig:f7}
\end{figure}

One way to characterize this effect is to plot the ratio of the
integrated to peak fluxes as a function of redshift. Here we use
$\theta^2 = (f_{\rm{int}}/f_{\rm{peak}})$ as defined by
\citet{Ivezic}, where $\theta > 1$ for an extended source. In Figure
\ref{fig:f7}, we see the effects of surface brightness dimming which
goes as $(1+z)^4$. Some of the apparent fall-off with redshift is
simply due to the declining number of sources, but it does appear that
at the highest redshifts ($z\gtrsim1.5$), extended sources are being preferentially
lost. 
However, we emphasize that the (relatively)
high frequency of the FIRST observations already biases the sample
towards unresolved objects and reiterates the claim by \citet{Ivezic}
that the fraction of complex sources is small within the FIRST
sample. Thus, while we are not complete to quasars with extended radio
emission, those objects are not dominating our sample, even at low
redshift, and should not influence any trends with redshift.  See the
next section and both \citet{Bondi2008} and \citet{Hodge2011} for
further discussion of resolution incompleteness.

\subsubsection{FIRST Detection Limit}
\label{sec:firstlim}

Our final demographical analysis involves the FIRST detection
limit, specifically how much this limit varies from the nominal limit
of 1\,mJy and why.

The depth to which FIRST can detect a source depends on sky position;
being in the proximity of a bright object and the systematic increase
in noise for lower declinations complicate FIRST's sensitivity
\citep{FIRST95}. In addition, the radio detection limit
for FIRST is calculated using peak fluxes; this makes it difficult to
accurately account for extended sources whose radio emission could be
distributed throughout various components that may or may not exceed FIRST's detection threshold \citep{FIRST95, White07}. Therefore, the
source counts with radio fluxes near the 1 mJy detection limit are
incomplete, with extended sources being the most incomplete.

\begin{figure}[h!]
\epsscale{0.5}
\plotone{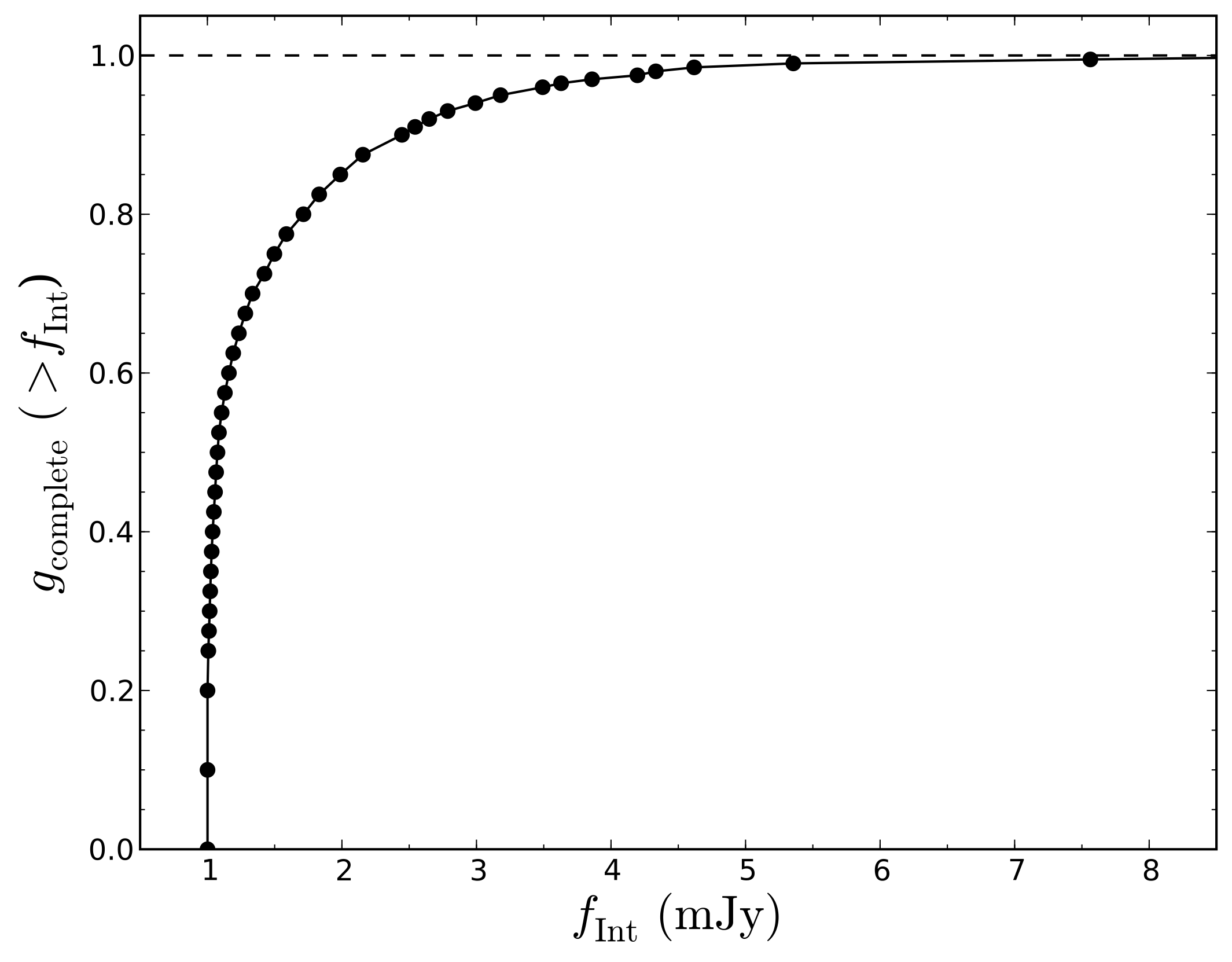}
\caption{FIRST completeness (provided by R. L. White 2013, private communication) as a function of integrated radio flux (mJy).  The incompleteness at the NVSS flux limit estimated here is consistent with that determined by \citet{Ivezic}.}
\label{fig:f8}
\end{figure}

The completeness of FIRST is shown in Figure \ref{fig:f8} as a
function of integrated flux. The dots represent discrete values
communicated by R. L. White (2013), and the solid line shows the
linear fit between adjacent points that we used to interpolate
completeness percentages. To compute the completeness efficiency,
R.L. White (2013) used the measured size distribution of detected
quasars.  Based on this figure, we can see that FIRST suffers
significant incompleteness above what is normally considered the
``detection limit''.

A concern is that any analysis probing to fluxes close to the nominal
detection limit will suffer due to the relative uncertainty of the
incompleteness correction near the limit.  That said,
\citet[Section~3.9]{Ivezic} found that FIRST is not more than 13\%
incomplete at the NVSS (NRAO VLA Sky Survey; \citealt{NVSS}) flux
limit of $\sim$2.5\,mJy, which is consistent with
Figure~\ref{fig:f8}. We will further discuss how this incompleteness
could affect our results in Section~\ref{sec:RLF}.

\subsection{Radio-Loud Definition}
\label{sec:aro}

\begin{figure}[H]
\epsscale{0.75}
\plotone{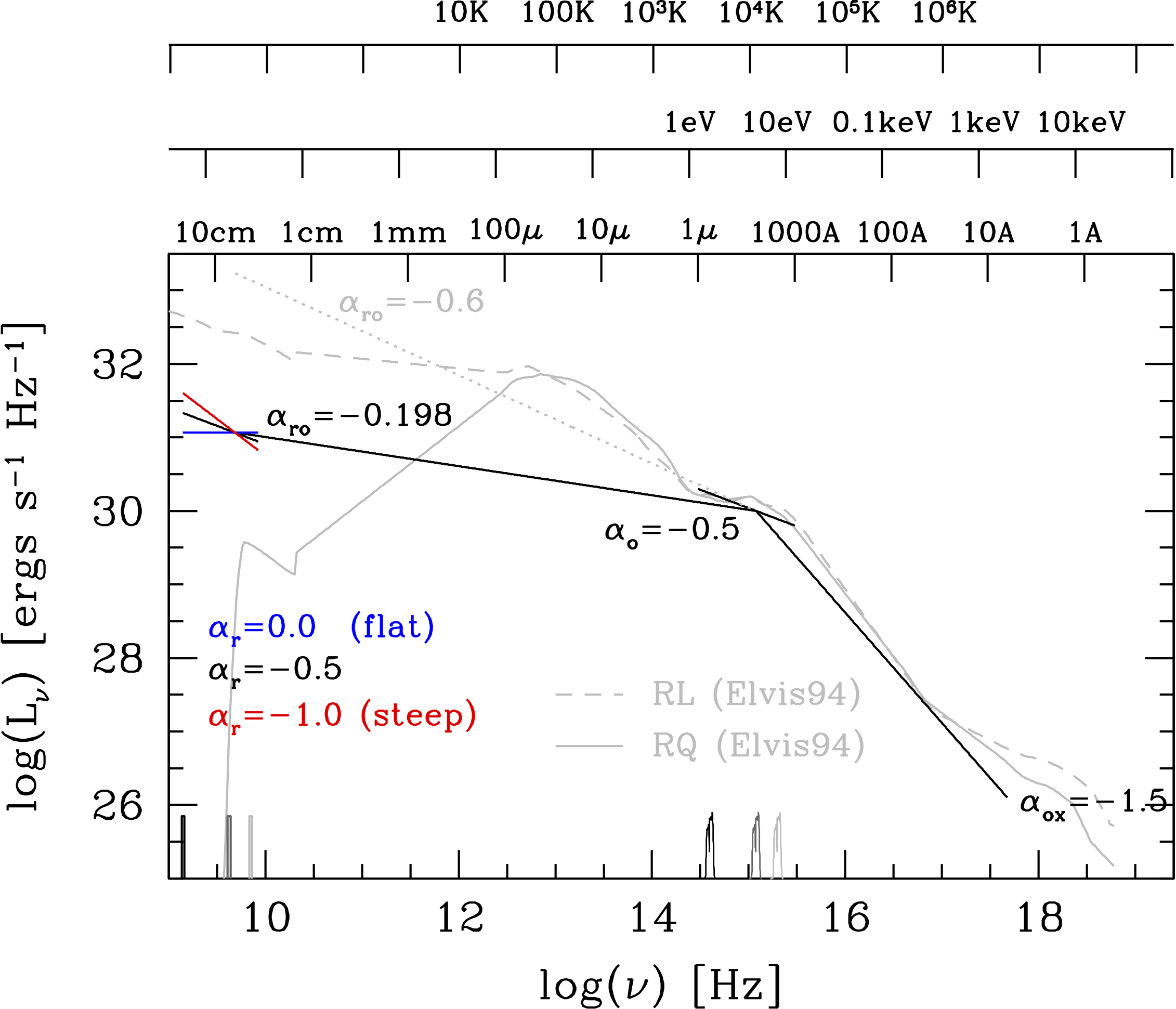}
\caption{Spectral energy diagram comparing the distribution of power in the radio, optical, and X-ray regimes.  The black lines show the mean radio-to-UV-to-X-ray SED of a quasar with \ltwofive\ $= 30$ ergs $\rm{s}^{-1} \, \rm{Hz}^{-1}$. \aro\ and \aox\ give the slope of the SED between the radio and optical and the optical and X-ray. \aro\ is not universal for quasars; here we have shown the slope corresponding to the traditional division between RL (steeper slopes) and RQ (flatter slopes). The red and blue lines show the range of radio slopes in that part of the SED. The solid and dashed gray curves show the mean RQ and mean RL SEDs, respectively, from \citet{Elvis94}. The dotted gray line shows a slope equivalent to $\log R=3$---particularly radio-loud. At the bottom of the panel we show the transmission of the 1.4GHz and i-band bandpasses at $z=0$, 2, and 4 (as black, dark gray, and light gray, respectively), demonstrating that, at $z=2$, the 1.4GHz and i-band bandpasses are close to 5GHz and $2500$\AA.}
\label{fig:f9}
\end{figure}

Before we begin our analysis of the data, it is worthwhile to review
the mean spectral energy distribution (SED) of quasars and to consider the
definition of a radio-loud quasar in the context of the broader
quasar SED. Figure~\ref{fig:f9} shows multiple quasar SEDs to help illustrate the difference between RL and RQ. A radio-loud definition based on luminosity would
mean simply making a cut along some constant value of the $y$-axis. A
typical value would be at $\log L_{\rm rad} = 32.5$ ergs
$\rm{s}^{-1} \, \rm{Hz}^{-1}$. However, as discussed by
\citet[Appendix C]{Ivezic} and \citet{Balokovic}, the radio luminosity
is the best indicator of radio-loudness only if the radio and optical
luminosities are {\em not} correlated.  As \citet{Balokovic}
demonstrates that these properties are indeed correlated, it means
that it is arguably more appropriate to consider the {\em ratio} of
the radio and optical luminosities.

Indeed, as noted above, the most common criterion used to classify
quasars as RL or RQ is the $R$ parameter \citep{Kellermann89}, which is
just the ratio of the radio (6 cm) and optical (4400 $\rm \AA$)
fluxes. While $R$ and $\log R$ \citep{Ivezic} have a long history in the
literature and are familiar to radio astronomers, the quasar field has
become much more dependent on multi-wavelength data. As such, it is
important to adopt terminology that is not specific to certain
wavebands (e.g., $\log R$ in the radio or the energy index, $\Gamma$,
in the X-ray), but rather terminology that spans the entire electromagnetic
spectrum.  Given the common usage of units that are related to ergs
$\rm{s}^{-1} \, \rm{cm}^{-2} \, \rm{Hz}^{-1}$, a logical choice is the
slope in $\log f_{\nu}$ vs.\ $\log \nu$ space, $\alpha$, where
$f_{\nu} \propto \nu^{\alpha}$ (as shown in Figure~\ref{fig:f9},
except in luminosity units).

In our work we will consider the radio-to-optical spectral index, \aro, rather than $\log R$, where we define \aro\ according to 
\begin{equation}
\alpha_{\rm ro}  = \frac{\log(f_{\rm 5\,GHz}/f_{\rm 2500\,\AA})}{\log(\nu_{\rm 5\,GHz}/\nu_{\rm 2500\,\AA})}= \frac{\log(f_{\rm 5\,GHz}/f_{\rm 2500\,\AA})}{\log(\lambda_{\rm 2500\,\AA}/\lambda_{\rm 5\,GHz})}= -0.186\log(f_{\rm 5\,GHz}/f_{\rm 2500\,\AA}),
\label{eq:alpha}
\end{equation}
or more practically by considering the ratio of radio luminosity to
optical luminosity. We have chosen the wavelength of $2500$\,\AA
\,because it is the same as is used in X-ray investigations for
comparisons with the optical/UV and represents the $i$-band at
$z=2$. We have also chosen the frequency of 5\,GHz because it is the
historical value used in the radio and roughly corresponds to the
frequency of the 1.4GHz (20 cm) FIRST data at $z=2$ (see below).

The values of $\alpha_{\rm ro}$ and $\log R$ are effectively
equivalent if the frequencies sampled are the same, but using a slope
(rise over run) instead of just the flux ratio (rise only) allows the
use of data at other wavelengths/frequencies without having to apply
significant corrections.  In other words, $\alpha_{\rm ro}$ is more
flexible than $\log R$.  For the sake of backwards compatibility with
previous work, the radio-to-optical spectral index, $\alpha_{\rm ro}$,
can be related to the traditional $\log R$ parameter as follows
\citep[e.g.][]{Wu12}:

\begin{equation}
R = \frac{f_{\rm 5\,GHz}}{f_{\rm 4400\,\AA}}
\end{equation}
where
\begin{equation}
\frac{f_{\rm 5\,GHz}}{f_{\rm 2500\,\AA}} = \frac{f_{\rm 5\,GHz}}{f_{\rm 4400\,\AA}}\frac{f_{\rm 4400\,\AA}}{f_{\rm 2500\,\AA}}
\end{equation}
and
\begin{equation}
\frac{f_{\rm 4400\,\AA}}{f_{\rm 2500\,\AA}} = \left(\frac{\nu_{\rm 4400\,\AA}}{\nu_{\rm 2500\,\AA}}\right)^{\alpha_{\rm opt}}=\left(\frac{\lambda_{\rm 2500\,\AA}}{\lambda_{\rm 4400\,\AA}}\right)^{\alpha_{\rm opt}}
\end{equation}
so that 
\begin{equation}
\alpha_{\rm ro}  = \frac{\log\left[R\left(\frac{\lambda_{\rm 2500\,\AA}}{\lambda_{\rm 4400\,\AA}}\right)^{\alpha_{\rm opt}}\right]}{\log(\nu_{\rm 5\,GHz}/\nu_{\rm 2500\,\AA})}
\end{equation}
which simplifies to
\begin{equation}
\alpha_{\rm ro}  = -0.186\left[\log R + \alpha_{\rm opt}\log\left(\frac{\lambda_{\rm 2500\,\AA}}{\lambda_{\rm 4400\,\AA}}\right)\right] = -0.186\left[\log R -0.246\alpha_{\rm opt}\right].
\end{equation}
For the mean optical spectral index from \citet{VDBerk01}
($\alpha_{\rm opt}=-0.44$, needed to extrapolate between 2500\AA \,and
4000\AA) this corresponds to
\begin{equation}
\alpha_{\rm ro}  = -0.186\log R - 0. 020.
\end{equation}
To help calibrate $\alpha_{\rm ro}$ to the $\log R$ system, it may help to note that the traditional
loud-quiet division ($\log R=1$) would be roughly $\alpha_{\rm
  ro}=-0.2$ and that $\alpha_{\rm ro}=-0.6$ would correspond to a very
radio-loud source ($\log{R} \sim 3$).

Throughout the rest of this work we will assume that the radio and
optical luminosities of quasars are correlated and, as such, will use
the radio-to-optical flux ratio as given by $\alpha_{\rm ro}$ (rather
than $\log R$) to distinguish between RL and RQ
sources with $\alpha_{\rm ro}<-0.2$ as the definition for RL
quasars.

\section{Methods}
\label{sec:methods}

Our analysis considers both the median radio properties of quasars
(through a stacking analysis) and the extreme radio properties of
quasars (using the fraction of objects in the radio-loud tail of the
distribution).  Here we explain in detail the methods used
in these analyses before comparing the results of these two methods in
Section~\ref{sec:results}.

\subsection{Radio Properties in the Extreme: the Radio-Loud Fraction (RLF)}
\label{sec:RLF}

We begin our analysis by investigating the radio-loud fraction (RLF),
which is the the percentage of quasars that have $\alpha_{\rm ro}<
-0.2$ ($\log R > 1$).  \citet{Jiang07} used a sample of more than
30,000 quasars to determine that the RLF increases with decreasing
redshift and increasing optical luminosity. Their results may mean
that the amount of radio emission with respect to that of the optical
may change as a function of these two parameters; however, it could
also suggest that the population densities of RL and RQ quasars evolve
with respect to one another.

\begin{figure}[t!]
\epsscale{1.0}
\plottwo{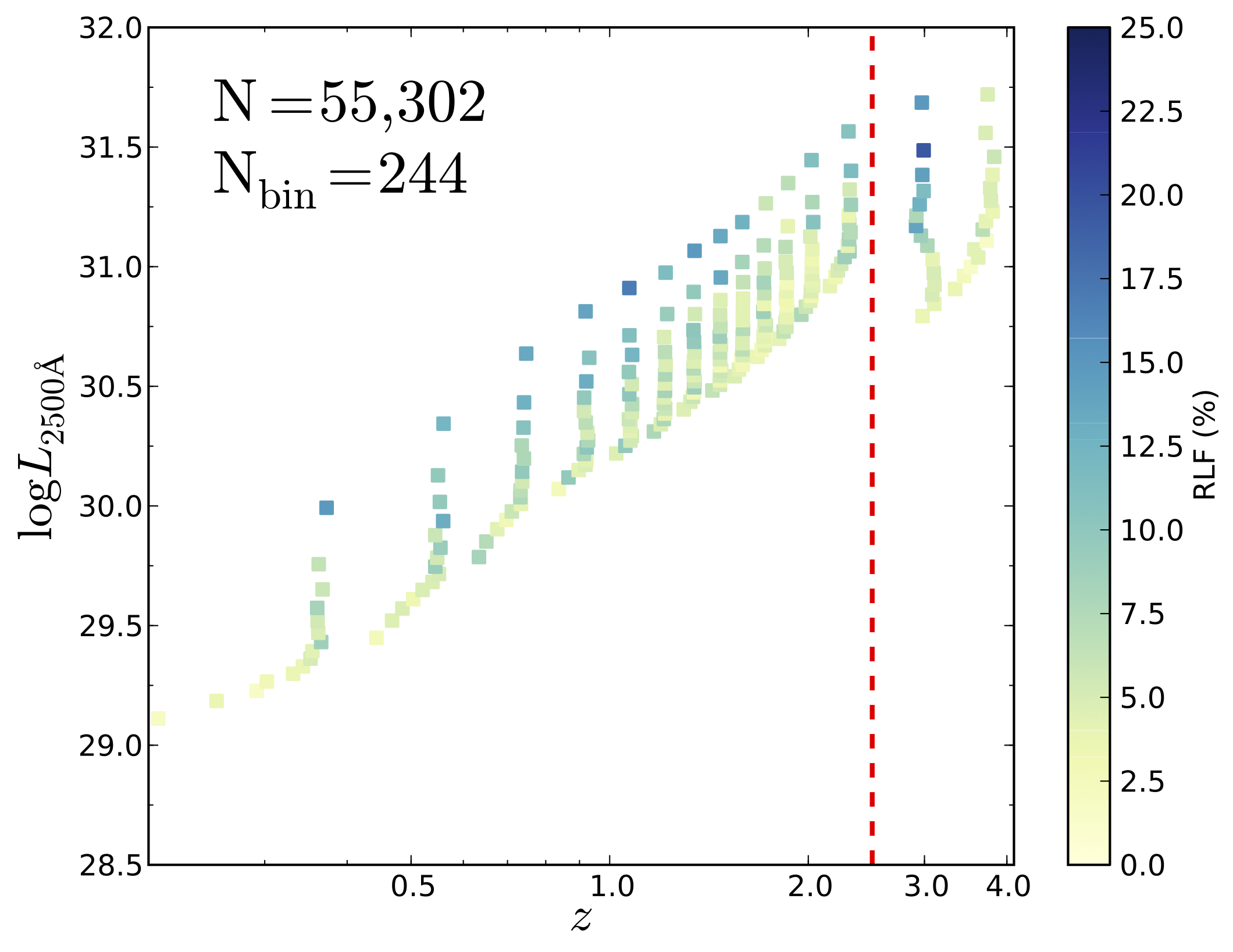}{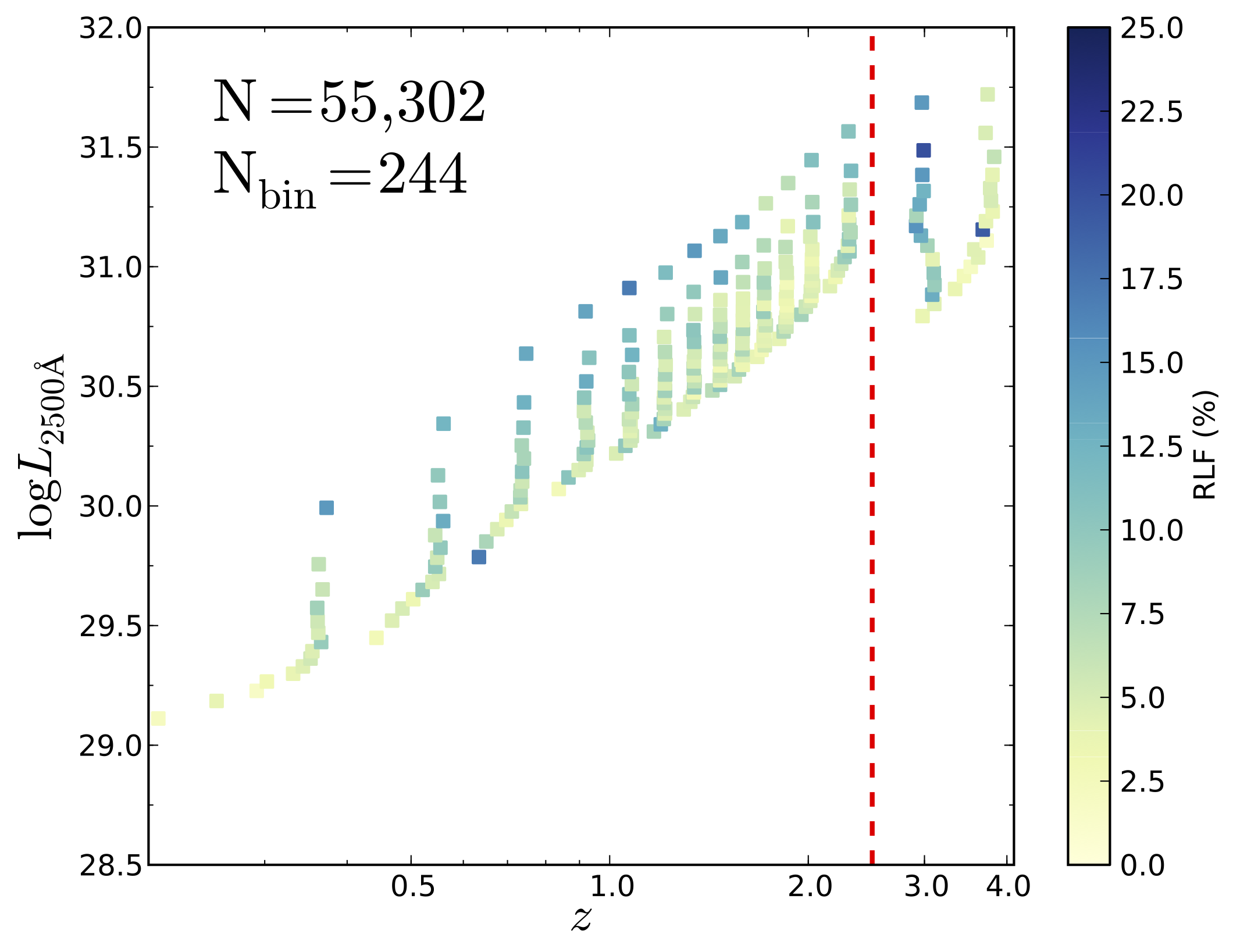}
\caption{RLF as a function of both \ltwofive\ and redshift for optically detected quasars within the FIRST observing area from Sample B. The plot on the left uses no correction when computing the number of RL objects within a bin, while the plot to the right uses the RL completeness correction (see Figure \ref{fig:f8}). Comparing the two, both plots show a declining RLF with increasing redshift and decreasing luminosity with the completeness corrected version (right) showing a less pronounced, yet still present, trend. The data to the right of the dashed red lines in both plots suffer the most from selection effects.}
\label{fig:f10}
\end{figure}

\citet{Jiang07} showed that examining the RLF in 2-D $L-z$ space
rather than the marginal distribution of $L$ and $z$ separately lead
to very different results.  We will perform the same analysis here
with a larger, more uniform sample.  Since redshift and luminosity are
degenerate properties in flux-limited surveys, we divide our samples
into equally populated bins within \ltwofive-$z$ space; this process
allows us to isolate changes due to just one of the variables.
Specifically, we first sort the quasars by redshift, dividing them
into a number of slices with an equal population of quasars contained
within each slice.  Then we sort the objects in each redshift slice by
luminosity and further bin the objects so that there are an equal
number of objects in each $L-z$ bin.  Quasars within a bin were
flagged RL if \aro\ $< -0.2$, and, initially, the RLF for each bin was
calculated by dividing the number of RL quasars by the total number of
objects for that bin; see Figure \ref{fig:f10} (left). The median $z$
and \ltwofive\ for each bin were used to plot the results, and the
color of each bin represents the RLF.

In order to correct for the incompleteness discussed in
Section~\ref{sec:firstlim}, we weigh each RL quasar that has a
measured integrated flux less than 10 mJy by its corresponding
completeness on our best fit function (Figure~\ref{fig:f8}; solid
line). For example, a RL object with an integrated radio flux of 1.075
mJy ($g_{\rm complete} = 0.500$) counts as two RL objects since
$1/g_{\rm complete} = 1/0.500 = 2$. Because the completeness function
drops off so quickly for integrated fluxes less than 1 mJy, all
detected RL objects with values smaller than this are scaled by
$g_{\rm complete} = 0.10$. RL quasars with integrated fluxes greater
than or equal to 10 mJy always count as one RL object, and the total
number of objects within each bin remains unchanged when computing the
RLF.

The plot on the right of Figure \ref{fig:f10} shows the dependence of
the RLF on both redshift and \ltwofive\ for Sample B after applying
the completeness correction. We see that the trend in RLF from the
upper-left to the lower-right is reduced but still present. Everything
to the right of the vertical red line located at $z \sim 2.5$ in both
plots denotes where the SDSS optical selection was very inefficient;
further care is required to fully understand our analysis beyond this
redshift. In short, the optical selection is less complete at this
redshift (compare panels a and b in \citealt[Fig. 6]{Richards06}), so
the quasars discovered at $z \sim 2.5$ are more likely to be radio
sources.

This type of analysis allows us to see how the RLF is changing as a
function of multiple parameters; we can then compare these results with the
mean radio properties of quasars in order to see if the direction of
change is the same for both methods.  Our analysis in
Section~\ref{sec:Lz} starts with the $L-z$ plane as shown here.  In later sections,
we will construct similarly binned samples using other observed
quantities, plotting some third parameter as a color-scale at the
median value of the $x$ and $y$ quantities.  Specifically, we also
consider the \ion{C}{4} blueshift and equivalent width (EQW) (Section~\ref{sec:CIV}), the
so-called ``Eigenvector 1'' parameter space (Section~\ref{sec:CIV}),
and the combination of BH mass and accretion rate
(Section~\ref{sec:BH}).

\subsection{Radio Properties in the Mean: Stacking Analysis}
\label{sec:Stacks}

\subsubsection{Image Stacking}
\label{sec:ImageStacking}

By stacking the radio images of all known quasars covered by the FIRST
survey, we hope to learn about the mean radio properties of these
objects.  We can then contrast these findings with the properties
identified using formally radio-loud quasars.  Our stacking analysis
follows that of \citet{White07}. For a more detailed explanation, see
that paper, but the process is briefly described here.

First, using the optical coordinates of our target quasar populations,
0$\farcm$5 x 0$\farcm$5 radio images were downloaded from the FIRST
data extraction website{\footnote
  {http://third.ucllnl.org/cgi-bin/firstcutout}}.  As with the RLF
analysis, we wish to explore the mean radio characteristics of quasars
as a function of various properties.  As such we will stack the radio
images in bins based on these parameter spaces (e.g., $L-z$,
Section~\ref{sec:Lz}; CIV and EV1, Section~\ref{sec:CIV};
BH properties, Section~\ref{sec:BH}; color,
Section~\ref{sec:StacksColor}).

After assigning each quasar to a 2-D parameter bin, all of the FIRST radio
images within each bin were added using a median stacking procedure
\citep[see][]{White07}: a pixel in the final stacked image corresponds
to the median value of the pixels occupying that same location from
the set of radio images within a bin. 
Since \citet{White07} show that the median converges to the mean for
distributions such as we consider herein, we will generally refer to
our median stacking results as the mean.

After combining the cutouts into stacked images, the peak flux values
of our stacked sources need to be corrected for what \citet{White07}
designate as ``snapshot bias", which appears to be related to the
well-known problem of ``clean bias'' associated with FIRST sources
\citep{FIRST95}.   \citet{White07} 
found that a correction of the form:
\begin{equation}
f_{\rm peak, \, \mathrm{corr.}} = \mathrm{min}(1.40 \, f_{\rm peak}, \, f_{\rm peak} + 0.25 \; \mathrm{mJy}),
\label{eq:Snapshot}
\end{equation}
is needed, where $f_{\rm peak}$ is the peak flux density (mJy) of the median
stack. The flux boundary that determines which part of the equation to
implement is 625 $\mu$Jy.  As that value is more than 200 $\mu$Jy
greater than the largest median peak flux density we achieve, we will
only need to multiply our measurements by 1.40 for the
entirety of our analysis to correct for this bias.

\subsubsection{Median Stacking Diagnostics}
\label{sec:MedStackDiag}

\begin{figure}[h!]
\epsscale{0.6}
\plotone{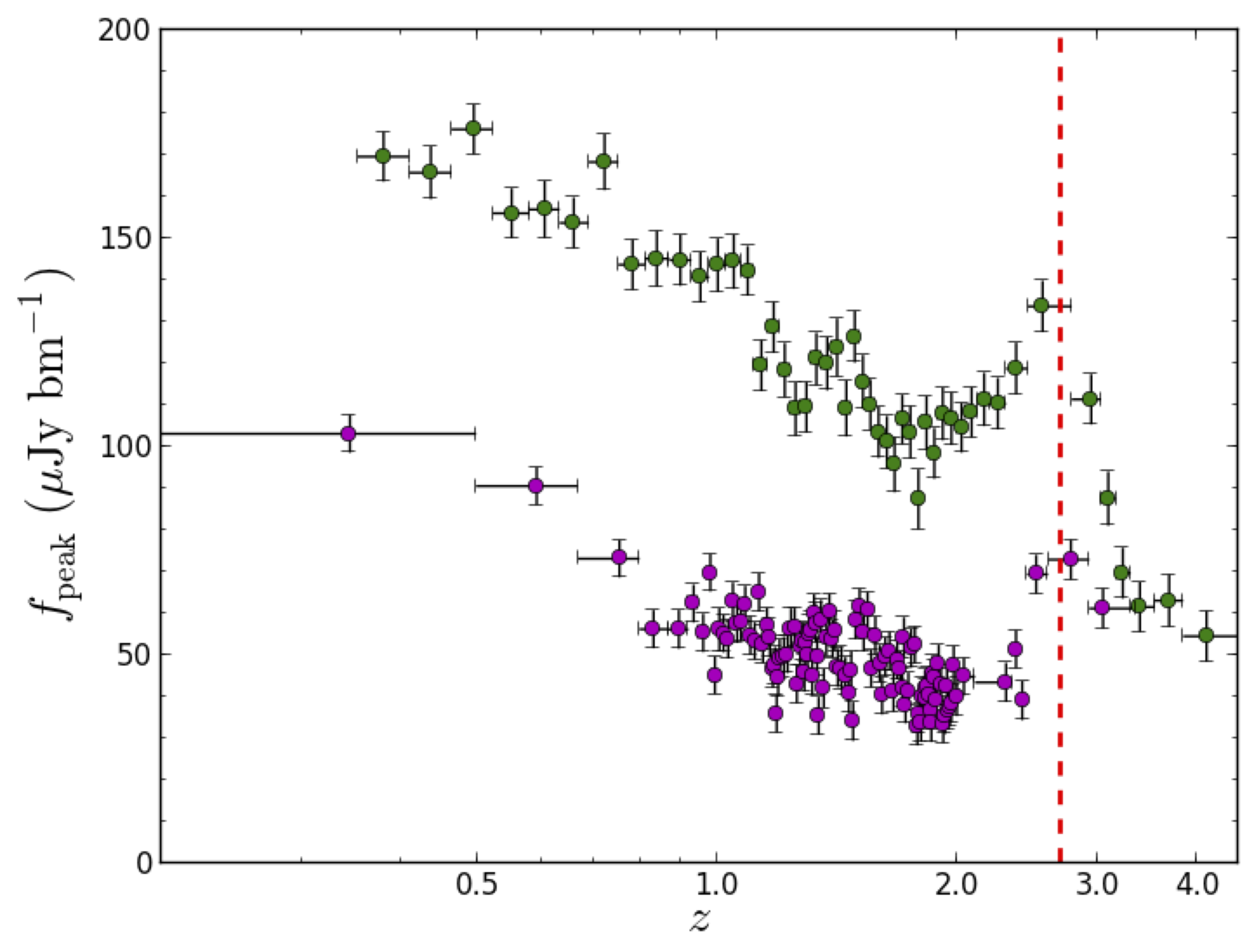}
\caption{Peak flux density ($\mu$Jy bm$^{-1}$) of median stacked quasars as a function of redshift \citep[see][Fig. 6]{White07} (Sample B: green, 1116 quasars per point; Sample D: purple, 1981 quasars per point).  The Sample D sources are fainter in the optical than the Sample B sources.
The vertical dashed lines represent $z=2.7$, which is the upper limit of efficient SDSS optical selection.}
\label{fig:f11}
\end{figure}

Before we can interpret the results of the stacking analysis, we must
first understand what biases are inherent to the process by looking at
some diagnostic information.  We first explore the distribution of
mean radio flux density by stacking in redshift bins (Figure
\ref{fig:f11}), breaking Sample~B (D) into 50 (100) redshift bins with
1116 (1981) quasars per bin.
After applying the median stacking
procedure described above, we get the same basic results as
\citet{White07}: the median flux density declines up to $z = 2$.
This trend of decreasing flux density with
redshift is expected based on inverse square law dimming. Note
Sample~B includes 10,000 more quasars than considered by
\citet{White07} (41,295 SDSS DR3 quasars) and should be clean of
selection effects up to $z\sim2.2$.

We observe an increase in median flux density starting at roughly
$z=2.2$ for all our samples (typically peaking at $z\sim2.7$). This
increase can be attributed to selection effects whereby the SDSS
optical selection was very inefficient at $z \sim 2.7$ while the
radio selection is more complete (compare panels a and b in
\citealt[Fig. 6]{Richards06}). As such, the quasars discovered at $z
\sim 2.7$ are more likely to be radio sources, thus biasing the
observed mean flux and requiring that a robust analysis be limited to
$z<2.2$.  

\begin{figure}[h!]
\epsscale{0.6}
\plotone{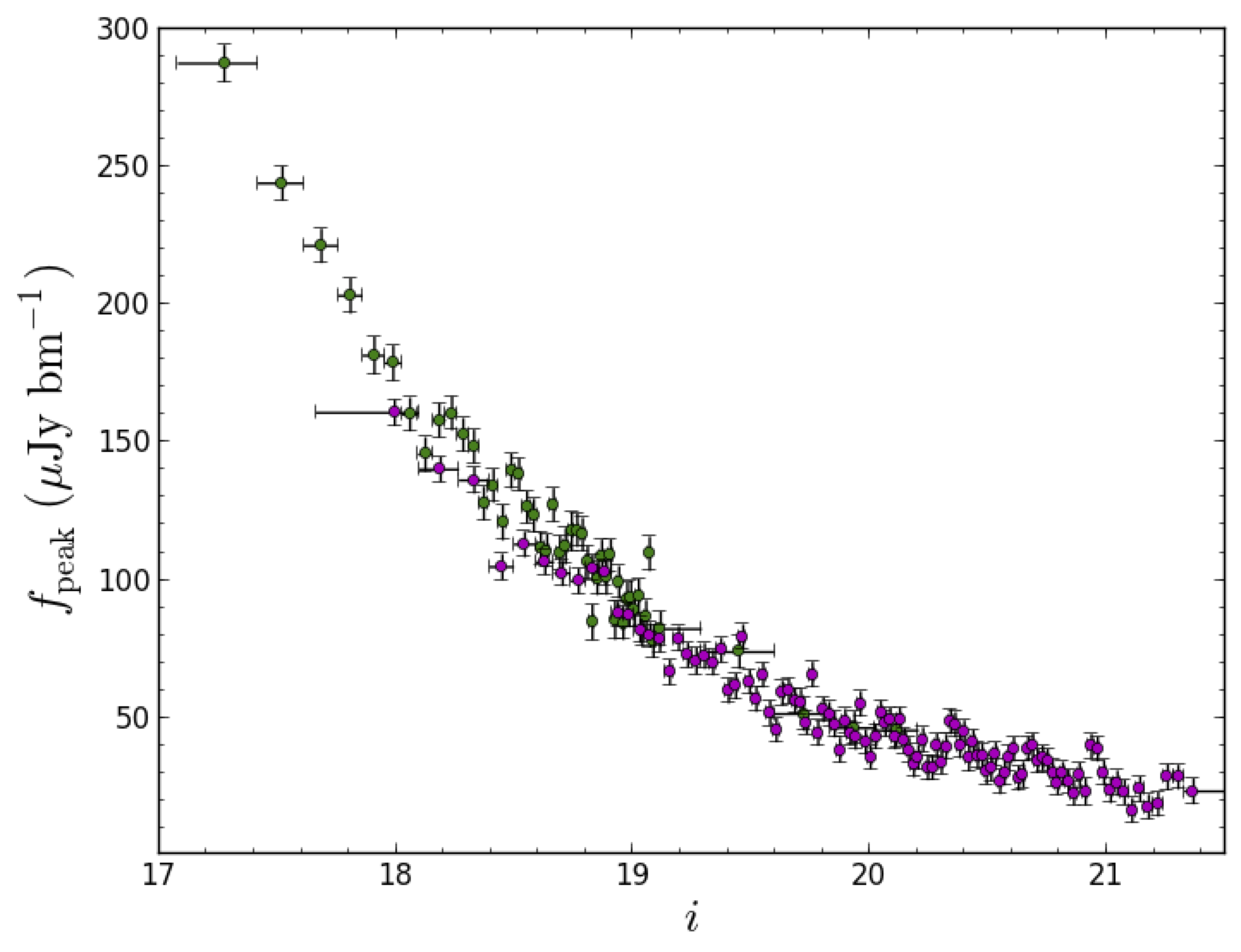}
\caption{Peak flux density ($\mu$Jy bm$^{-1}$) of median stacked quasars as a function of $i$ band color \citep[see][Fig.~7]{White07} (Sample B: green; Sample D: purple). As before, the strongest radio emitters are associated with the optically brightest sources. This trend is consistent with dimming of both the radio and optical with increasing redshift.}\label{fig:f12}
\end{figure}

\begin{figure}[h!]
\epsscale{0.6}
\plotone{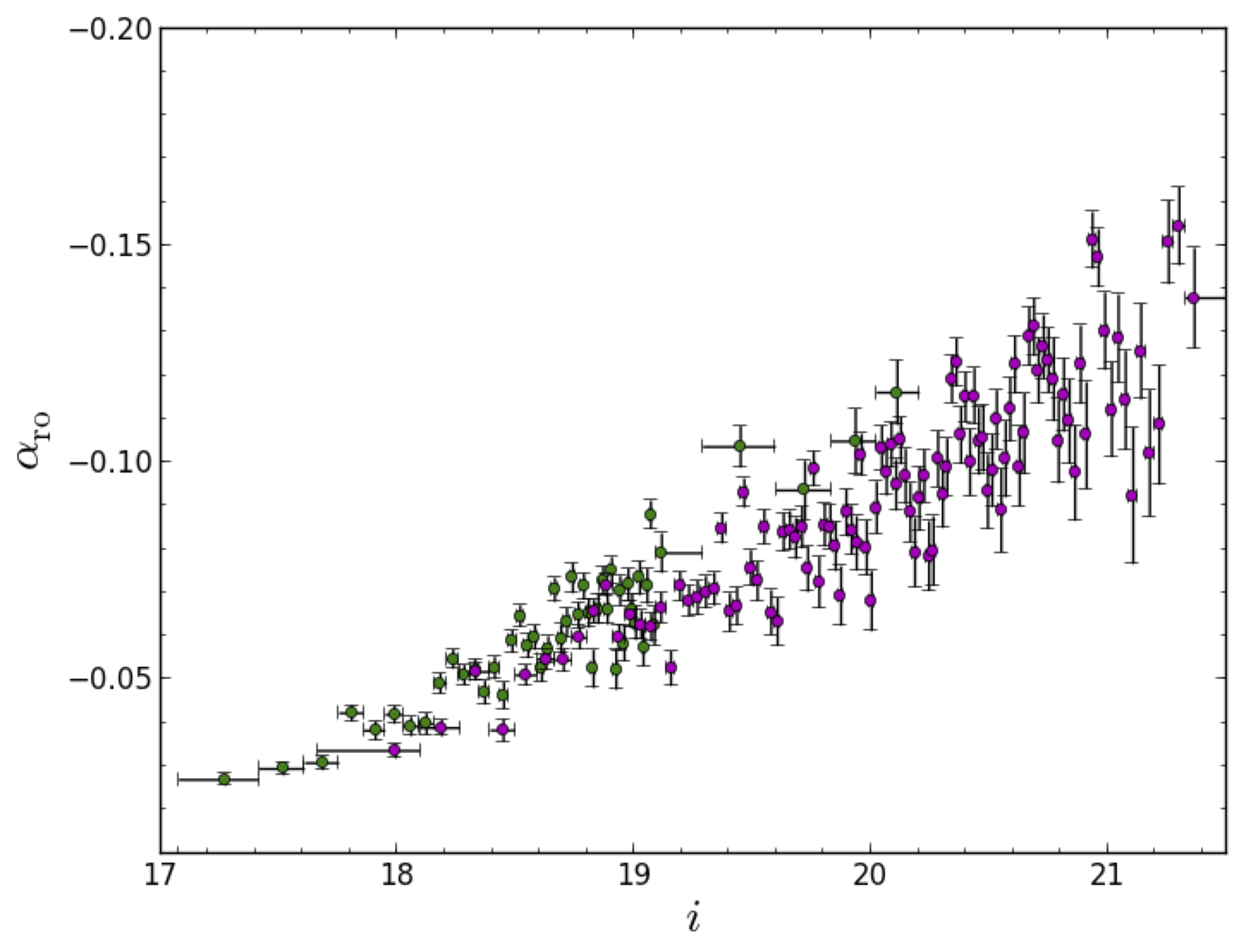}
\caption{\aro\ of median stacked quasars as a function of $i$ band color \citep[see][Fig.~12]{White07} (Sample B: green; Sample D: purple).  While Figure \ref{fig:f12} showed that the brightest sources in the optical are the brightest in the radio, the ratio of radio to optical flux is such that the radio-loudest objects (more negative values of \aro) are associated with the faintest optical sources.}
\label{fig:f13}
\end{figure}

We next investigate the mean radio flux density and \aro\ 
as a function of $i$-band magnitude to explore the
correlation between radio and optical brightness. Figure~\ref{fig:f12}
shows that the strongest radio emitters are also the optically
brightest, while Figure~\ref{fig:f13} shows that the optically
faintest sources are the most radio-loud, consistent with
\citet[Figs. 7 and 12]{White07}.  These trends mean that, as for the
RLF, some caution is needed in interpreting trends of radio properties
that follow trends with apparent magnitude.

\subsubsection{Choice of Radio Loudness Metric}

While the RLF is our metric for the extreme radio properties, we must
decide what metric to use for comparison of the mean radio properties.
We will conclude that the \aro\ is the parameter of choice.  Knowing
that, the reader can skip to Section~\ref{sec:results} if desired;
however, it is worth spending some time looking at the trends with
radio flux and luminosity in $L-z$ parameter space and reviewing how
we made the choice of \aro\ as our comparison metric before comparing
the results to the RLF.

\begin{figure}[h!]
\epsscale{1.0}
\plottwo{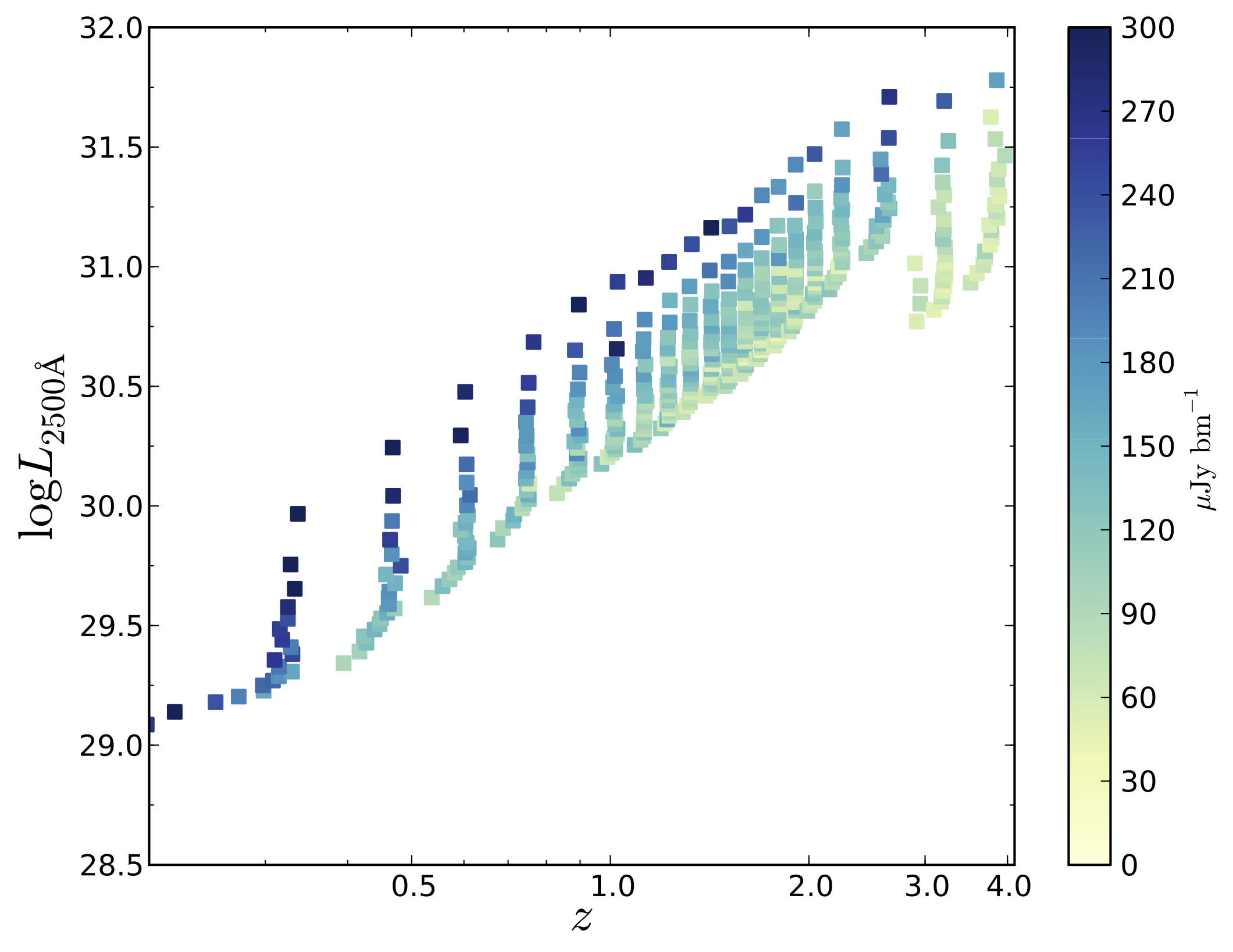}{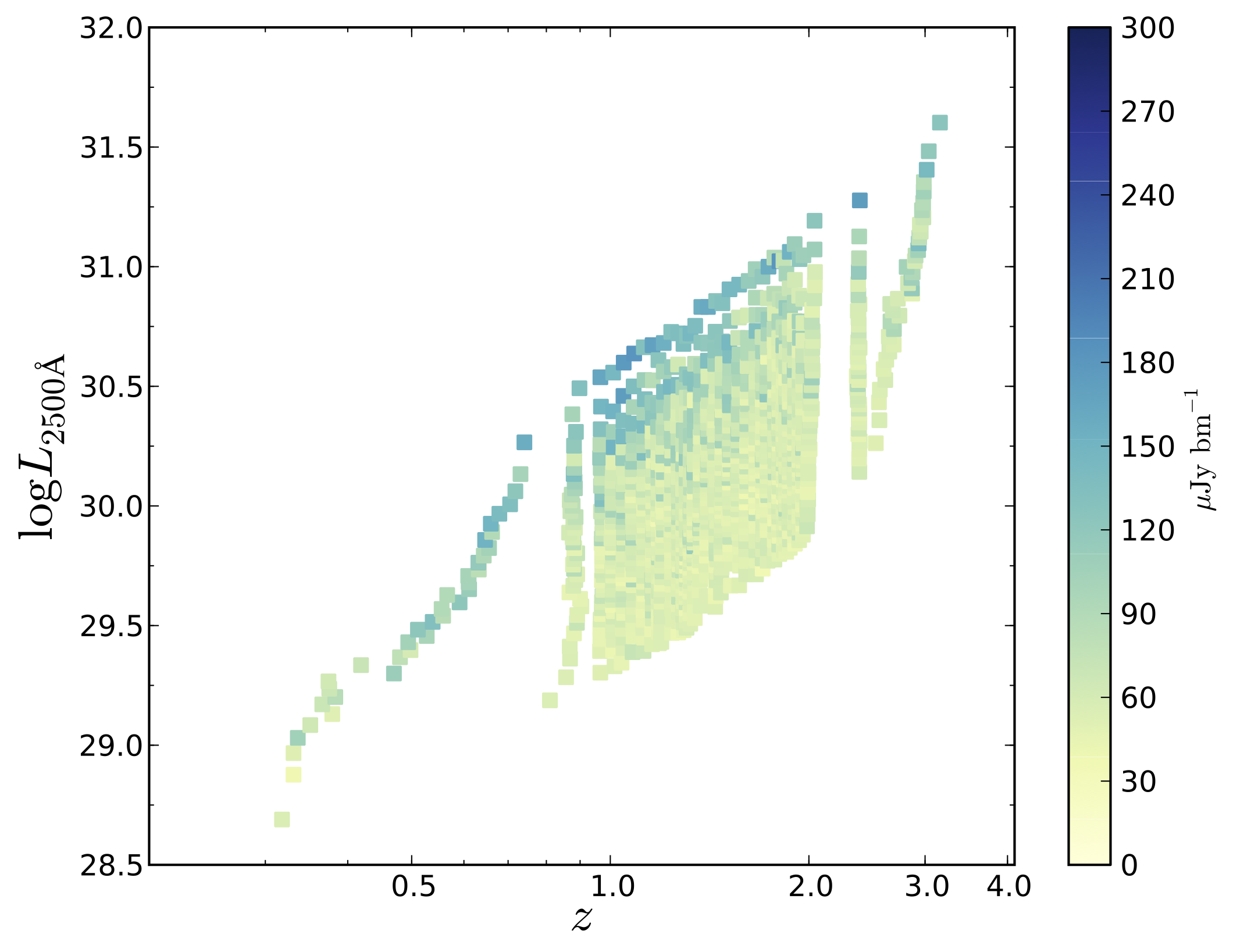}
\caption{Peak flux density ($\mu$Jy) of median stacked quasars as a function of both redshift and \ltwofive.  ({\em Left:}) Sample B, 138 objects per bin; ({\em Right:}) Sample D, 151 objects per bin. The trend roughly follows the $i$-band magnitude with brighter quasars in the optical being brighter in the radio.} 
\label{fig:f14}
\end{figure}

In Figure \ref{fig:f14} we show the median radio flux density (colored squares) as a
function of $L$ and $z$.
Here we see that, at a
fixed redshift, quasars that are optically more luminous have higher
radio fluxes, and at a fixed optical luminosity, lower redshift quasars
have higher radio fluxes. The trend is roughly consistent with the mean
radio flux being primarily dependent on the optical magnitude: 
optically brighter quasars are radio brighter, on average; see also
Figure~\ref{fig:f12}.


\begin{figure}[t!]
\plottwo{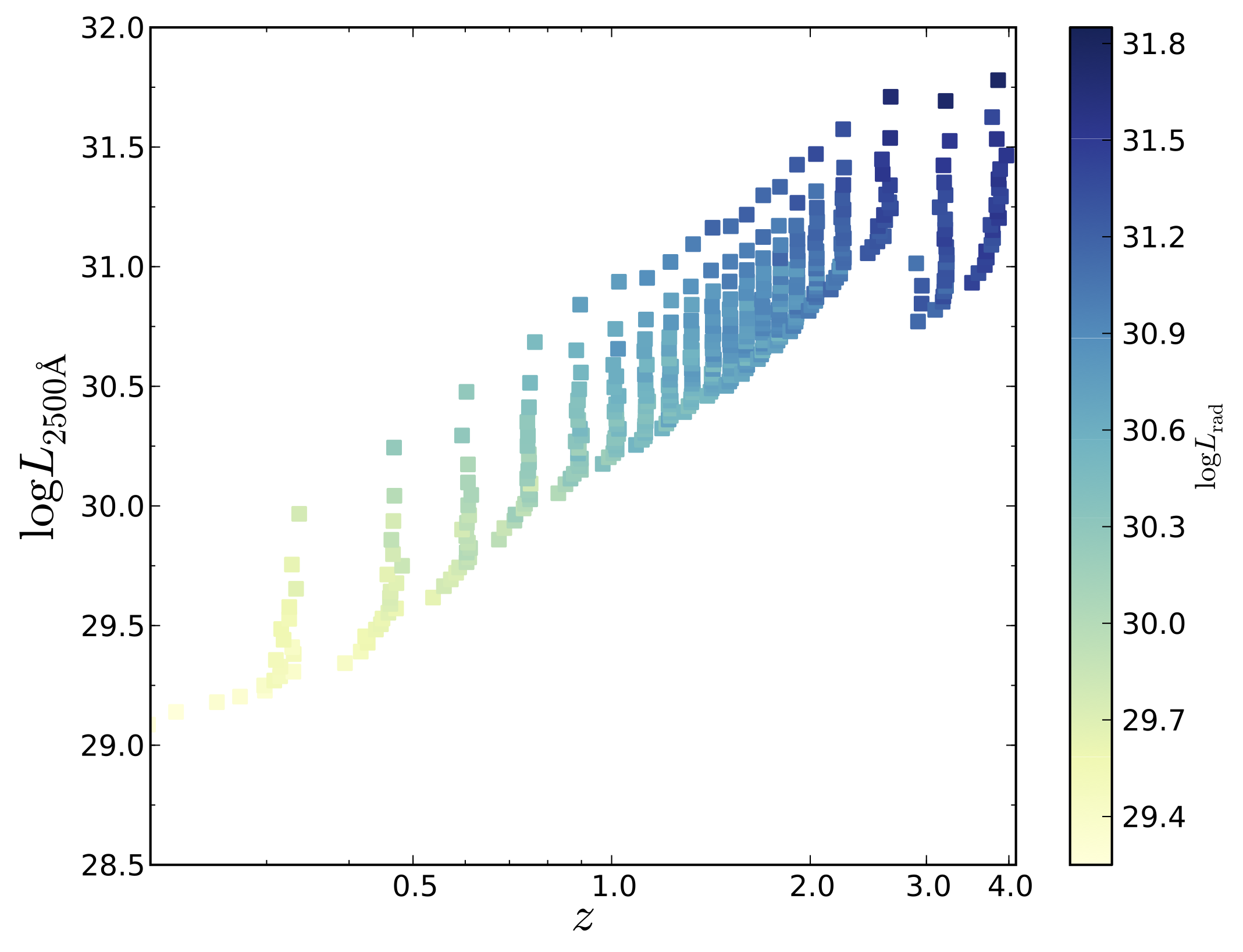}{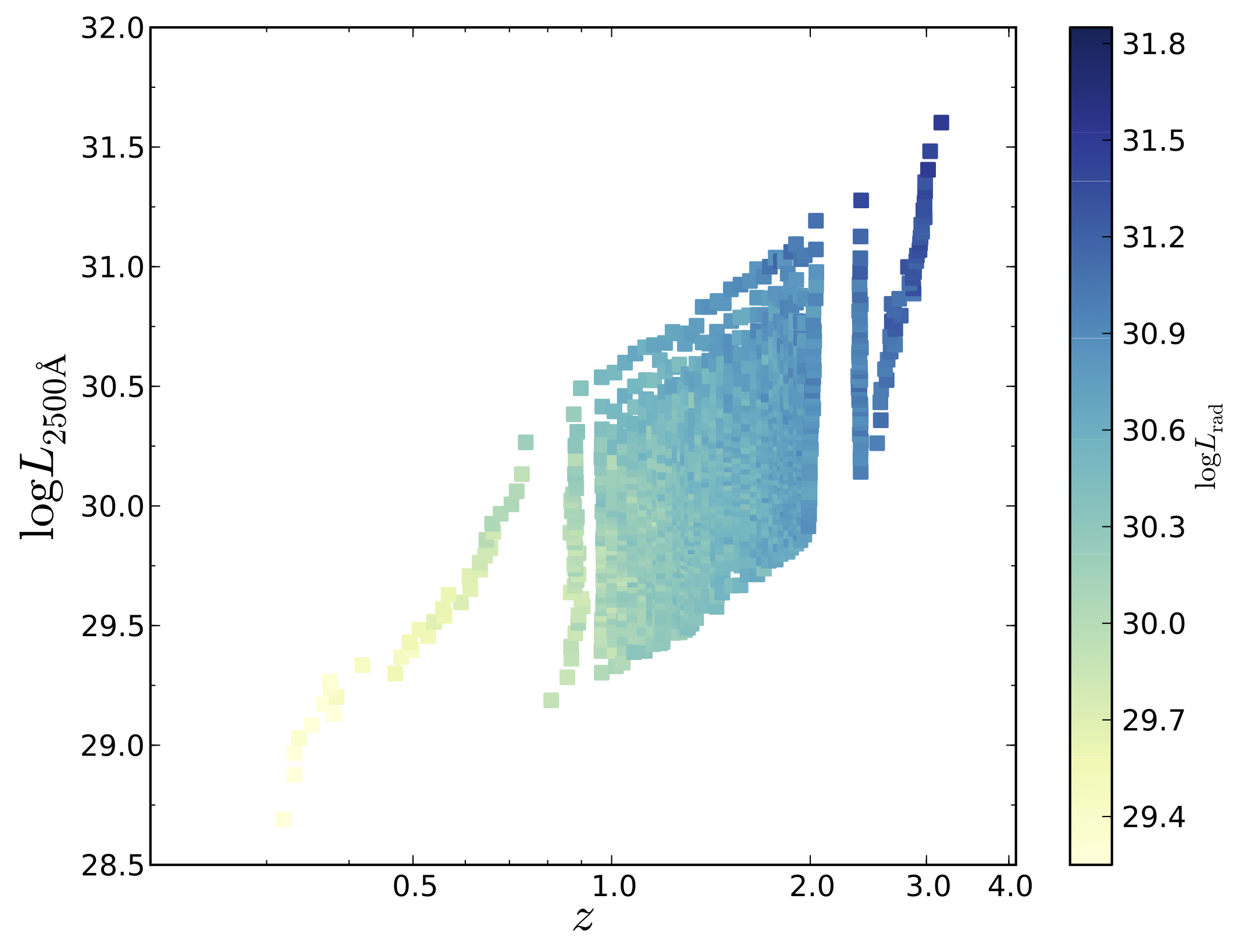}
\caption{Radio luminosities ($\rm{erg} \, \rm{s}^{-1} \, \rm{Hz}^{-1}$) of median stacked quasars as a function of both redshift and \ltwofive. ({\em Left:}) Sample B; ({\em Right:}) Sample D. The bins are the same as those used in Figures \ref{fig:f14}. As expected, the most radio luminous sources are at high-$z$ and have high optical luminosities with radio luminosity decreasing from the upper-right to lower-left.}
\label{fig:f15}
\end{figure}

Since we really want to understand the intrinsic radio properties, we
now convert from apparent brightness to luminosity, where the
conversion from radio flux to radio luminosity is determined according
to Equation \ref{eq:RadLum} as discussed in Section \ref{sec:RadLum}.
Figure \ref{fig:f15} shows the results of stacking the radio
{\em luminosities} in the \ltwofive-$z$ plane.  Again, more luminous sources
in the optical tend to be more luminous in the radio, but a larger
effect is seen with redshift, where a small radio flux at high-$z$ can
translate to a high radio luminosity.  The most radio luminous sources
are at high-$z$ and have high optical luminosities.  Objects with
roughly equal radio luminosities span a diagonal from the upper-left to
the lower-right while radio luminosity decreases from the upper-right
to lower-left.

As noted in Section \ref{sec:aro}, looking at the radio luminosity as
a measure of radio-loudness is correct only if there is no correlation
between the optical and the radio. If there is a correlation, then it is more
appropriate to consider the ratio of the two, or equivalently, the
spectral index between the radio and optical, \aro, which is defined
in Section \ref{sec:aro}.

\begin{figure}[h!]
\plottwo{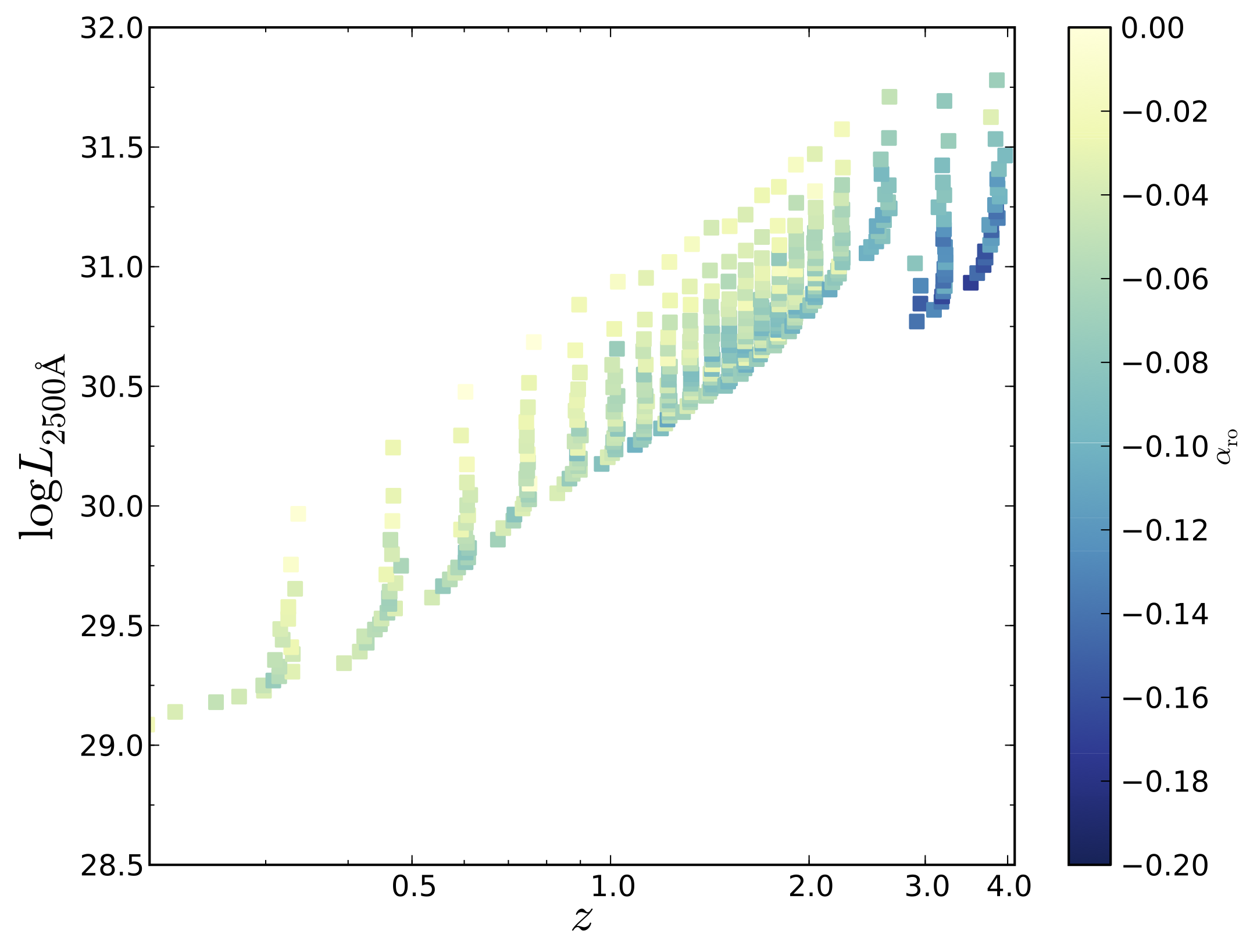}{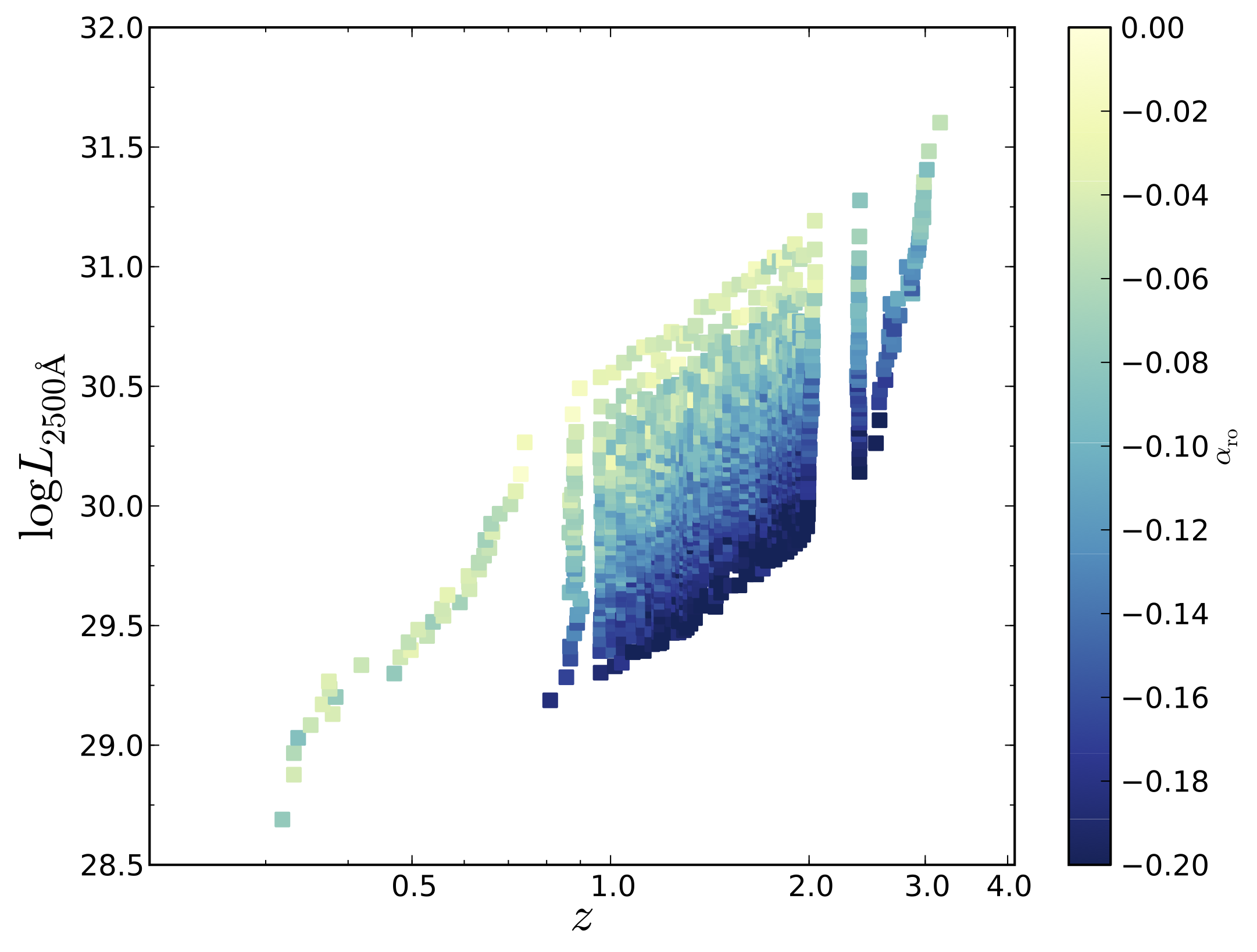}
\caption{Radio to optical spectral indices ($\alpha_{\rm ro}$) of median stacked quasars as a function of both redshift and \ltwofive. ({\em Left:}) Sample B; ({\em Right:}) Sample D. These bins are the same as those used in Figures \ref{fig:f14} and \ref{fig:f15}. This trend is completely opposite to that found for the RLF (see Figure \ref{fig:f10}). The median stacking shows stronger radio sources (relative to the optical) with decreasing optical luminosity (at fixed redshift) and with increasing redshift (at fixed optical luminosity).}
\label{fig:f16}
\end{figure}

Figure~\ref{fig:f16} shows the resulting distribution in \aro, which
is a measure of the {\em slope} of the SED between the radio and
optical. We see that normalizing by the optical luminosity has
produced a significantly different trend than we saw in
Figure~\ref{fig:f15}. That trend is for quasars to be stronger radio
sources (relative to the optical) with decreasing optical luminosity
(at fixed redshift) and with increasing redshift (at fixed optical
luminosity).
This trend is perhaps unexpected but is indeed consistent
with Figure~\ref{fig:f15}, where we saw that equal radio luminosities
occupied roughly diagonal tracks in \ltwofive-$z$ space.  Along one of
those diagonals, the objects with the lowest optical luminosity will
have the largest radio-to-optical ratio, so we expect radio-dominance
from the objects along the lower boundary of the distribution.

As there is precedent for the slope of the spectral energy
distribution in quasars to be a function of luminosity, it is also
important to consider how \aro\ may change with \ltwofive. In
particular, it has been repeatedly shown \citep[e.g.,][]{Avni82, Steffen06,
  Just07, Lusso10} that there is a non-linear relationship between the
X-ray and UV luminosity in quasars: quasars with double the UV
luminosity do not have double the X-ray luminosity.  Failure to
correct for any similar systematic trends in \aro\ with optical
luminosity could lead to biased conclusions. As such we investigate
the relationship between $L_{\rm rad}$ and $L_{\rm opt}$ in terms of
the behavior of \aro\ as a function of \ltwofive\ by separating the
quasars into bins of optical luminosity with 1000 objects in each bin.

\begin{figure}[h!]
	 \centering
        \begin{subfigure}[b]{0.3\textwidth}
                \centering
                \includegraphics[width=\textwidth]{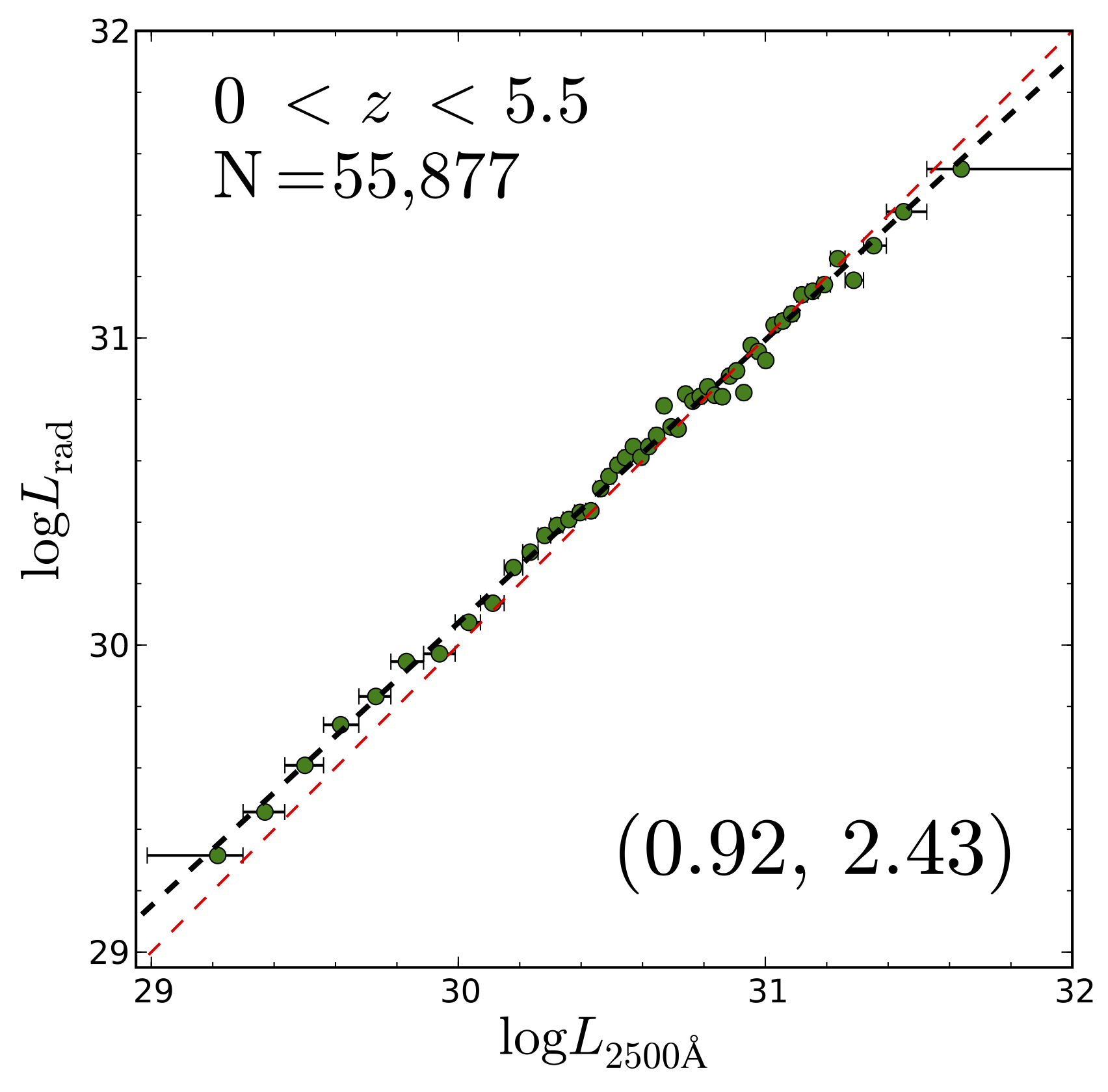}
        \end{subfigure}
        ~
        \begin{subfigure}[b]{0.3\textwidth}
                \centering
                \includegraphics[width=\textwidth]{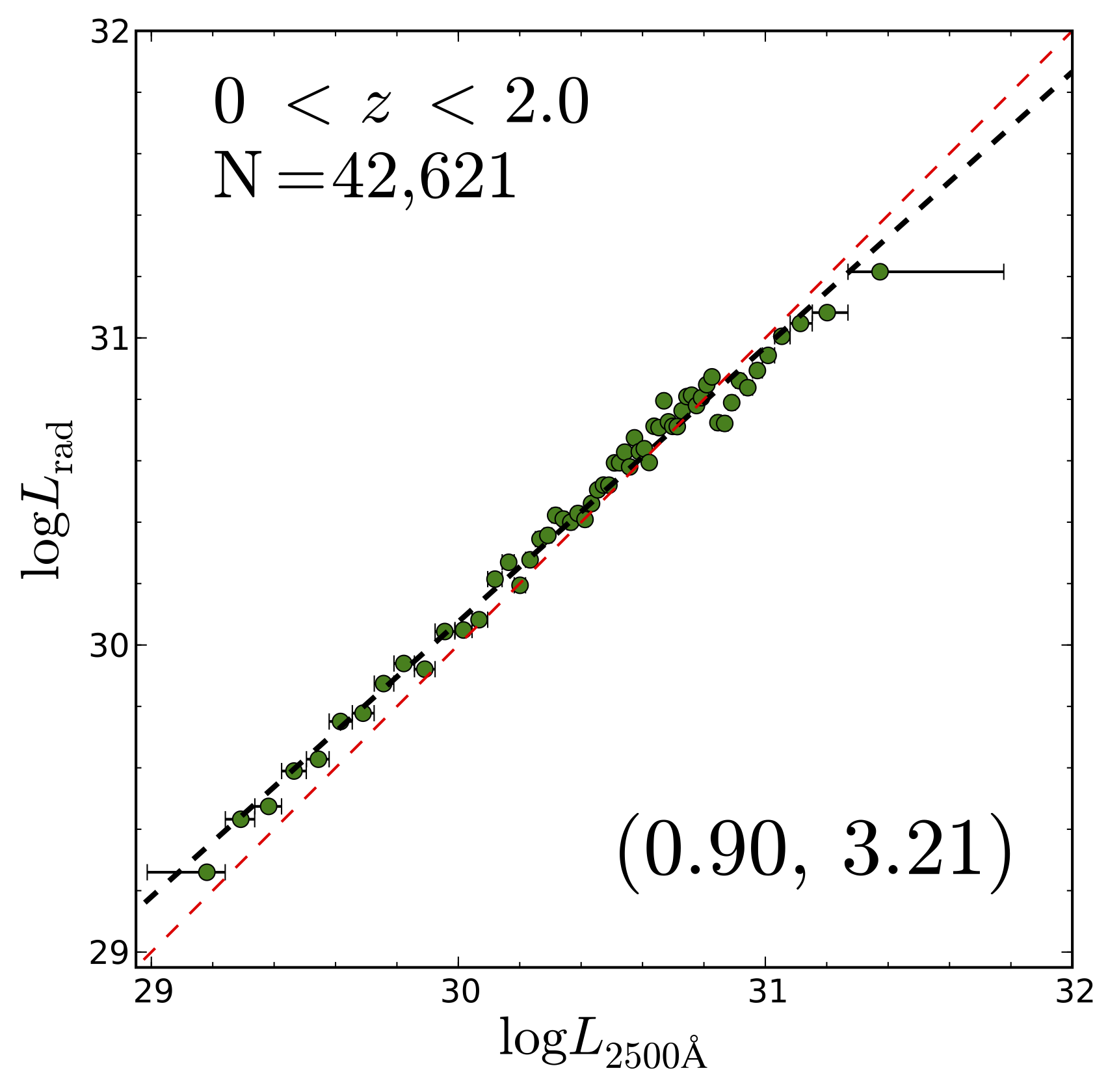}
        \end{subfigure}
        ~
         \begin{subfigure}[b]{0.3\textwidth}
                \centering
                \includegraphics[width=\textwidth]{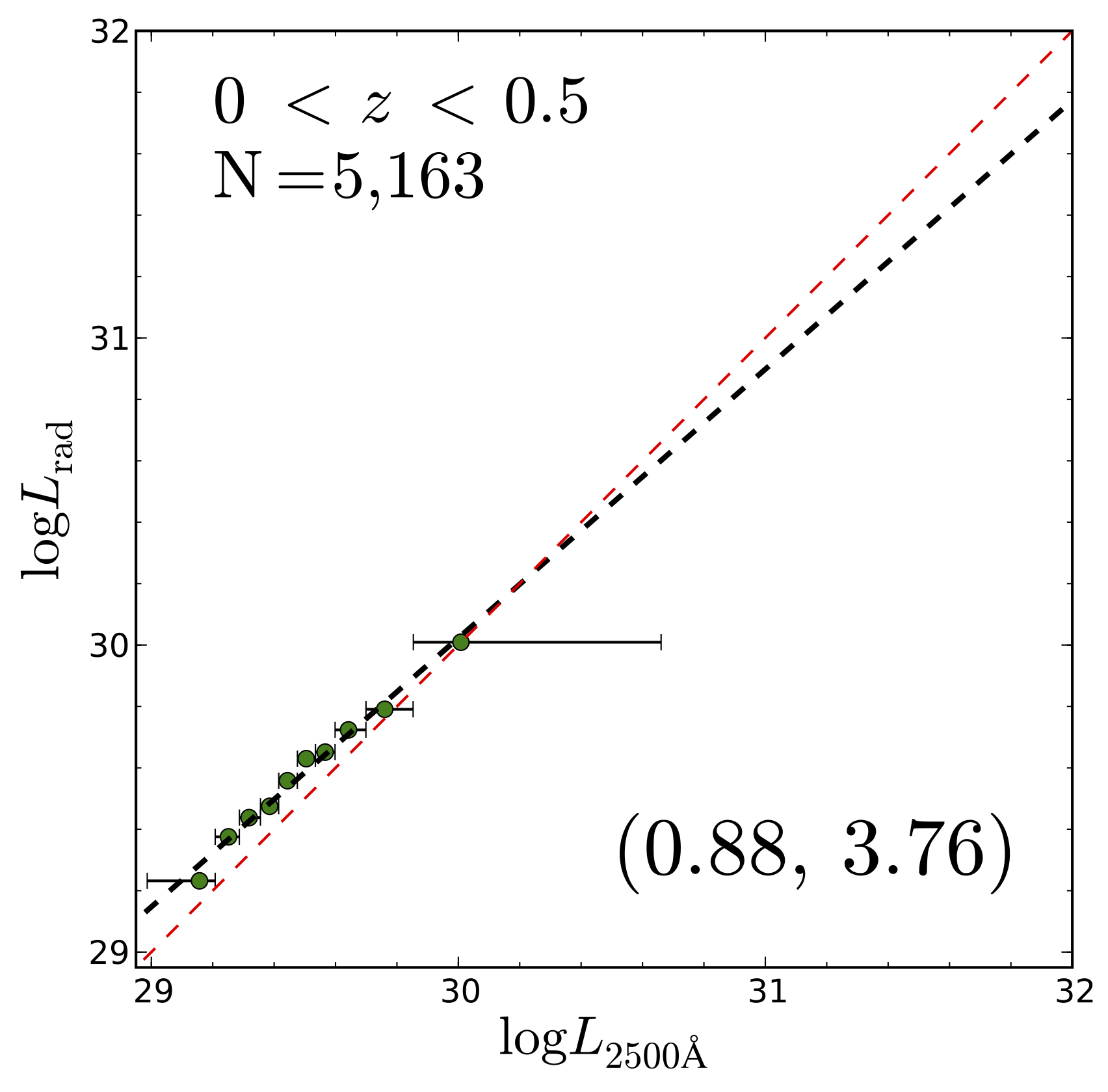}
        \end{subfigure}
        
         \begin{subfigure}[b]{0.3\textwidth}
                \centering
                \includegraphics[width=\textwidth]{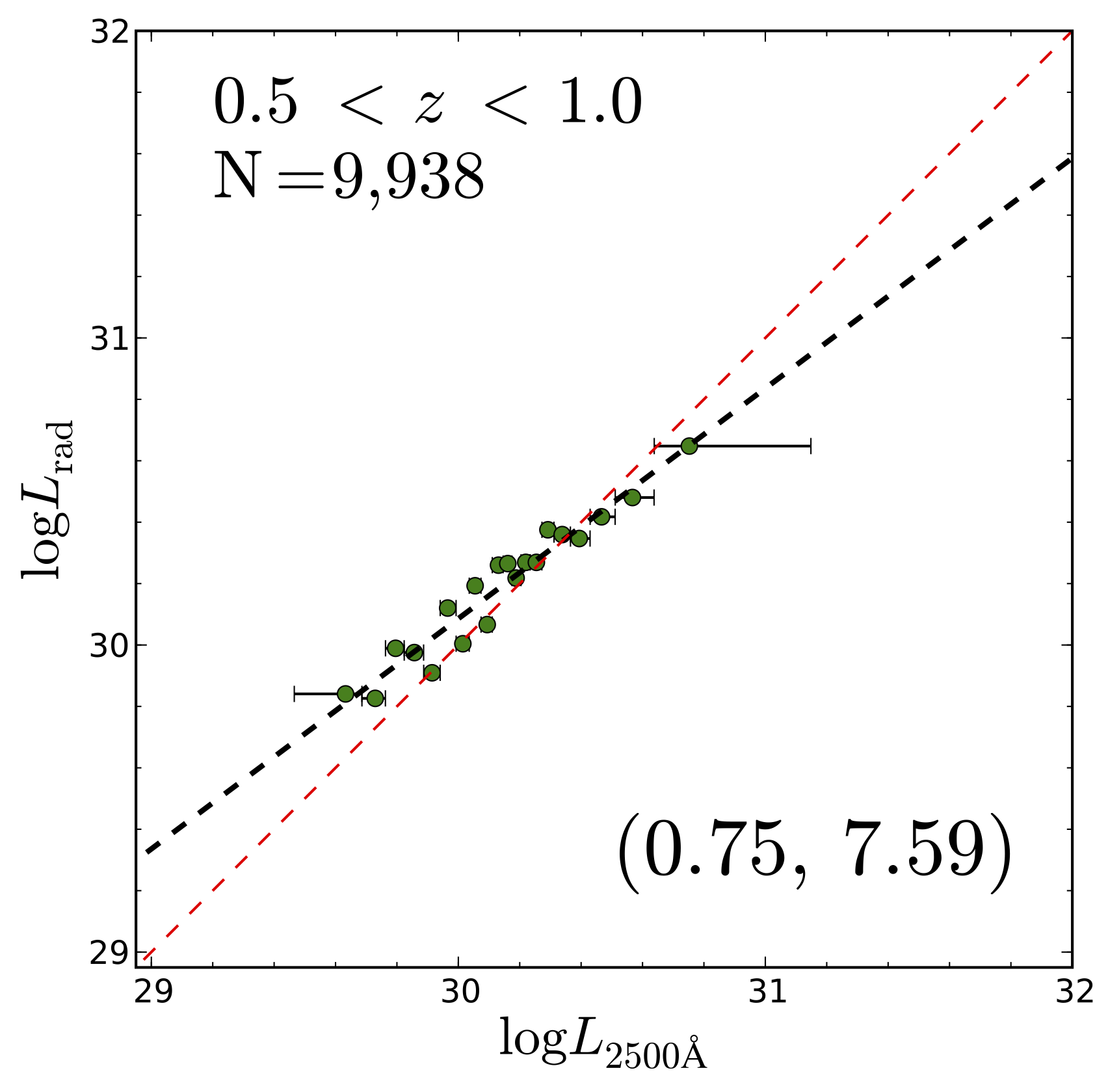}
        \end{subfigure}
        ~
        \begin{subfigure}[b]{0.3\textwidth}
                \centering
                \includegraphics[width=\textwidth]{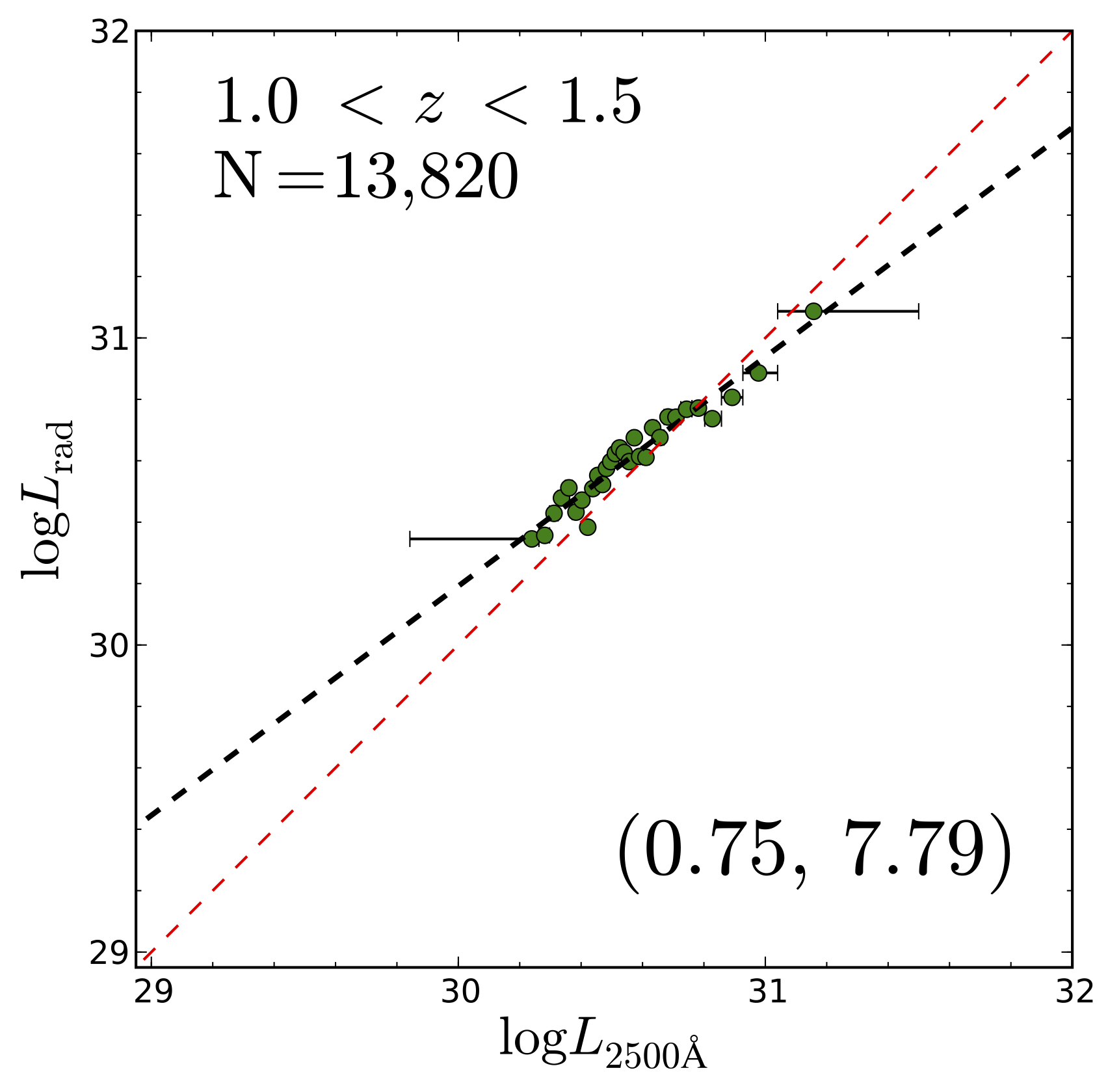}
        \end{subfigure}
        ~
         \begin{subfigure}[b]{0.3\textwidth}
                \centering
                \includegraphics[width=\textwidth]{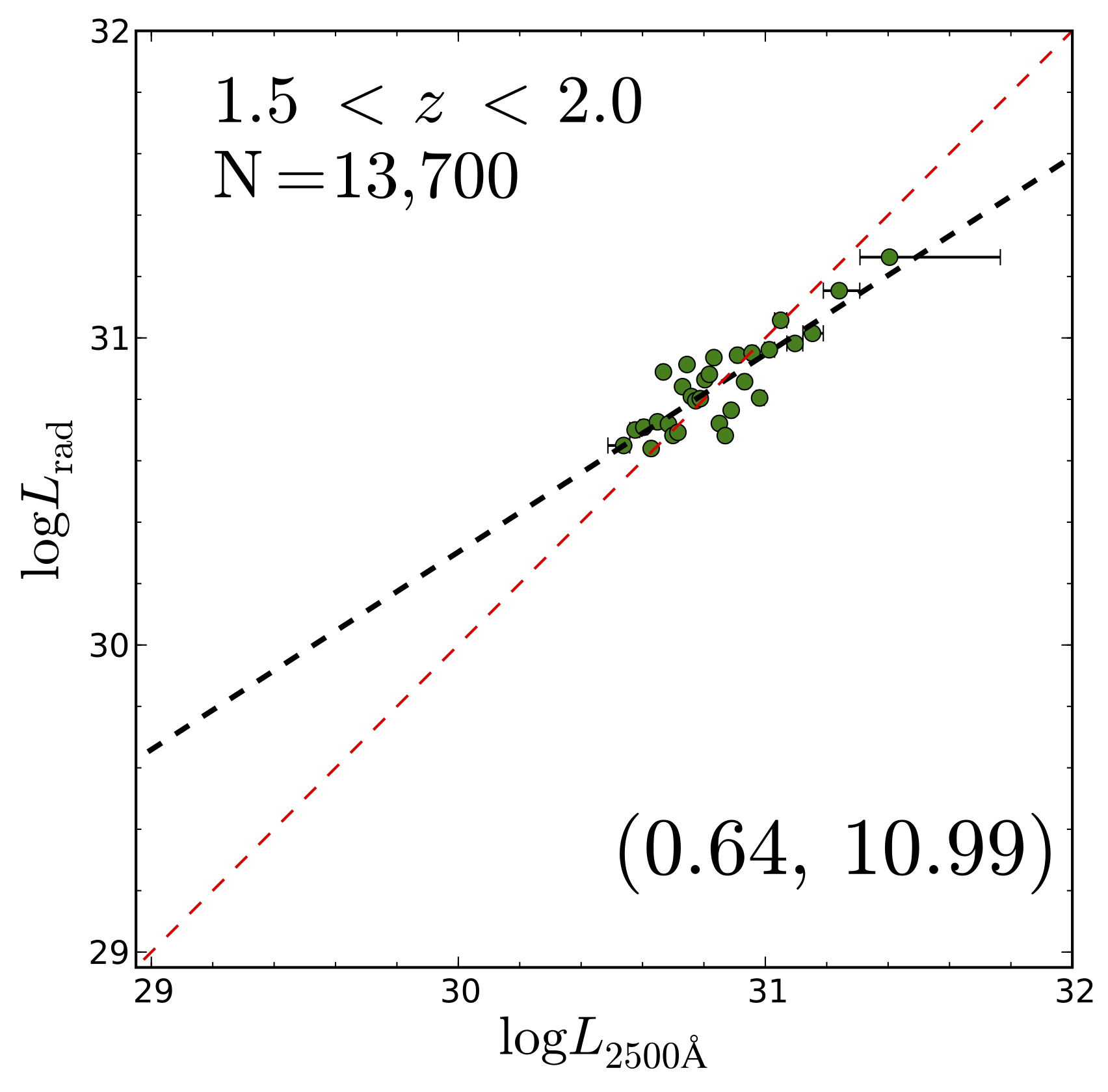}
        \end{subfigure}
        
          \begin{subfigure}[b]{0.3\textwidth}
                \centering
                \includegraphics[width=\textwidth]{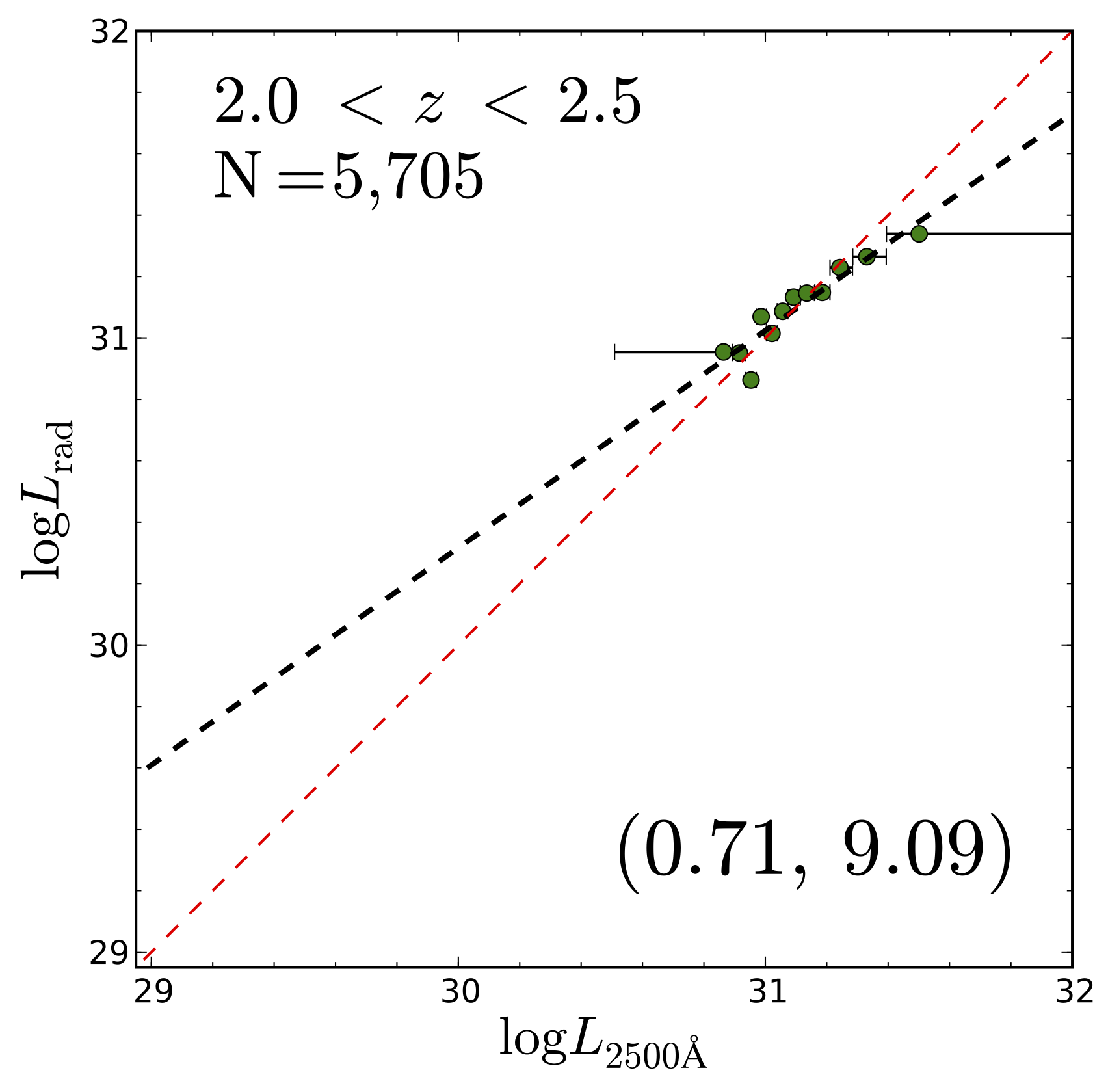}
        \end{subfigure}
        ~
        \begin{subfigure}[b]{0.3\textwidth}
                \centering
                \includegraphics[width=\textwidth]{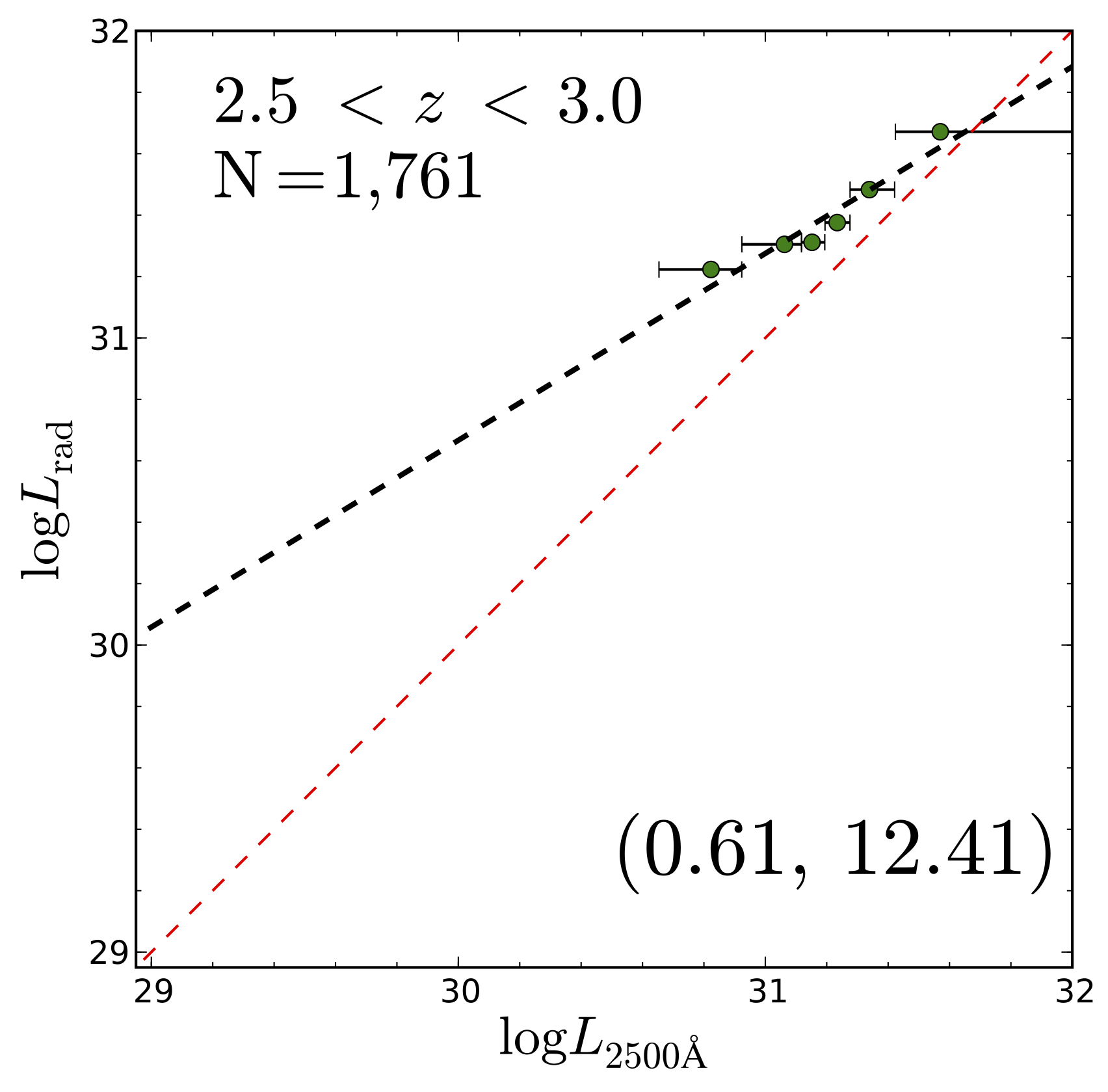}
        \end{subfigure}
        ~
         \begin{subfigure}[b]{0.3\textwidth}
                \centering
                \includegraphics[width=\textwidth]{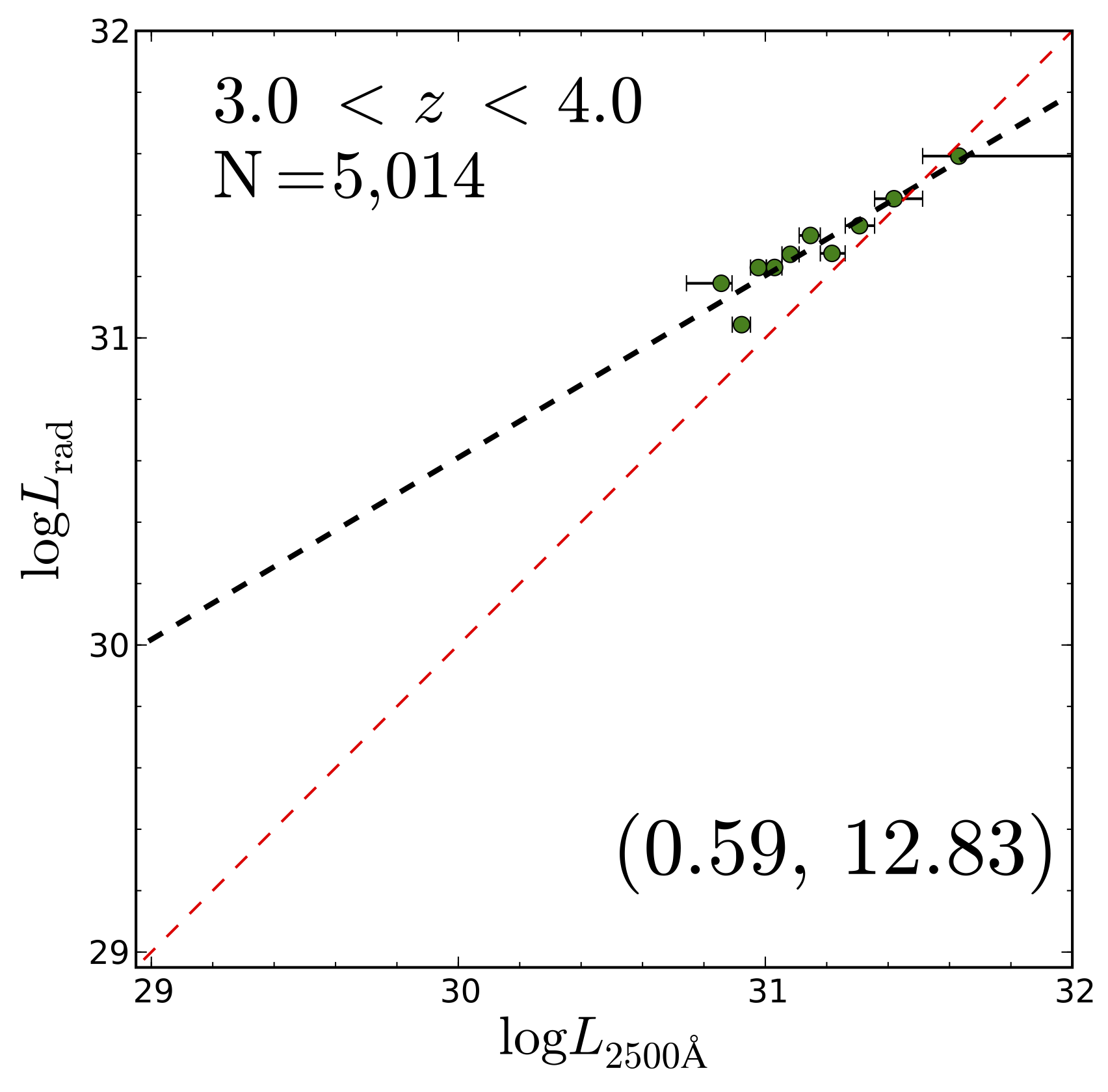}
        \end{subfigure}
	\caption{Radio luminosity dependence on \ltwofive\ within different redshift intervals \citep[see][Fig.~10]{White07} for Sample B. The dashed red line in each plot shows where $L_{\rm rad} = {L_{\rm opt}}$. The best fit lines for each redshift range are shown as the different dashed black lines, and the values in the lower-right corner indicate the slopes and intercepts of these lines. Sample B shows a non-linear relationship between $L_{\rm rad}$ and ${L_{\rm opt}}$, as $L_{\rm rad} \sim {L_{\rm opt}}^{.92}$.}
\label{fig:f17}
\end{figure}

The top-left plot of Figure \ref{fig:f17} shows the correlation between
$L_{\rm rad}$ and \ltwofive\ \citep[see][Fig.~9]{White07} for the
entire range of redshifts within Sample B, whereas the other panels
show restricted redshift ranges. Here it is important to have limited
our analysis to Sample B as using a less homogeneous sample can
imprint biases onto the distribution in \aro-\ltwofive\ parameter
space. We have further limited our analysis to point sources to avoid
contributions from the host galaxy to the optical luminosity and to
$z<2.2$ to avoid the known bias towards radio sources in the SDSS
selection function at higher redshifts.

The best fit line is computed as $\log(L_{\rm{rad}}) = m \,
\log(L_{\rm 2500 \, \AA}) \, + \, b$, where the median \ltwofive\
value for each bin was used and the coordinate pairs $(m,b)$ represent
the slope and y-intercept for the linear best fit models.  Just as in
\citet{White07}, the radio luminosities for our four samples do not
increase linearly with the optical luminosities.  For low redshift ($z<2.2$) point-source quasars within Sample B, we find that the relationship is $L_{\rm rad} \sim
{L_{\rm opt}}^{0.92}$. This corresponds to a factor of $~2.5$ in radio
luminosity between the least and most luminous quasars in the optical,
similar to what was found by \citet{White07}.  This deviation from a
linear relationship is not as strong as it is in the X-ray (exponent
of $\sim0.72$ in \citealt{Steffen06}); however, the lever arm in
extrapolating from the optical to the radio is longer than that between the
optical and X-ray, and any deviation from linearity is still important
to account for.

\begin{figure}[h!]
\plottwo{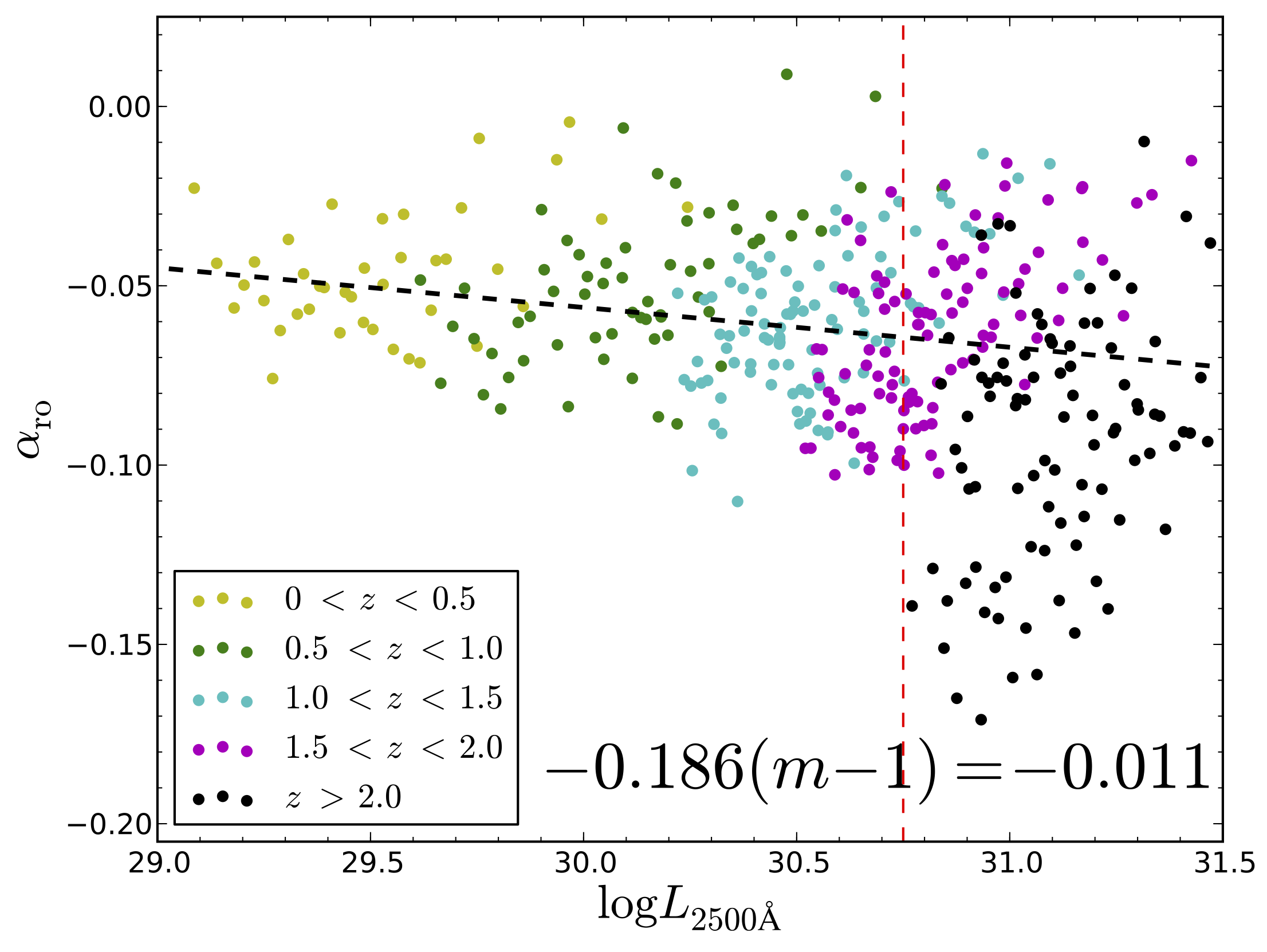}{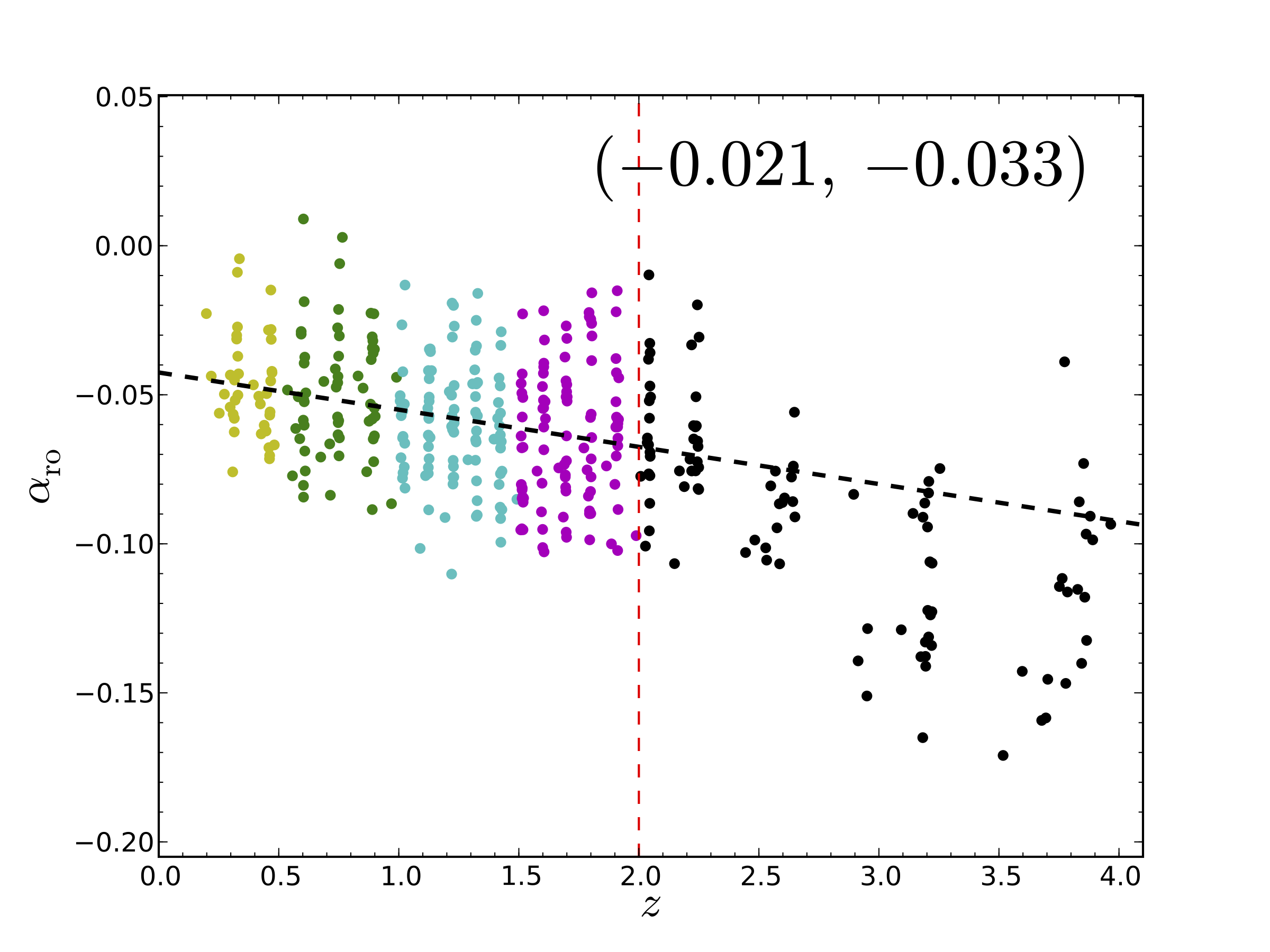}
\caption{\emph{Left:} The dependence of $\alpha_{\rm ro}$ on \ltwofive\ \citep[see][Fig. 5, top]{Steffen06}. The dotted black line is the linear best fit for Sample B. Recall that more negative values of $\alpha_{\rm ro}$ mean more RL; thus, the higher luminosity objects are biased to a more RL SED. Different redshift bins are highlighted with different colors. We have removed sources to the right of the dashed red line when computing the linear best fit so as to not artificially skew it based on selection effects. \emph{Right:} The dependence of $\alpha_{\rm ro}$ on redshift \citep[see][Fig. 7, top]{Steffen06}. The discrete appearance of the points in this panel is an artifact of initially binning our samples with respect to redshift.}
\label{fig:f18}
\end{figure}

In Figure \ref{fig:f18}a, we effectively show the same information as
is given in Figure~\ref{fig:f17}, but we have color-coded the
different redshift regions and are now plotting \aro\ on the $y$-axis.
To compute our stacked \aro values, each of the radio cutouts (in Jy) is first divided by
its corresponding quasar's optical flux. Then, the cutout ratios
within a bin are median stacked and the maximum pixel value in the
stacked image is taken to be $f_{5 \, \rm GHz}/f_{\rm 2500 \,
  \AA}$. Finally, the median \aro\ value is found by plugging into
Equation \ref{eq:alpha}. Here we find that the redshift regions occupy
wedge-shaped distributions that are consistent with the
flux-limited nature of the quasar sample. As such, our best-fit line
(for $z<2$) removes quasars above $\log{L_{2500 \, \rm \AA}}=30.75$
beyond which there is an artificial bias in the sample.

The equation reported for the linear best fit model is $\alpha_{\rm
  ro} = -0.186 (m-1) \, \log(L_{\rm 2500 \, \AA}) \, -0.186 \, b$,
such that the values of $m$ and $b$ have the same meaning in both
Figure~\ref{fig:f18}a and Figure~\ref{fig:f17}. All four of our
samples (although only Sample B is pictured) show that $\alpha_{\rm
  ro}$ {\em decreases} (gets more radio-loud) as \ltwofive\ increases.
This is opposite to what we found in Figure~\ref{fig:f17} and would
seem to be due to the biased nature of the redshift slices in
Figure~\ref{fig:f17} as highlighted by the color-coding of
Figure~\ref{fig:f18}a.  Indeed, Figure~\ref{fig:f18}a suggests that
there is a small {\em increase} in radio luminosity with optical
luminosity (consistent with $L_{rad} \sim {L_{opt}}^{1.011}$).  


Since our samples are flux limited, any evolution in $L$ could instead
be an evolution in $z$.  As such, we reproduced Figure \ref{fig:f18}a
with redshift instead of \ltwofive\ to be sure that there were no
additional biases. Figure \ref{fig:f18}b shows the dependence of
$\alpha_{\rm ro}$ on $z$ \citep[see][Fig.~7, top]{Steffen06}. The
coordinate pair $(a,B)$ represents the slope and y-intercept for the
linear best fit model such that $\alpha_{\rm ro} = a \, z \, + \,
B$, where we have limited the fitting to data with $z < 2.0$ as we did
in Figure~\ref{fig:f18}a. All four of our samples (Sample B,
pictured) show that $\alpha_{\rm ro}$ slightly decreases with increasing
$z$.  This trend with redshift is larger than that seen in the X-ray
\citep{Steffen06}.

It is an open question as to whether we should be using \aro\ (see Equation
\ref{eq:alpha}) or \daro\ $= \alpha_{\rm{ro}, \, \rm obs.} \, - \,
\alpha_{\rm{ro}, \, \rm best \, fit}$ (i.e., \aro\ corrected for luminosity
and/or redshift) in our analysis. The {\em shape} of the SED is
measured by \aro\ whether or not \aro\ has any luminosity or redshift
dependences. If it is the shape that matters (as it is for the dependence of radiation line-driven winds on \aox), then we should be using \aro. In that case, our
current analysis will suffice. If, on the other hand, we care more
about the shape relative to the mean at a given $L$ or $z$, then
we should be using \daro.  For example, if dust reddening were
causing a trend in \aro\ with $L$, we might prefer to use
\daro. Indeed, absorption is an issue for \aox; however, in our case,
the relative deficit of optical flux is for the most luminous sources,
not the least luminous. Therefore, it is unlikely that dust reddening is
causing the increase in radio-loudness with luminosity.  

\begin{figure}[t!]
\plottwo{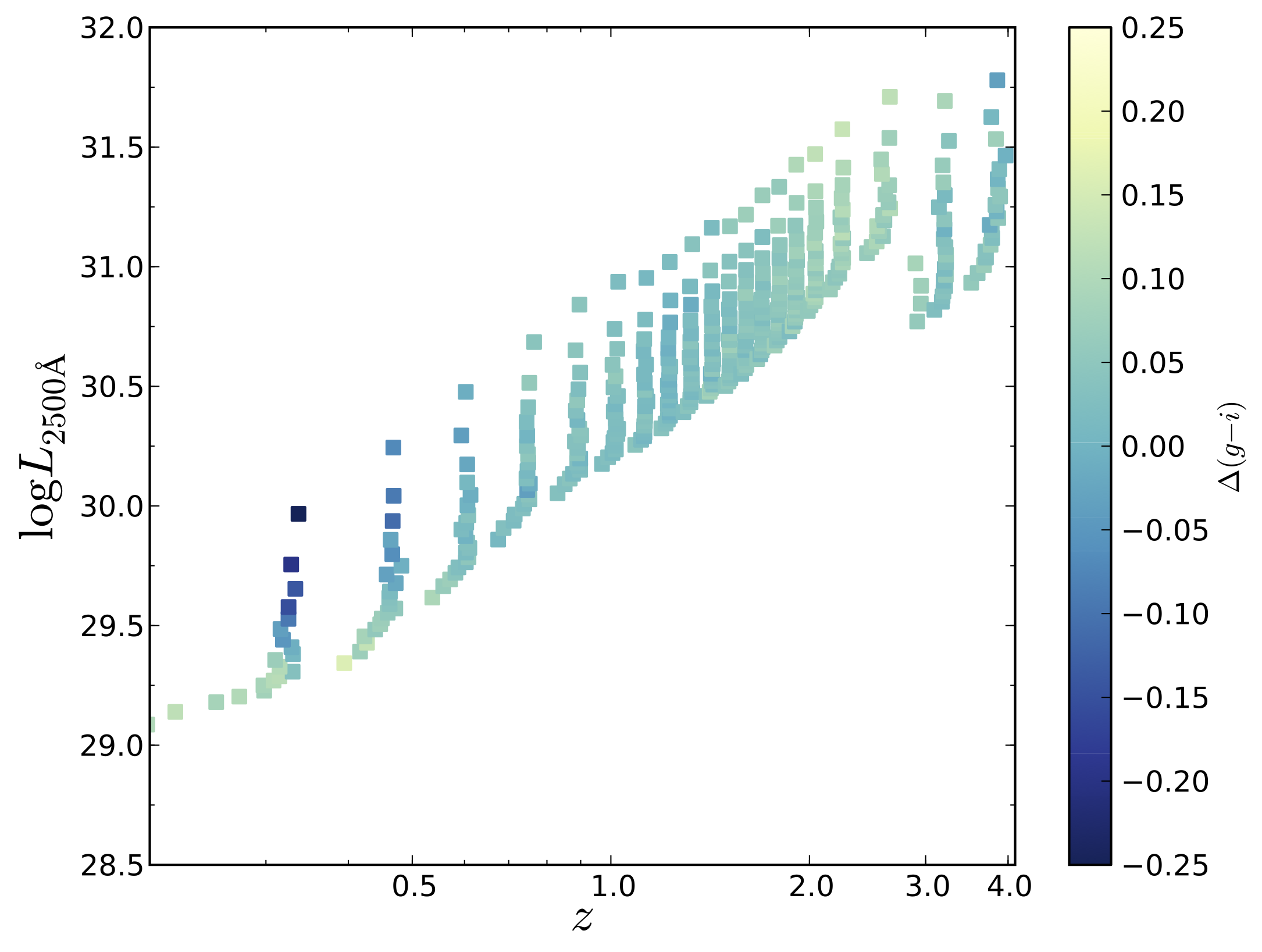}{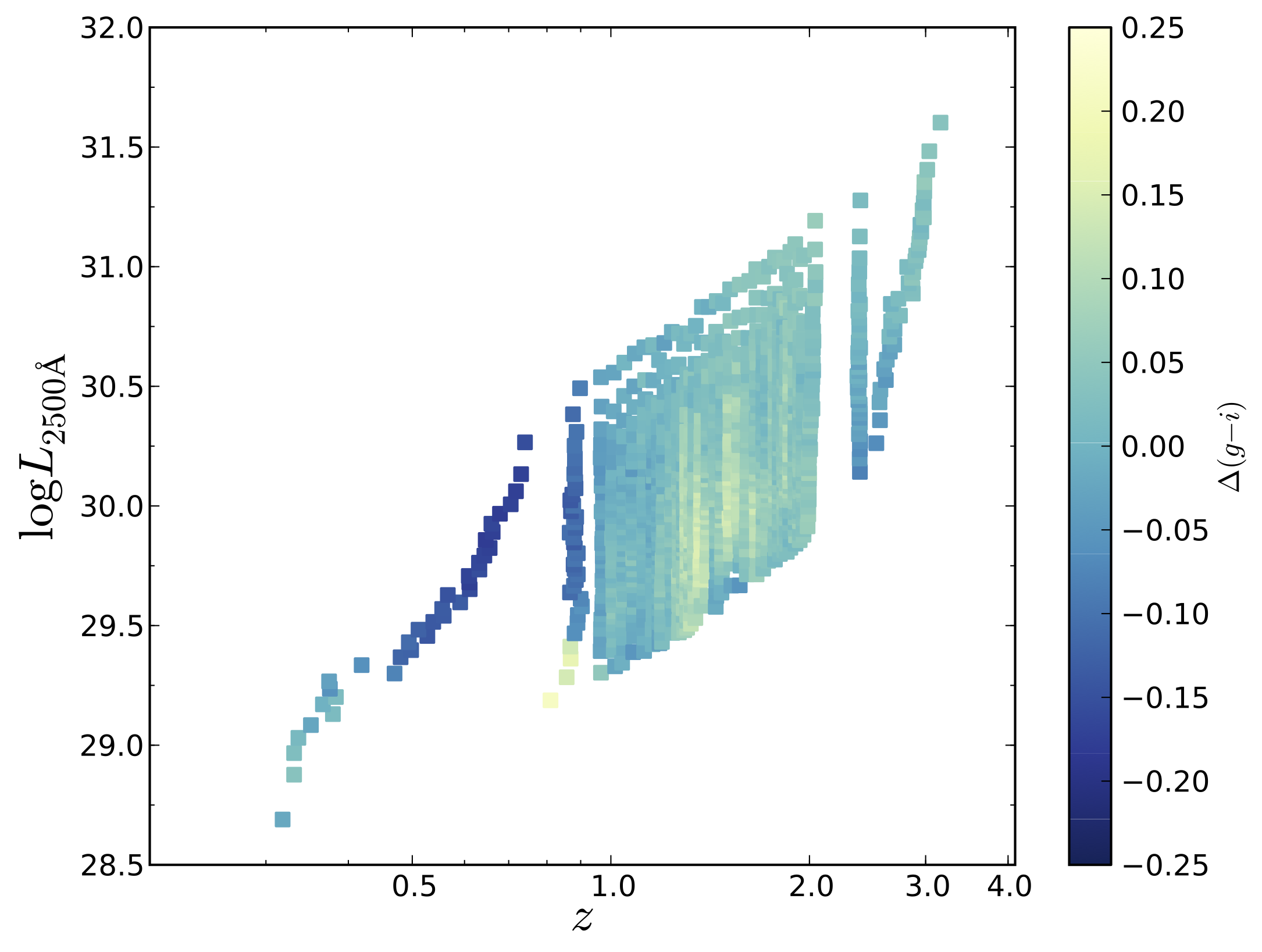}
\caption{The relative color, $\Delta(g-i)$, of median stacked quasars as a function of both redshift and \ltwofive. These bins are the same as those used in Figures \ref{fig:f14}, \ref{fig:f15}, and \ref{fig:f16}. The distribution of $\Delta(g-i)$ values in the $L-z$ plane shown suggest that dust is not the cause of the trends in mean radio loudness with $L$ and $z$.}
\label{fig:f19}
\end{figure}

We can see this in another way in Figure~\ref{fig:f19} which shows the
mean relative color in each of the bins. Ignoring the lowest redshift
quasars (where host galaxy contamination makes determining the
relative colors difficult), we see that there is no strong trend
towards redder colors with fainter
magnitudes.  
As such, it would appear that dust is not the explanation the cause of
the trend of \aro\ in $L-z$ space.

A more accurate determination of the $L$ and $z$ dependence of \aro\
is a question suitable for its own investigation (e.g.,
\citealt{Steffen06}).  We will leave our analysis in terms of \aro,
noting that the trends could change with \daro.

\section{Results}
\label{sec:results}

\subsection{The $L$ and $z$ Distributions}
\label{sec:Lz}

\begin{figure}[h!]
\plottwo{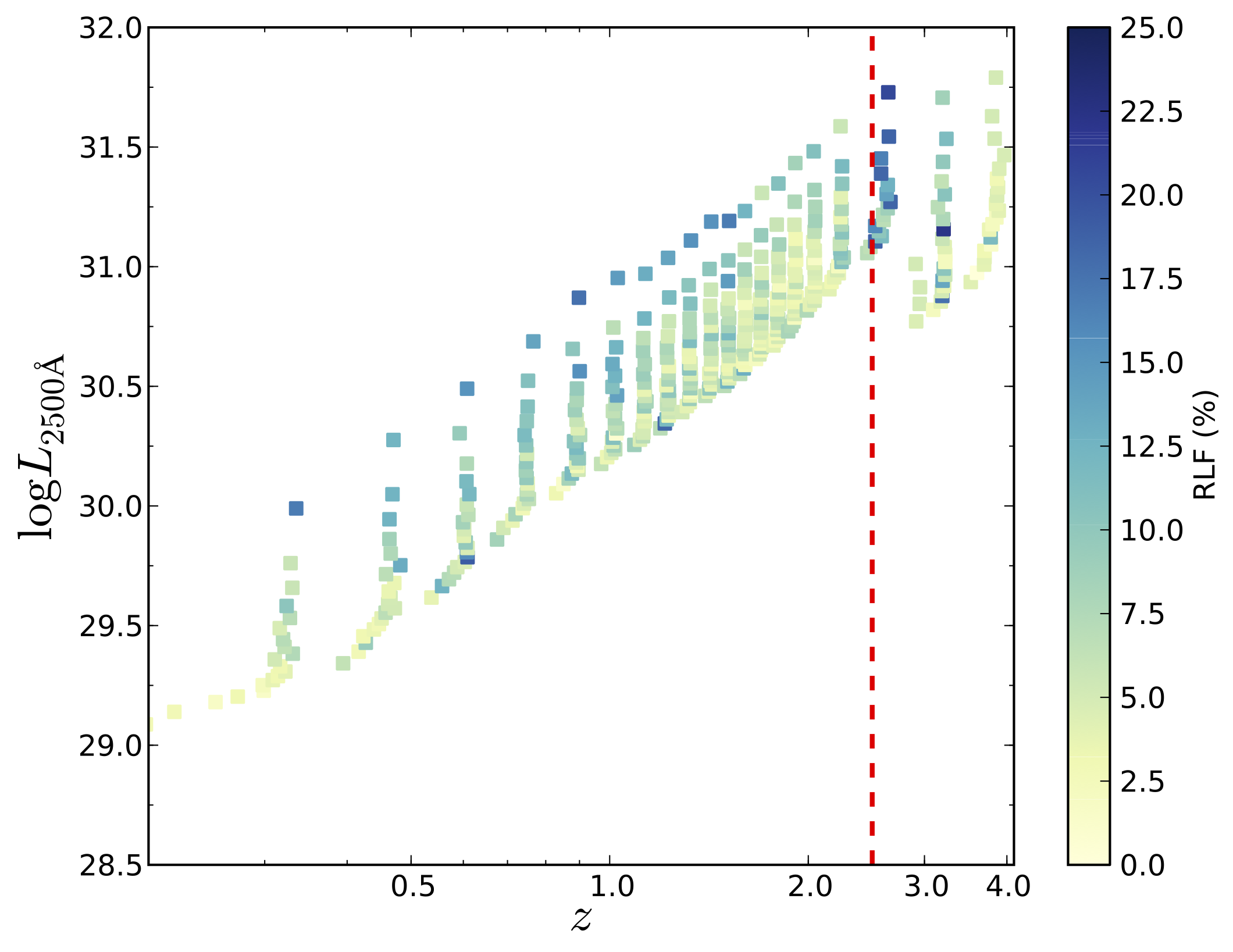}{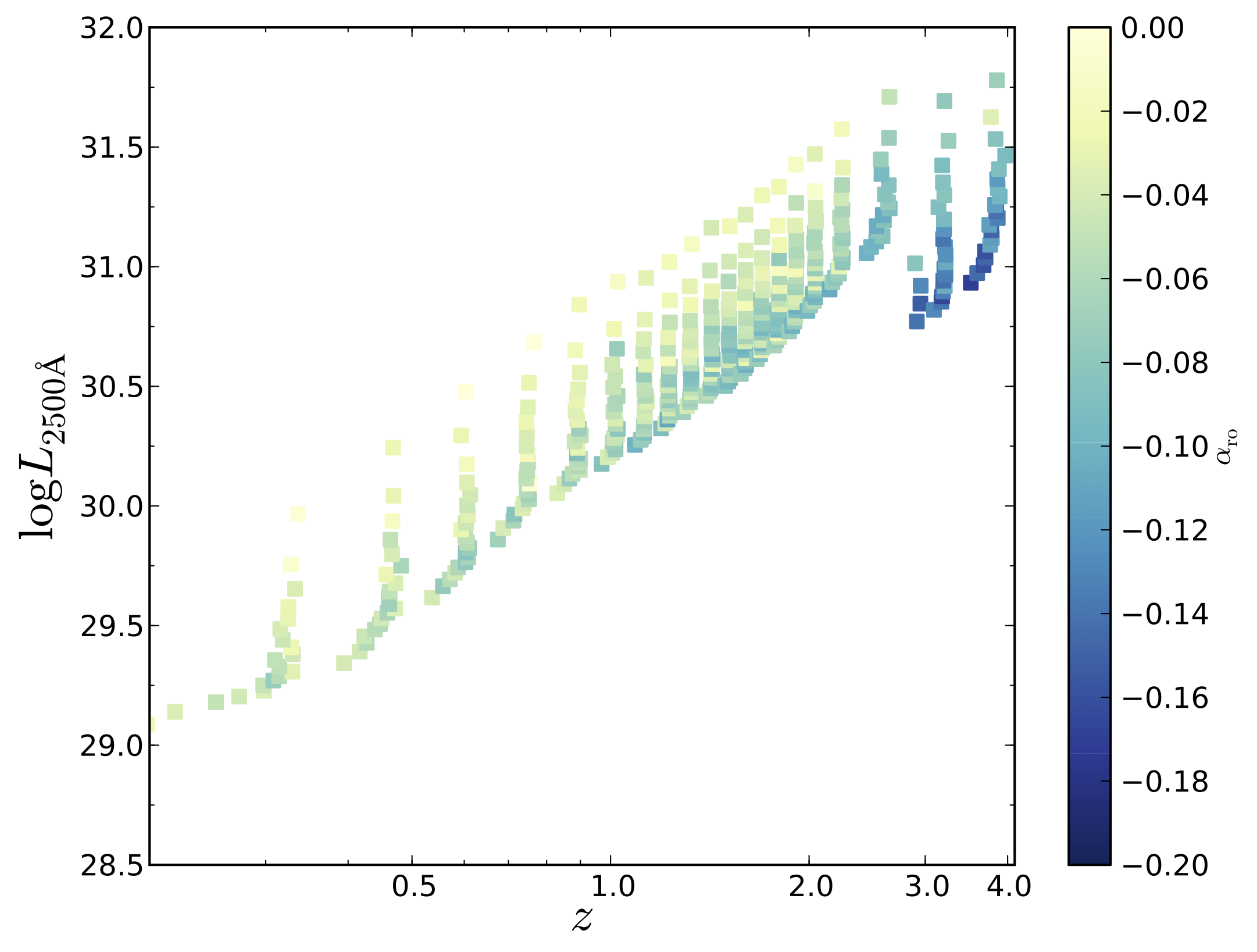}   
\caption{({\em Left:})  RLF as a function of both \ltwofive\ and redshift for optically detected quasars within the FIRST observing area for Sample B. The RL completeness correction (see Section \ref{sec:firstlim}) has been applied.  The boundary on the lower edge represents the SDSS flux limit of $i<19.1$ for $z<3$ and $i<20.1$ for $z>3$.
The trend seen confirms the results of \citet{Jiang07} by demonstrating a decrease in RLF with increasing redshift and decreasing luminosity.
({\em Right:})  Radio to optical spectral indices ($\alpha_{\rm ro}$) of median stacked quasars as a function of both redshift and \ltwofive. 
This trend is completely opposite to that found for the RLF in the left panel. The median stacking shows stronger radio sources (relative to the optical) with decreasing optical luminosity (at fixed redshift) and with increasing redshift (at fixed optical luminosity).
}
\label{fig:f20}
\end{figure}

We begin our comparison of mean and extreme radio properties of
quasars in the $L-z$ parameter space.  
Figure \ref{fig:f20} (left) shows how the RLF of
equally populated bins depends on both redshift and \ltwofive\ for
Sample B.
We find that the RLF declines with increasing redshift (for a given
luminosity) and decreasing luminosity (for a given redshift). These
results would appear to confirm the findings of \citet{Jiang07} by
using at least twice the number of sources. 

Contrasting the RLF trend in the left-hand panel of
Figure~\ref{fig:f20} is the \aro\ trend in the right-hand panel.
Specifically, we find that quasars are stronger radio sources
(relative to the optical) with decreasing optical luminosity (at fixed
redshift) and with increasing redshift (at fixed optical luminosity).
Thus,
it appears that the {\em mean} radio properties of quasars are not
following the same trends as the {\em extreme} RL population.
\citet{Singal11} similarly find increasing radio loudness with
increasing redshift. Their apparent discrepancy with the results of
\citet{Jiang07} can be explained by the difference between the mean
radio-loudness (as is shown in Figure~\ref{fig:f20} [right] and in
\citealt{Singal11}) and the RLF (as is shown in Figure~\ref{fig:f20} [left]
and in \citealt{Jiang07}). Our results are consistent with both papers
when considered in this light. Thus, for these two parameters, the
{\em mean} radio properties of quasars are not following the same
trends as the {\em extreme} RL population. Indeed, \citet{Balokovic}
also find that, as redshift increases, quasars become both more RL on
average but also less likely to inhabit the formally RL tail of the
distribution.

We note that the trends in Figure~\ref{fig:f20} are such that the RLF
declines (and the mean radio loudness increases) in the direction
following decreasing $i$-band magnitude (see also
\citealt[Figure~7a]{Jiang07} and
\citealt[Figure~10]{Balokovic}). As it is not clear why an intrinsic
quasar property should be a strong function of the apparent magnitude,
these results must be taken with a grain of salt. As noted in
Section~\ref{sec:firstlim}, the completeness correction should be good
down to a radio flux of $\sim$2.5 mJy. However, plugging that value
into Equation~5 of \citet{Ivezic}, we find that our analysis is only
robust to $i=17.9$ which is not deep enough to determine if the
separate RLF trends in \ltwofive\ and $z$ are real or due to
incompleteness (or some other selection effect).  Thus, for the case of
the RLF, there must be concern that incompleteness could be
causing that dependence.  A radio survey covering a significant
fraction of the FIRST area and to at least 3$\times$ the depth of
FIRST would be needed to test this effect.  However, our stacking
analysis should be independent of the completeness of FIRST, which
argues that the \aro\ trend with optical magnitude may be real.



Another issue with this type of analysis is that if the radio
distribution does indeed require two components (or if it is bimodal),
then it may be the case that the dividing line between the populations
should change with luminosity as noted by \citet{Laor03}. Thus, it is
possible that we could be under- (or over-) stating the trends with
RLF in Figure~\ref{fig:f20}. As we cannot establish to what extent
these trends are robust, we move on to looking for other demographics
to provide further constraints on the nature of radio-loud emission in
quasars.  We will discuss the interpretation of Figure~\ref{fig:f20}
further in Section~\ref{sec:Discuss}.

\subsection{Accretion Disk Winds: Principal Component and CIV Analyses}
\label{sec:CIV}

As noted by \citet{White07}, the problem is essentially that there is
no practical way to identify from optical properties of quasars which
{\em individual} quasars are likely to be radio-loud. We hope that
extending our analysis to more detailed spectral properties of quasars
in the optical/UV will offer more insight. Arguably the most in-depth
analysis of quasar spectral properties has come from the Principal
Component Analysis (PCA) first carried out by \citet{BG92}.

\citet{BG92} showed that significant new insight could be gained by
examining the range of differences in quasar continua and emission
lines using a PCA (or ``eigenvector'') analysis. They found that the
properties of the H$\beta$, \ion{O}{3}], and \ion{Fe}{2} emission
lines were well correlated with other differences seen in quasar
spectra. Moreover, they found that these differences were correlated
with the radio continuum in a way that suggested that RL and RQ
quasars are not ``parallel sequences'' due to a lack of RQs
matching the extremes of the RL sample. \citet{Boroson02} extended
this work with a larger sample and \citet{BF99} and
\citet{Sulentic00a} added additional line and continuum features to
this matrix of quasar ``eigenvectors''.

\citet{Sulentic00b} showed that much of the information from the first
eigenvector of quasar properties is captured by simply looking at the
FWHM of H$\beta$ and the strength of optical \ion{Fe}{2} emission
relative to H$\beta$, R$_{\rm Fe \, II}$ = W(\ion{Fe}{2} $\lambda$4570
blend)/W(H$\beta_{\rm BC}$). They used this diagram to divide quasars
into two populations (A/B). While there is a continuum between the
populations, it is useful to think of the extrema in this context, and
they found that RL quasars are generally isolated to Population B,
whereas RQ quasars appear in both; see also \citealt{Zamfir2008}.

\begin{figure}[h!]
\plottwo{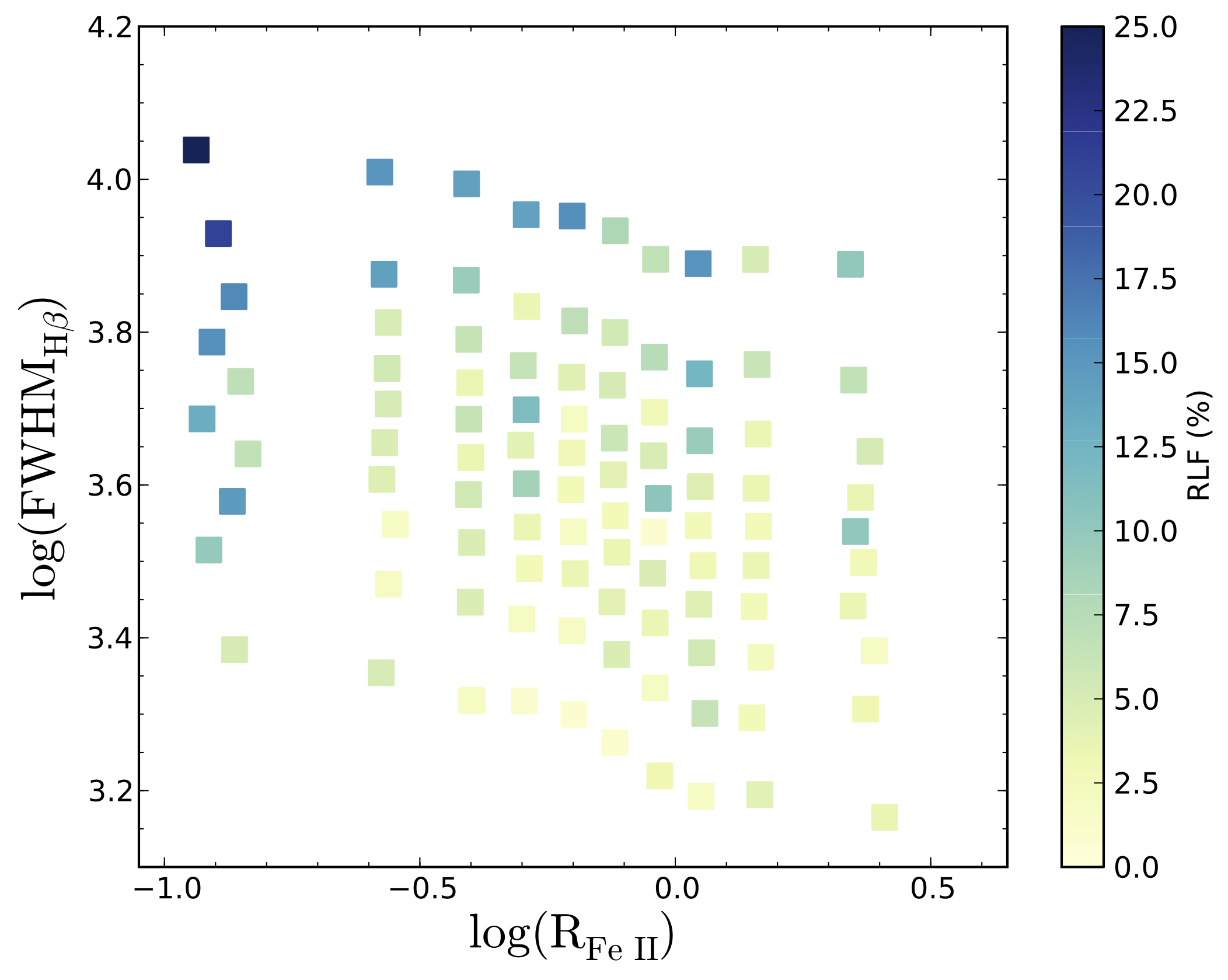}{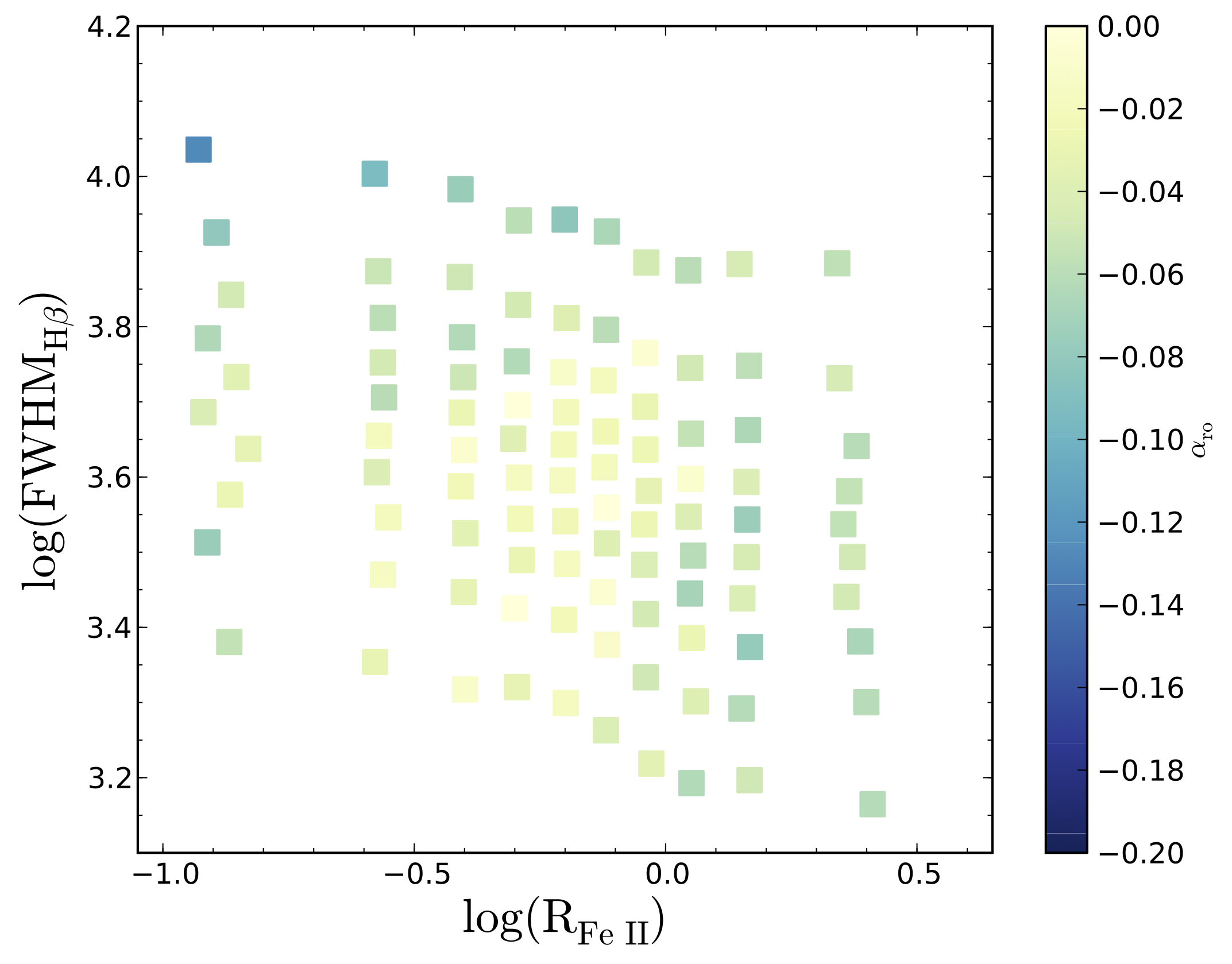} 
\caption{Radio properties of quasars in the low-$z$ EV1 (FWHM H$\beta$ vs.\ R$_{\rm Fe \, \rm II}$) parameter space for optically detected quasars within the FIRST observing area. ({\em Left:}) The RLF, where the RL completeness correction (see Section \ref{sec:methods}) has been applied. The highest RLF bins are concentrated to the high FWHM H$\beta$-low R$_{\rm Fe \, II}$ corner of this low-$z$ EV1 parameter space (in agreement with \citealt{Sulentic00b}). ({\em Right:}) Radio to optical spectral indices ($\alpha_{\rm ro}$) of stacked quasar cutouts in low-$z$ EV1 parameter space.  Here there is not as clear a trend, but the bin with the most radio-loud sources also has the largest H$\beta$ FWHM and the weakest \ion{Fe}{2}.}
\label{fig:f21}
\end{figure}

In that context, we consider the radio properties of the quasars in
our samples in this simplified ``Eigenvector 1 (EV1)'' parameter
space. The left panel of Figure~\ref{fig:f21} shows how the RLF of
equally populated bins evolves in low-redshift EV1 parameter space
(FWHM H$\beta$ vs.\ R$_{\rm Fe \, II}$) for all samples, while the
right panel gives the median \aro\ values. The highest RLFs are found
in the top-left (typically hard spectrum) corner of each panel,
consistent with the findings of \citet{Sulentic00b}. Importantly,
there is no gradient in optical magnitude in this parameter space, so
this result must be more fundamental than our analysis of the $L-z$
distribution, which was subject to vagaries of FIRST completeness
corrections.

There is less of a discernible trend in the mean radio properties as
shown in the right-hand panel of Figure~\ref{fig:f21} than for the RLF
in the left-hand panel. However, we note that the most RL sources
(most negative \aro) are still those in the top-left corners with the
broadest H$\beta$ and weakest \ion{Fe}{2} emission lines.

While EV1 encodes the largest differences in otherwise similar quasar
spectra, the objects which this type of analysis is based upon are
necessarily low-redshift (as the spectrum must cover
H$\beta$). However, the mean SDSS quasar has a redshift closer to
$z\sim1.5$, where the EV1 parameters are no longer included in the
optical. To this end, it has been shown that the \civ\ emission line
can be used to isolate extrema in quasar properties at high-redshift
in a manner similar to EV1 at low-redshift
\citep[e.g.,][]{Sulentic00a, Richards02b, Sulentic07, Richards11}.  It
would appear that high-redshift quasars occupy a broader parameter
space than low-redshift quasars, presumably due to a larger diversity
of black hole masses and accretion rates. \citet{Richards11} argue that
this diversity can be connected to the ability of a quasar (through
its intrinsic SED) to power a strong radiation line-driven wind and
that the \civ\ line represents an EV1-like diagnostic.

Specifically, \citet{Richards11} argue that the \civ\ emission line
properties of a quasar, particularly the equivalent width (EQW) and
the ``blueshift'' (the offset of the measured rest-frame line peak
from the expected laboratory value), can provide an understanding of
the trade-off between the disk and wind parameters of quasars
\citep[see][]{Murray95,Elvis00,Proga00,Leighly04,Casebeer06,Leighly07}.
The \civ\ emission line is a good diagnostic for a variety of
reasons. Aside from Ly$\alpha$, it is the most conspicuous emission
line in high-redshift quasars, which allows for high S/N measurements
of this line in many objects. More importantly, the EQW and blueshift
of \civ\ have the largest range of emission line properties for all
high redshift quasars, increasing our ability to locate trends. Additionally, the
blueshifting of the \civ\ line with respect to the quasar's rest frame
\citep{Gaskell82,Wilkes84} is practically universally present in
spectra of luminous quasars \citep{Sulentic00b,Richards02b}.

In the context of a \civ\ analysis, \citet{Richards11} find that RQ
quasars span the full space occupied by both quasar types---in
contrast to findings of \citet{BG92} and \citet{Boroson02}. On the
other hand, the RL quasars were largely confined to that part of
parameter space with small \civ\ blueshifts (and large EQWs)
\citep[see][Fig.~7]{Richards11}.
\citet{Richards11} interpret this result in a disk-wind
framework and argue that, on average, RL quasars have weaker radiation
line-driven winds than RQs.

\begin{figure}[h!]
\plottwo{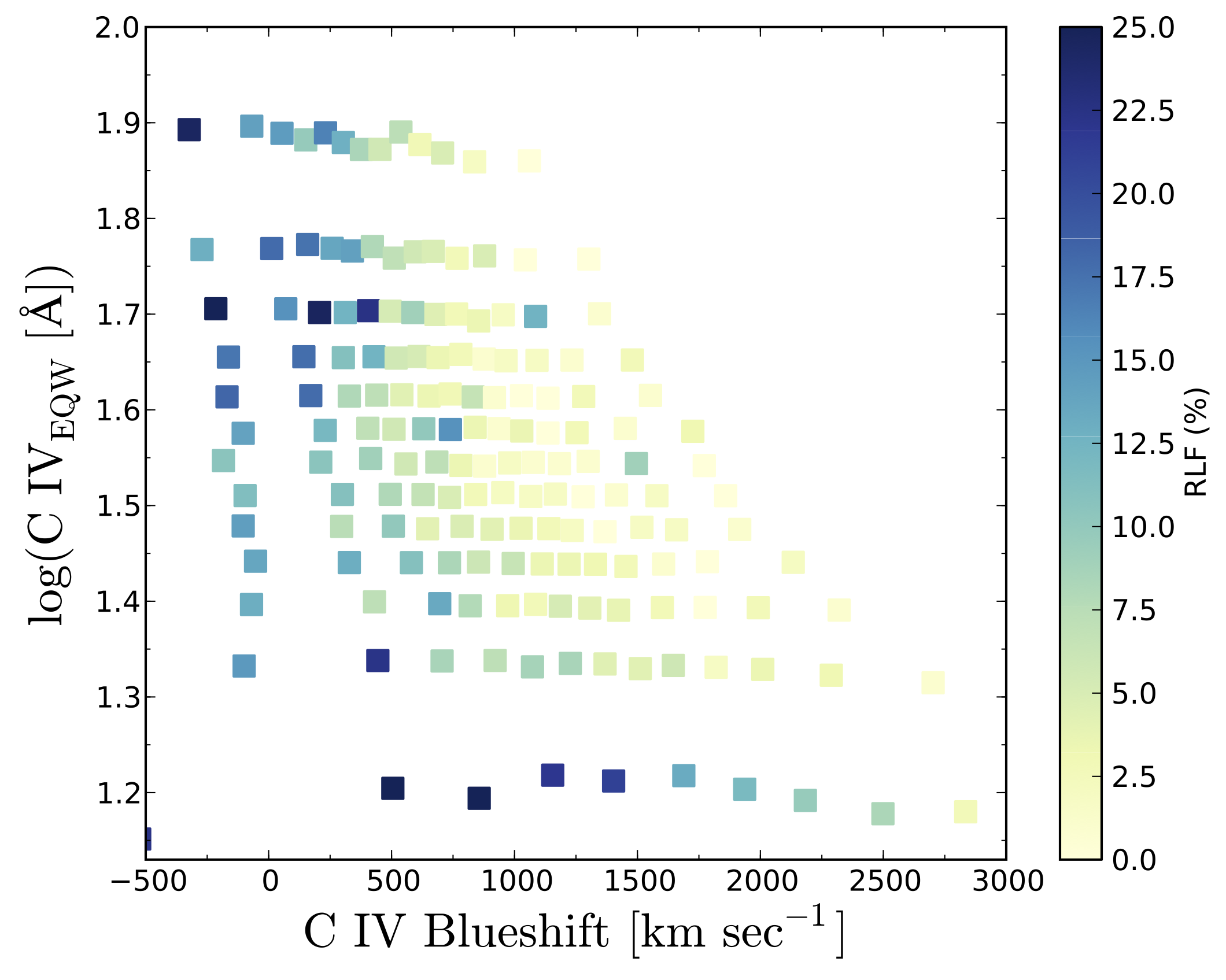}{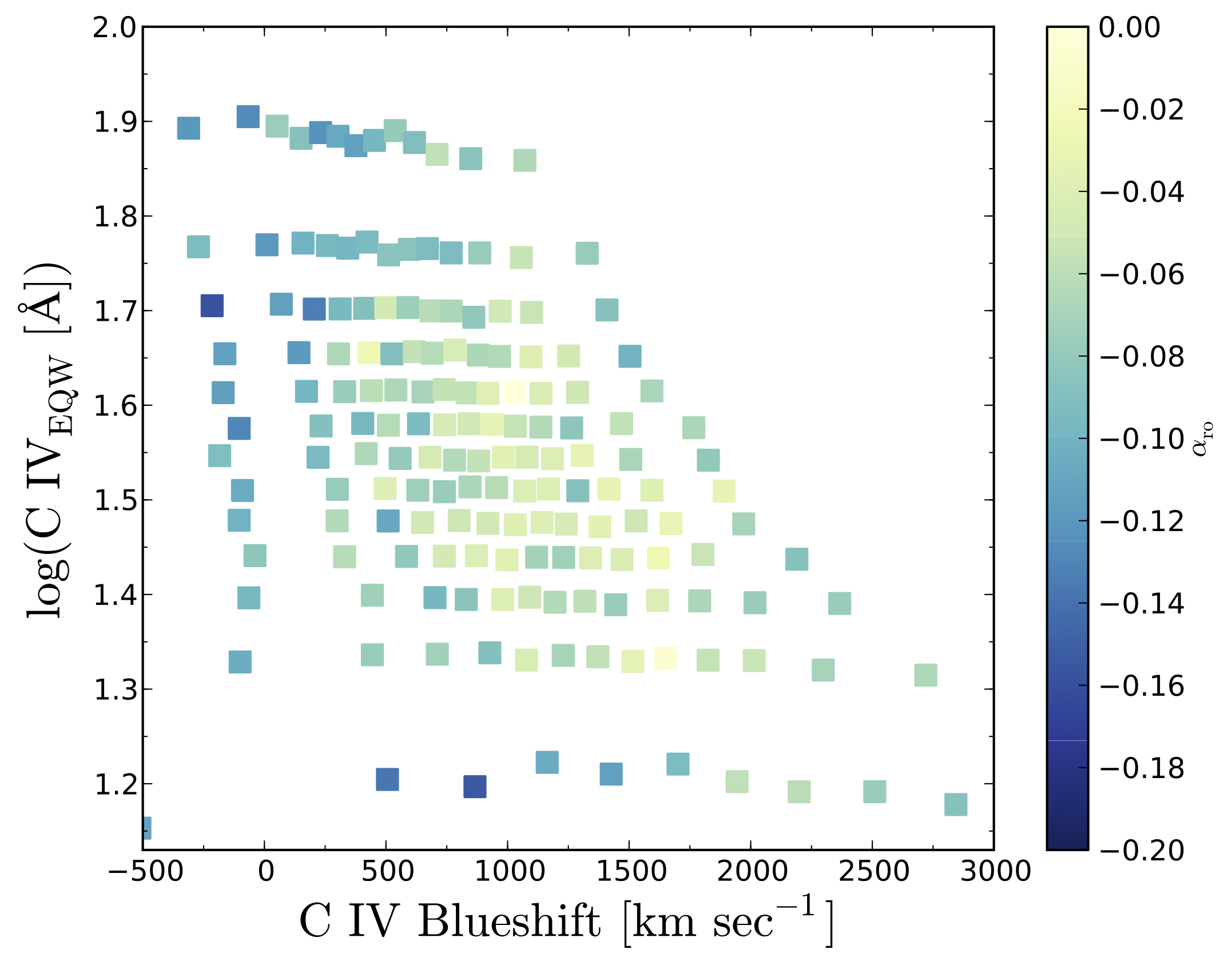}
\caption{Radio properties as a function of both \civ\ equivalent width (EQW) and blueshift for optically detected quasars within the FIRST observing area. ({\em Left:}) The RLF, including the RL completeness correction (see Section \ref{sec:methods}). The RLF primarily decreases from low to high blueshift with no discernible RLF trend in EQW; see also \citet{Richards11}. Objects in the row with the lowest EQWs likely have higher RLFs due to selection effects.  ({\em Right:}) Radio to optical spectral indices ($\alpha_{\rm ro}$) of stacked quasar cutouts in \civ\ parameter space.  As with the EV1 parameter space, the mean shows a weaker trend than the RLF that evolves in the same direction.
}
\label{fig:f22}
\end{figure}

Here we take the analysis of the radio properties of quasars in \civ\
parameter space one step further than \citet{Sulentic07} and
\citet{Richards11}.  Specifically, we have repeated our dual analyses
in \civ\ parameter space. Figure \ref{fig:f22} (left) shows that the
RLF primarily decreases from low to high blueshift. No discernible RLF
trend exists with respect to EQW\footnote{Objects with very low EQWs were examined by eye. These objects were found to be atypical, being mostly BALs, miniBALs, relatively featureless, highly reddened, etc. It may be that such sources are intrinsically more radio-loud, but it is more likely that such objects appear in the sample due to the bias towards radio-detected quasars in the parts of parameter space where optical selection is inefficient.}.  In terms of the mean
radio properties shown in the right-hand panel of
Figure~\ref{fig:f22}, the mean trend is in the same general direction
as the RLF trend, but is weaker---similar to the EV1 trends in
Figure~\ref{fig:f21}.  However, it does appear that
small-blueshift quasars are more radio loud, on average, than those with large \civ\ blueshifts.

Ideally our goal here is to be able to identify a UV emission line
parameter that would predict whether or not an {\em individual} quasar
is radio-loud.  Although we have not accomplished that goal, we {\em
  can} use this analysis to improve the statistical prediction from a
blanket $\sim10$\% to a fraction that ranges from $\sim$0\% to
$\sim$30\% as a function of \civ\ emission line properties. In one
sense this {\em does} allow a prediction of radio properties for at
least {\em some} quasars as it seems that quasars at the extreme end
of the \civ\ blueshift distribution (for a given \civ\ EQW) are
exceedingly unlikely to be radio-loud.

\subsection{Black Hole Mass and Accretion Rate}
\label{sec:BH}

Two of the most important properties that govern how quasars behave
are the mass of the central black hole (BH) and its accretion
rate. While we cannot measure the mass and accretion rate of a black
hole directly, we can derive {\em estimates} of these values by using
so-called BH mass scaling relations \citep[e.g.,][]{VP06} in
conjunction with emission line and continuum information. We now
explore how the RLF behaves as a function of these BH mass estimates.

\begin{figure}[h!]
\plottwo{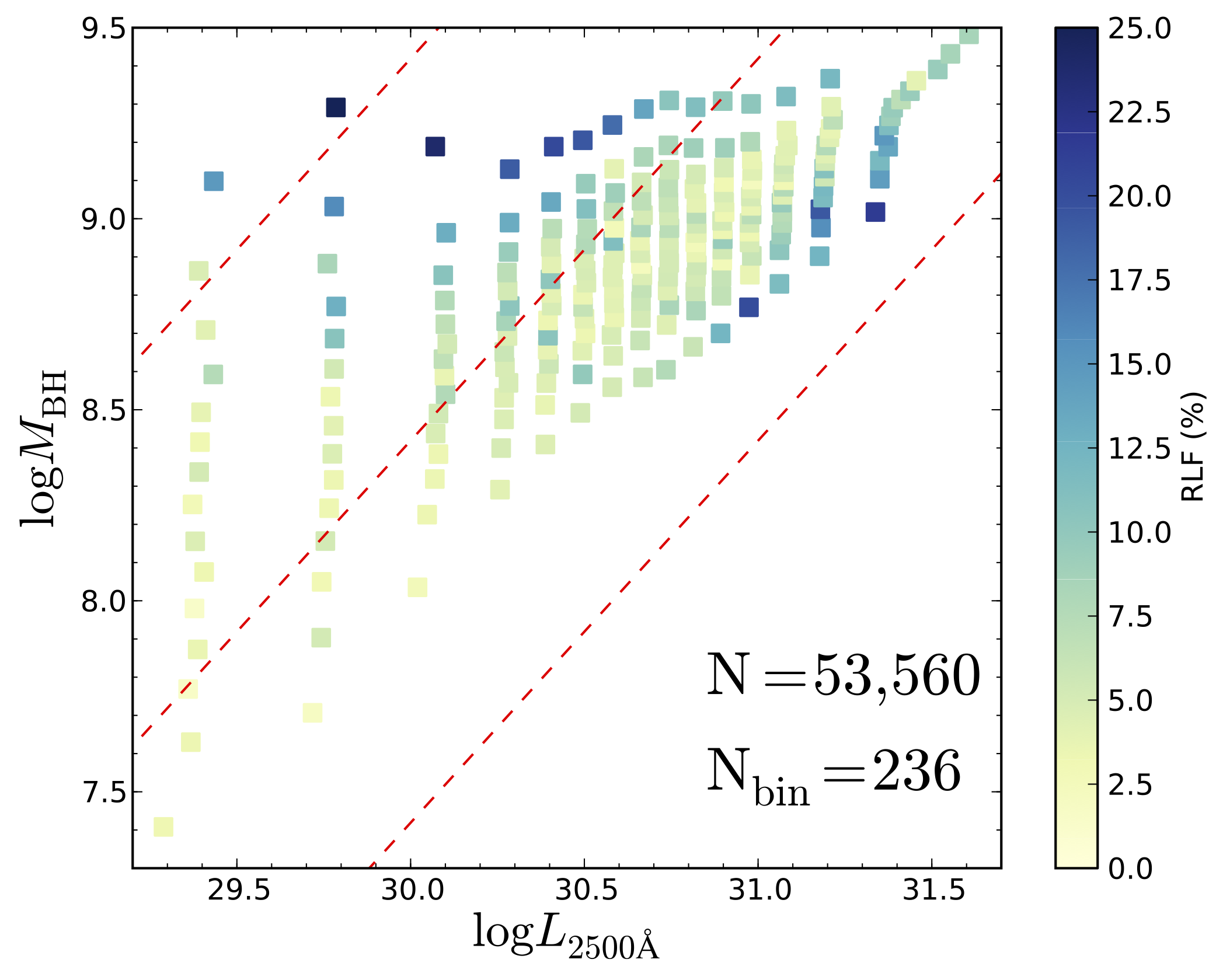}{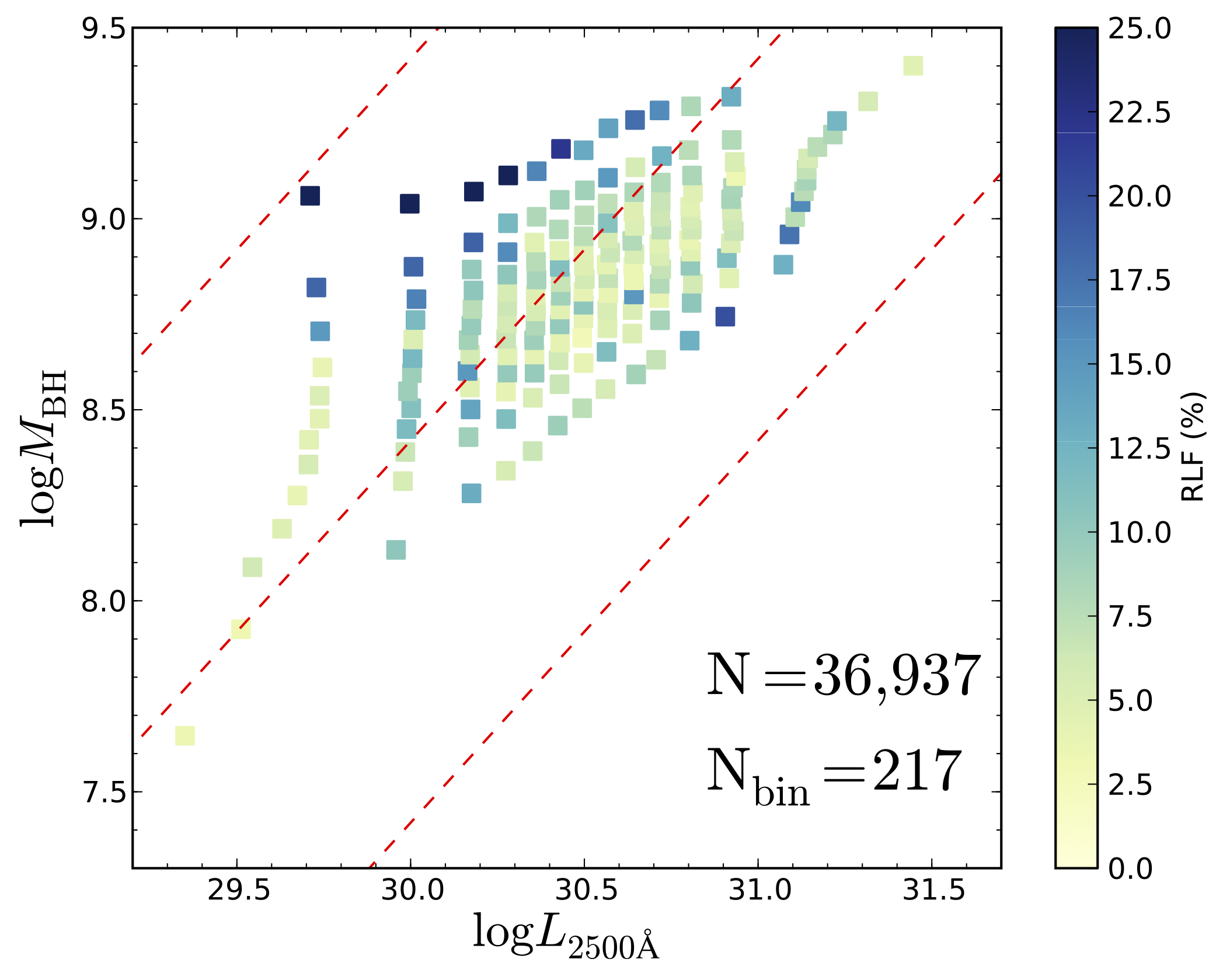} 
        \caption{RLF as a function of BH mass and accretion rate for quasars within the FIRST observing area for Samples B (left) and D (right). As with all of our RLF plots, we have corrected for RL completeness (see Section~\ref{sec:firstlim}). For both samples, the highest RLFs are positioned in the corners of parameter space: high BH mass for the lowest accretion rates and low BH mass for the highest accretion rates. The dashed red lines show where $L/L_{\rm Edd} = 0.01$, $0.1$, and $1.0$ in order from top-left to bottom right. We assume $L_{\rm Bol} = 2.75 L_{\rm 2500 \, \AA}$ \citep{Coleman13}.}
\label{fig:f23}
\end{figure}

BH masses were compiled by \citet{Shen11} and have been corrected as
described in Section~\ref{sec:DR7QC}. Assuming a bolometric correction
of $L_{\rm Bol} = 2.75 L_{\rm 2500 \, \AA}$ \citep{Coleman13}, we can
convert \ltwofive\ to $L_{\rm Bol}$ to determine the Eddington Ratio,
$L_{\rm Bol}/L_{\rm Edd}$, where $L_{\rm Edd}$ is derived directly
from the BH mass estimate. Figure \ref{fig:f23} shows how the
radio-loud fraction (of equally populated bins) depends on \ltwofive\
(effectively accretion rate) and black hole mass.  We have presented
the data in this way instead of plotting $L/L_{\rm Edd}$ directly;
\citet{Richards11} argue that it is optimal to investigate BH mass and
accretion rate separately in case there are any threshold effects (low
mass/low accretion rate can have the same $L/L_{\rm Edd}$ as high
mass/high accretion rate, but potentially very different
properties). Nevertheless, $L/L_{\rm Edd}$ appears as dashed red lines
in Figure \ref{fig:f23}. The line on the lower-right indicates
$L/L_{\rm Edd}=1$, or (theoretical) maximal accretion (per mass),
while the line on the top left is at $L/L_{\rm Edd}=0.01$ and the line
in the middle represents $L/L_{\rm Edd}=0.1$.

In Figure \ref{fig:f23}, we find that the bins with the highest RLFs
are situated in the corners of parameter space, specifically high BH
mass/lowest accretion rate and low BH mass/highest accretion rate. The
lowest RLFs exist in a diagonal band that stretches from low BH mass
for the lowest accretion rates to high BH mass for the highest
accretion rates.

\begin{figure}[h!]
       \centering
        \begin{subfigure}[b]{0.45\textwidth}
                \centering
                \includegraphics[width=\textwidth]{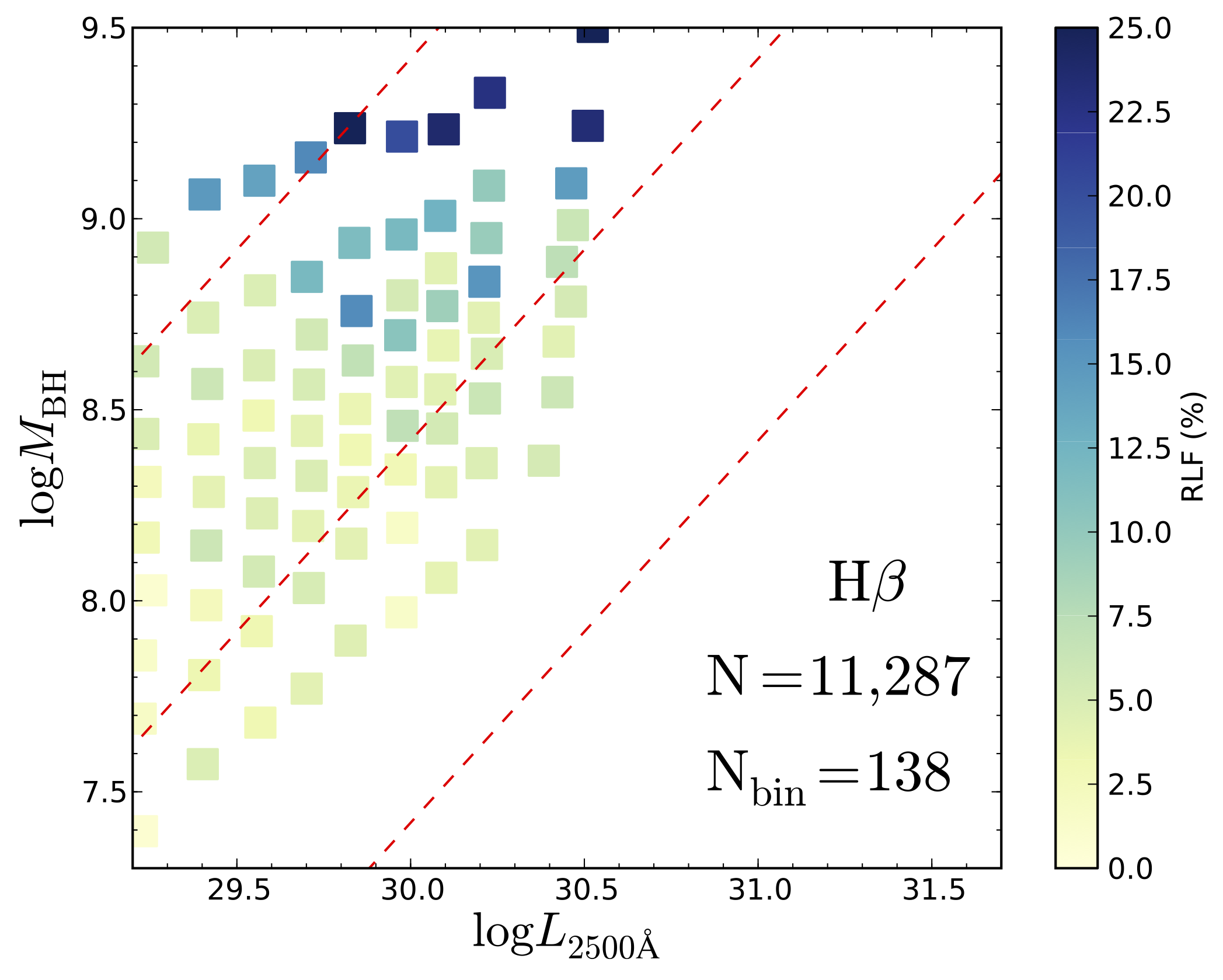}
        \end{subfigure}
        ~ 
                   \begin{subfigure}[b]{0.45\textwidth}
                \centering
                \includegraphics[width=\textwidth]{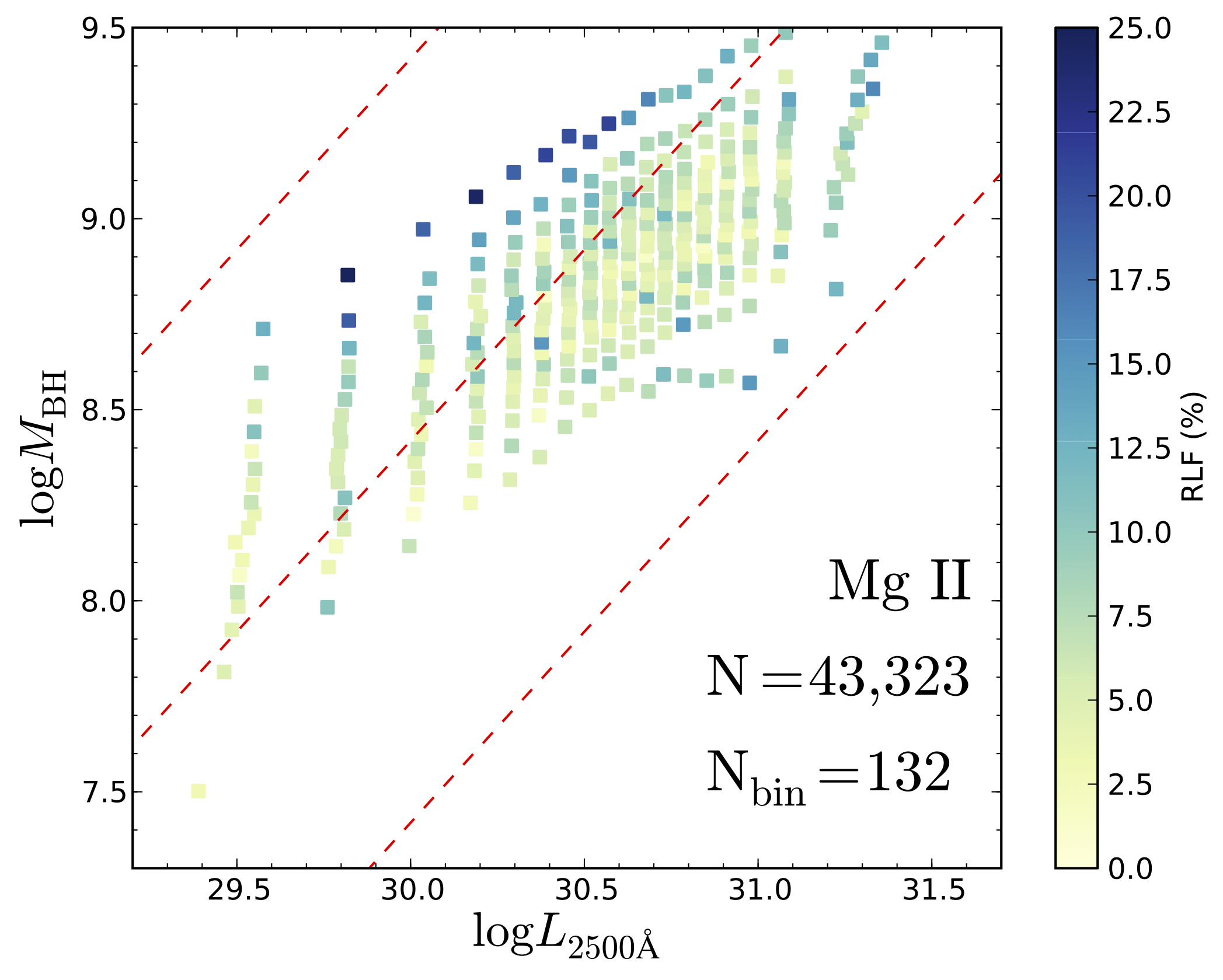}
        \end{subfigure}
        
            \begin{subfigure}[b]{0.45\textwidth}
                \centering
                \includegraphics[width=\textwidth]{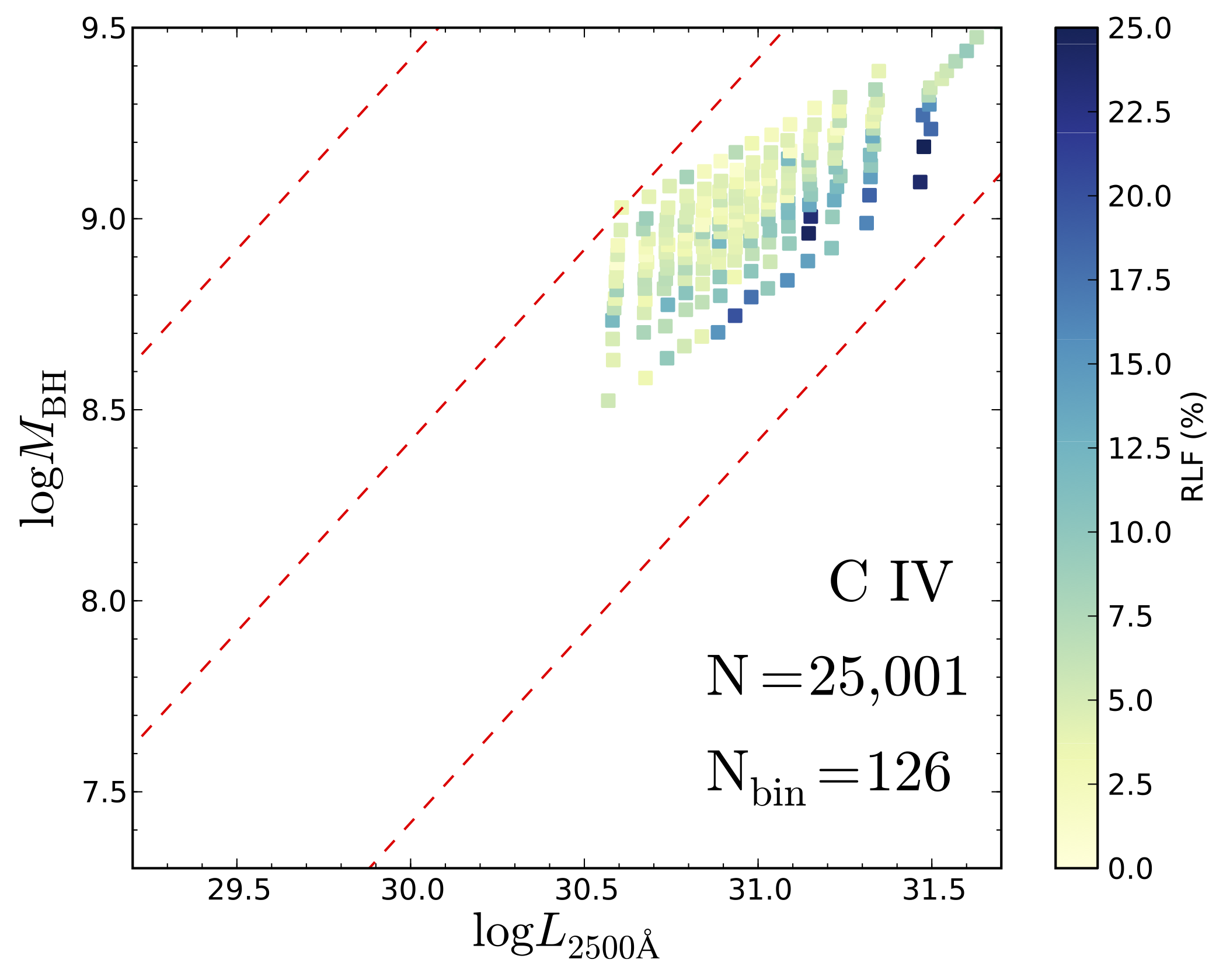}
        \end{subfigure}
\caption{Figure~\ref{fig:f23}, Sample B, split into different populations based on the spectral line used to calculate the BH masses. Each panel effectively represents quasars within different redshift bins \citep[see][Sec. 3.7]{Shen11}. For the two lowest redshift samples (H$\beta$ and \ion{Mg}{2}), the highest RLFs belong to the highest BH mass bins for these samples. The lowest BH mass bins in the highest redshift sample (\civ) are (apparently) the most RL. We suggest that this difference is indicative of errors in the \civ\ BH mass estimates.  The dashed red lines from top-left to bottom right show where $L/L_{\rm Edd} = 0.01$, $0.1$, and $1$, in order. We assume $L_{\rm Bol} = 2.75 L_{\rm 2500 \, \AA}$ \citep{Coleman13}.}
\label{fig:f24}
\end{figure}

Since the estimation of BH masses in high-redshift quasars by way of
scaling relations is not an exact science and is dependent on the
emission lines used, we also create subsamples based on the origin of
these masses (see Figure \ref{fig:f24}). Specifically, our sample
includes quasars whose masses are estimated using the H$\beta$,
\ion{Mg}{2}, and \civ\ emission lines; this roughly corresponds to
$z<0.7$, $0.7 \le z<1.9$, and $z \ge 1.9$, respectively.  The mass
estimates computed using the H$\beta$ and \ion{Mg}{2} emission lines
are thought to be reliable, shown to be within a factor of 2.5 of the
masses found using reverberation mapping
\citep{McLureJarvis02}. Indeed, for the two low-redshift bins we see
behavior that is consistent with what we might expect from past
work. Namely, that the quasars with the highest masses (at a given
luminosity) are the most likely to be radio-loud, though not
exclusively radio-loud; most quasars in the RL regions are still RQ
\citep[see][]{Lacy01}.

However, when we examine the high-redshift subsample, we see something
quite different. Here the {\em lowest} mass quasars appear to be the
most radio-loud. There are a number of potential explanations for this
observation. One possibility is that there is an actual physical
transition at high redshift such that we are more likely to find RL
quasars in high $L/L_{\rm Edd}$ systems. This could be magnified (or
perhaps caused) by a change in the selection of quasars from UV-excess
sources at $z\lesssim2$ to $u$-band dropouts at higher
redshifts. Alternatively, instead of a physical change, perhaps the
high-redshift quasar sample is simply biased against RL quasars with
high-mass (at high redshift)?  However, we consider this unlikely:
although the completeness of FIRST drops with redshift, radio sources
are explicitly targeted for spectroscopy as quasars candidates in the
SDSS surveys. While there are known incompletenesses in the quasar
selection at high-$z$, the most glaring of these has to do with the
presence (or lack thereof) of Lyman-limit absorption systems
\citep{Worseck11} and is independent of the radio properties of the
quasars. Indeed, estimates of the completeness of the SDSS quasar
survey \citep[e.g.,][]{VB05} are inconsistent with the extreme level
of incompleteness that would be required to induce this effect.

Instead we argue that the problem lies in the estimation of BH masses
using the \civ\ emission line. This could arise from more of the \civ\
line being emitted in a wind component than has previously been
thought or in the form of a wind-strength dependence to the
proportionality constant in the radius-luminosity relation
\citep{Richards11}. While it is well-known that determining BH mass
scaling relations from the \civ\ lines are the most challenging
\citep[e.g.,][]{Fine10,Shen11, Assef11, Denney12, Runnoe13, Park13,
  Denney13}, the corrections necessary for the radio-loudness trend to
match the low-redshift BH mass trends are not consistent with the
level of ``tweaks'' to the \civ\ BH mass scaling relations that are
generally advocated. Rather, these BH mass estimates must be {\em
  catastrophically} wrong.  Otherwise, these trends would suggest an
unlikely situation whereby high-redshift and low-redshift quasars have
their radio properties governed by two different process, where the
switch just happens to occur at redshifts where the BH mass estimates
transition from using \ion{Mg}{2} to using \ion{C}{4}.  Specifically,
high-redshift RL quasars would have to have high \lledd, while
low-redshift RL quasars have low \lledd\ \citep[c.f.][]{Shankar10}.
Thus this issue is not just a matter for our analysis but speaks to
the broader problem of the use of BH masses estimated from \ion{C}{4} emission lines.

To reconcile the low-redshift and high-redshift quasars, the RL quasar
masses at high-$z$ are too low by as much 0.5 dex or more and not
simply by $\sim 0.2$--0.3 dex as is usually assumed.  Technically this
statement applies only to the RL quasars.  However, as
\citet{Richards11} argue that there are negligible differences in the
emission line properties of RL and RQ quasars with small \civ\
blueshifts, it must also be true that a large fraction of the masses
computed for RQ quasars are similarly erroneous. Indeed, it would seem
that the \civ\ BH mass estimates are close to being inverted (large BH
mass should be small, and vice versa).  As there is only a weak
correlation between the BH masses estimated from \civ\ and \ion{Mg}{2}
in \citet[Fig.~10]{Shen11}, this is perhaps not surprising.  We will
consider this issue further in Section~\ref{sec:Discuss}.

\begin{figure}[h!]
\plottwo{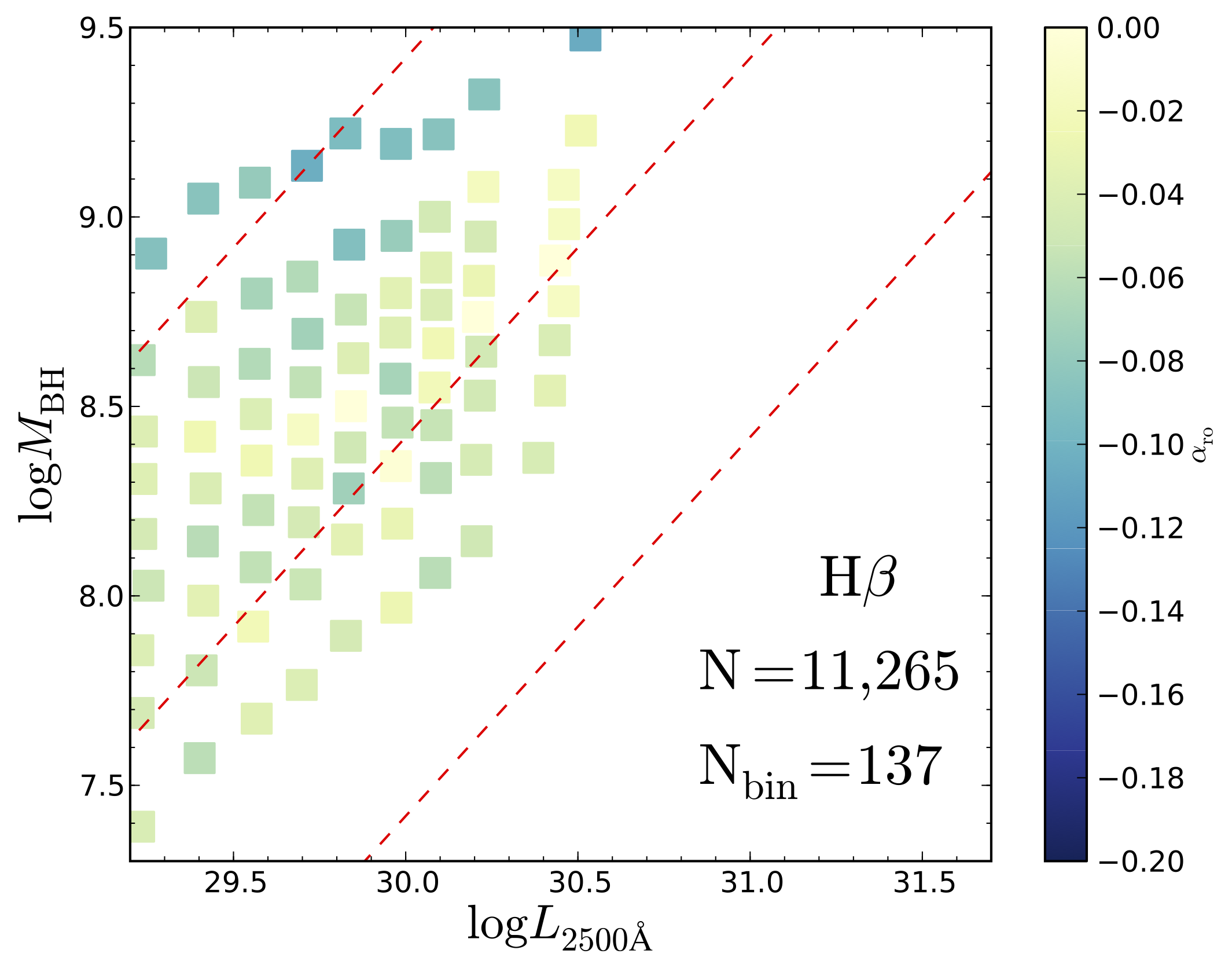}{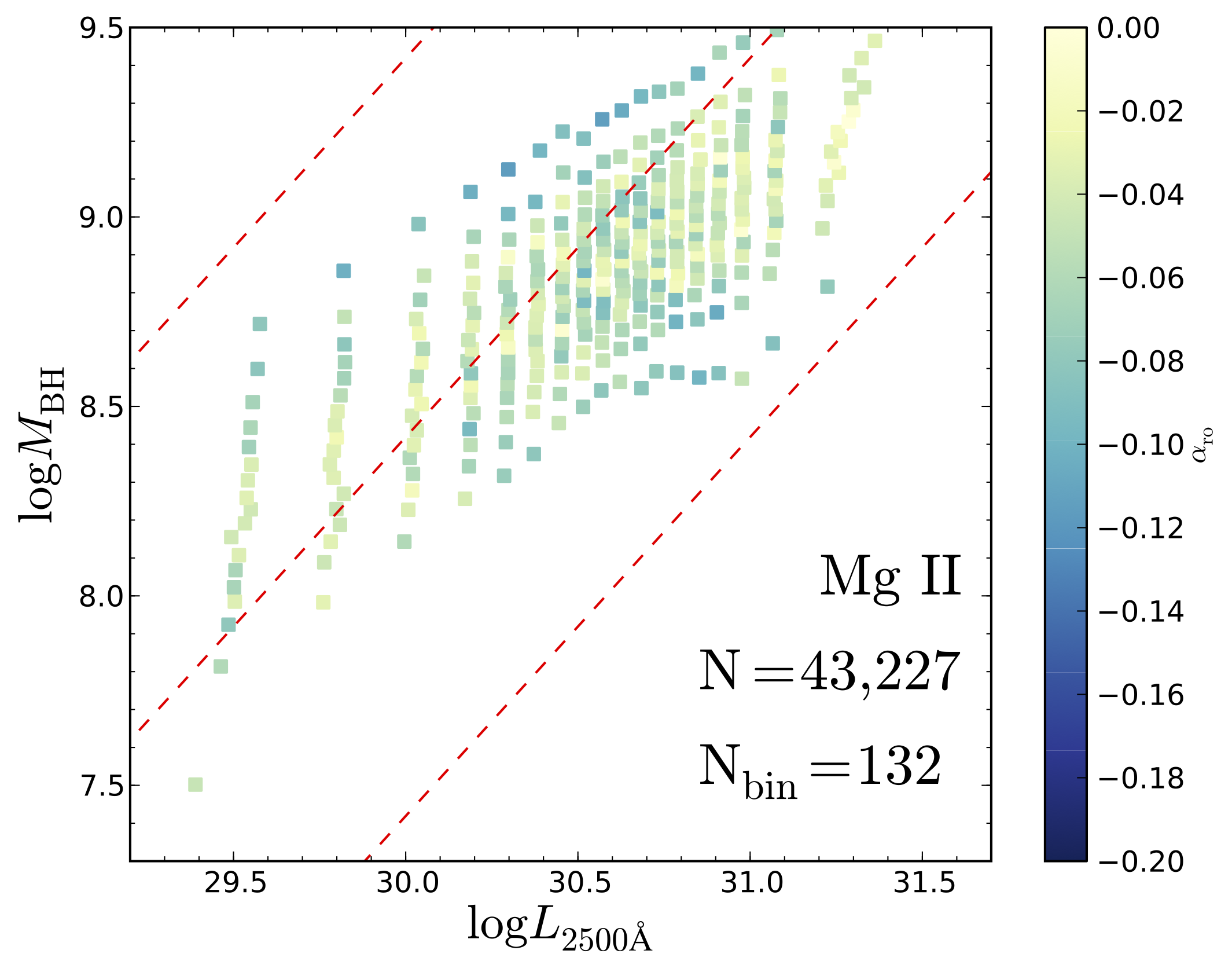}
\caption{Radio-to-optical spectral indices ($\alpha_{\rm ro}$) of stacked quasar cutouts in BH mass parameter space for optically detected quasars within the FIRST observing area for Sample B ({\em Left:} H$\beta$ masses; {\em Right:} \ion{Mg}{2} masses).  The dashed red lines from top-left to bottom right show where $L/L_{\rm Edd} = 0.01$, $0.1$, and $1$, in order. We assume $L_{\rm Bol} = 2.75 L_{\rm 2500 \, \AA}$ \citep{Coleman13}.}
\label{fig:f25}
\end{figure}

We now consider the mean radio-loudness as a function of mass,
accretion rate, and \lledd\ as we did in Figure~\ref{fig:f23},
plotting just the results for H$\beta$ and \ion{Mg}{2}.  Specifically,
Figure~\ref{fig:f25} shows mass vs. luminosity color-coded by \aro.
For H$\beta$, we find strong similarities between this analysis and the
RLF analysis with the most radio loud objects being towards the top
left of each panel such that the
mean radio-loudness increases towards lower \lledd, consistent
with the RLF.  There is no obvious trend for \ion{Mg}{2}.

\subsection{The Mean Radio Properties as a Function of Color}
\label{sec:StacksColor}

We extend our study of the mean radio properties of quasars by
exploring the correlation between the strength of radio emission and
optical color. As before, we split our samples into bins based on the
colors of optically-detected quasars and apply the median stacking
procedure described above. Similar to \citet[][Fig. 14]{White07}, we
will use $\Delta (g-i)$ for our measure of color.  As stated earlier,
$\Delta (g-i)$ is defined in such a way to remove the dependence of
color on redshift. It is roughly equivalent to $\alpha_{\rm opt}$,
the underlying continuum (excluding emission features) in the
optical-UV part of the SED.

\begin{figure}[h!]
\epsscale{0.6}
\plotone{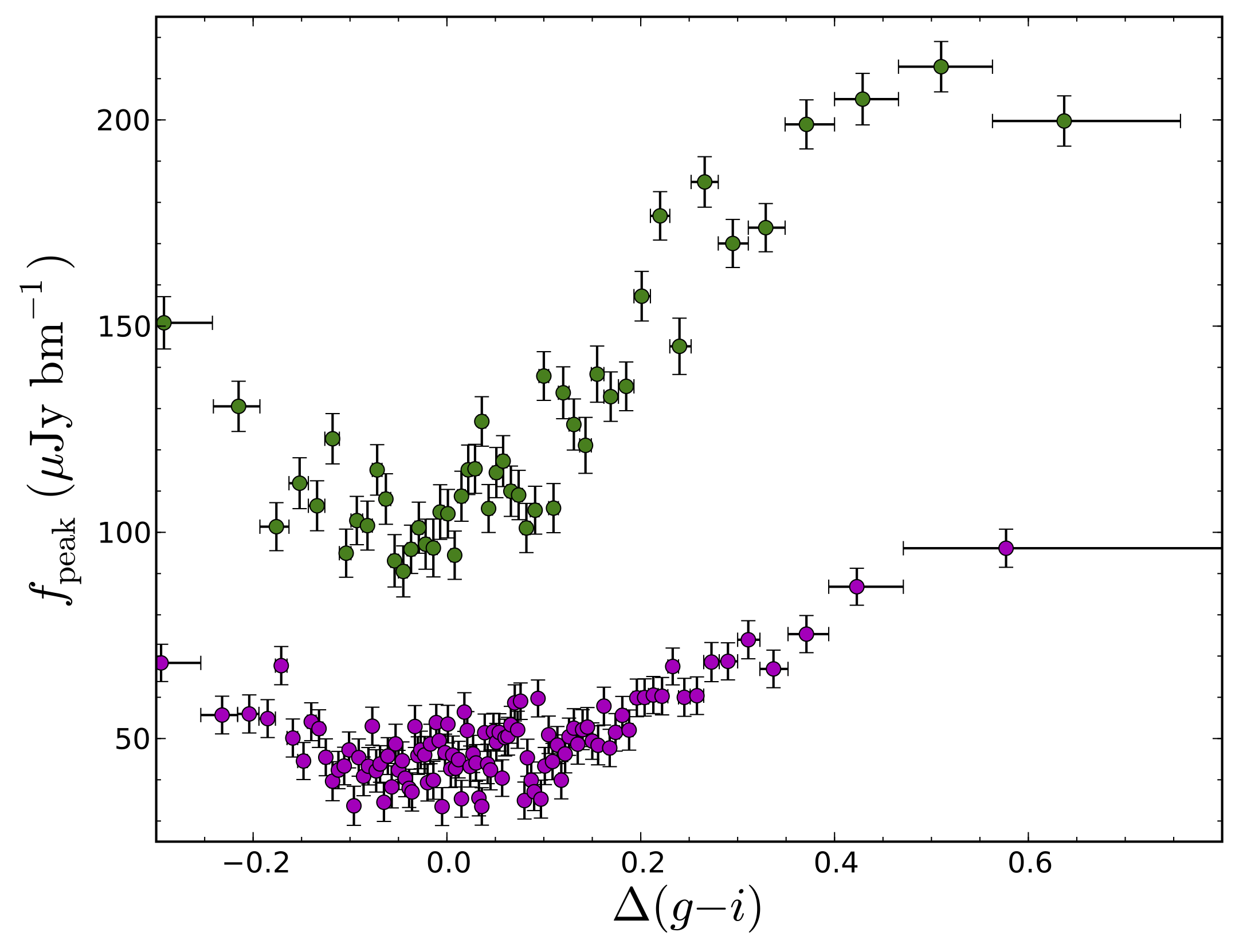}
\caption{Radio fluxes of stacked quasar cutouts as a function of $\Delta (g-i)$ with samples as in Fig.~\ref{fig:f11}.}
\label{fig:f26}
\end{figure}

Figure \ref{fig:f26} shows how median radio flux density varies as a
function of color. Just as in \citet{White07}, we find that bluer and
redder objects have higher radio flux densities with the reddest
objects being the brightest of all.  We find that objects with
$\Delta (g-i) > 0.6$ have peak flux densities 2-3 times larger than
quasars with average colors.
     
\begin{figure}[h!]
\epsscale{0.6}
\plotone{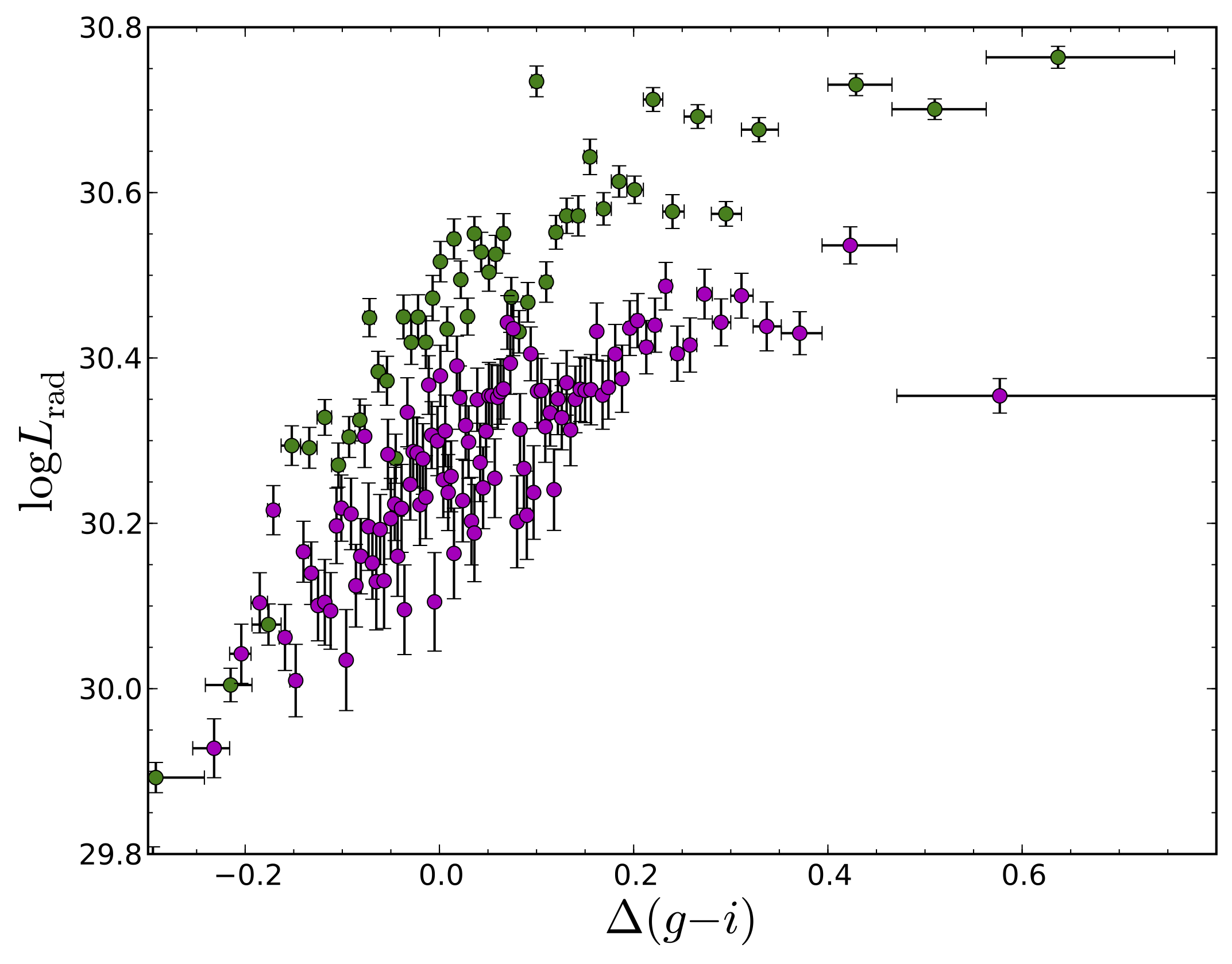}
        \caption{Radio luminosities ($\rm{erg} \, \rm{s}^{-1} \, \rm{Hz}^{-1}$) of stacked quasar cutouts as a function of $\Delta (g-i)$ (Sample B: green; Sample D: purple).}
\label{fig:f27}
\end{figure}

As we wish to analyze the intrinsic properties of
quasars, we naturally examined how radio luminosity changes
with optical color. Figure \ref{fig:f27} shows that, for each of our
samples, radio luminosity increases for redder objects. Thus,
the trend in flux density seen at both color extremes in
Figure~\ref{fig:f26} and in \citet{White07} appears to be artificial:
the blue objects simply have very different luminosities than the red
objects.


\begin{figure}[h!]
\epsscale{0.6}
\plotone{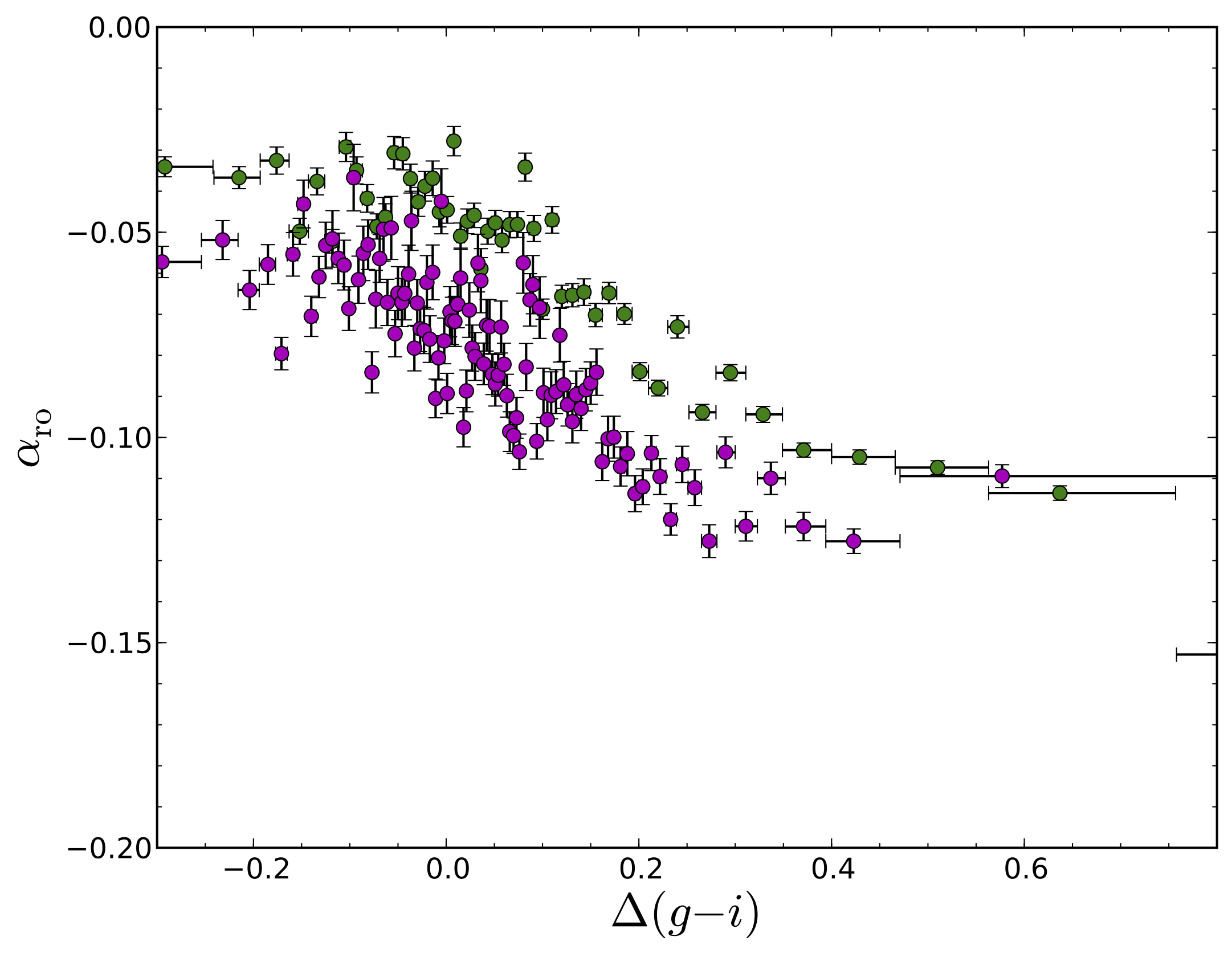}
\caption{Radio to optical spectral indices ($\alpha_{\rm ro}$) of stacked quasar cutouts as a function of $\Delta (g-i)$ (Sample B: green; Sample D: purple).}
\label{fig:f28}
\end{figure}

We can quantify the luminosity-corrected trend, instead, by plotting
$\alpha_{\rm ro}$ as a function of $\Delta (g-i)$.  Figure
\ref{fig:f28} shows that radio-loudness increases for redder quasars
(recall that more radio-loud corresponds to increasingly negative
values of $\alpha_{\rm ro}$). As \aro\ represents another way to
measure the $\log{R}$ parameter (Section~\ref{sec:aro}), our results
agree with those of \citet{White07}, who found an increase in the
$R$-parameter for objects with redder colors.

These trends match our results from Figures \ref{fig:f26} and
\ref{fig:f27} since objects with increasing radio flux and luminosity
should be radio-louder if optical luminosity remains the same.  While
we expect to measure less optical emission for redder objects because
of dust extinction (which does not reduce observed radio emission),
optical extinction by dust does not appear to be the lone cause of
this trend as it also applies to relatively blue quasars.

\section{Discussion}
\label{sec:Discuss}



\subsection{Mass, Accretion Rate, and Spin}

We first consider our results in the context of theoretical work on
the relationship between spin, mass, and accretion rate in
AGNs. \citet{BZ77} and \citet{BP82} provide the framework for how we
think the spin of the BH and the accretion disk, respectively, can be
tapped in the production of radio jets. \citet{Wilson95} argue that
$M_{\rm BH}$ and $\dot{M}$ are not that different for RL and RQ
quasars and that the difference must be related to the spin. Both
\citet{Zamfir2008} and \citet{Richards11} also argue for a similarity
of properties between RL and at least part of the RQ population.

Previous work has suggested that the radio-loudness is dependent on
either the black hole mass or the accretion rate. For example,
\cite{Laor00} and \citet{Lacy01} find that high BH mass is a necessary
(if insufficient) condition for being radio-loud and that there is
also a (weaker) correlation with \lledd. \citet{WooUrry02}, on the
other hand, argue that there is no dependence on BH mass.
\citet{Ho02} similarly finds little BH mass dependence and argues
instead for \lledd\ as the primary driver; however,
\citet{JarvisMcLure02} claim that the use of $R$, rather than $L_{\rm
  rad}$, by \citet{Ho02} to characterize quasars as RL or RQ led to
the uncorrelated results. \citet{JarvisMcLure02} also assert that
doppler boosting brightens the intrinsic $L_{\rm rad}$ of flat
spectrum sources and that the corrected data agrees with a dependence
of the form $L_{\rm rad} \propto M_{\rm BH}^{2.5}$
\citep{Dunlop03}. \citet{Sikora07}, using a sample that is largely
complementary to ours, found that radio-loudness increases with
decreasing \lledd\ but argued that there must also be a
secondary parameter (BH mass or spin) in effect. Based on a PCA
analysis, \citet{Boroson02} presents a schematic in which RL quasars
preferentially have both high BH mass and low \lledd.

Our results (Figures \ref{fig:f23}--\ref{fig:f25}) lead us to a
very different conclusion than that of \citet{Shankar10} (using much
of the same data) who find that high-redshift RL quasars have high
\lledd, while low-redshift RL quasars have low \lledd.  This
difference is because we consider the BH masses of the high-redshift
quasars to be in error.  After reconsidering the BH masses estimated
using the \civ\ emission line, we find that our results are in general
agreement with most of the investigations above: there is a clear
trend towards a higher RLF and a louder mean \aro\ with decreasing
\lledd. However, Figure~\ref{fig:f24} suggests that the most RL
sources at high luminosities do not have particularly low \lledd\ in
the absolute sense: they are simply the objects with the lowest
\lledd\ (and thus the highest BH masses) at that luminosity. This
finding may suggest that BH mass is the dominant effect and that low
\lledd\ is a consequence of the mass trend.  Moreover, the very
different RLF behavior at high and low redshift provide independent
evidence that BH masses determined from \ion{C}{4} should be used with
extreme caution.

Whatever controls the radio strength of quasars, we cannot lose sight
of the fact that, even where RL quasars are most prevalent, the vast
majority of the quasars would still be classified as RQ. Thus, whether
or not the radio-loudness distribution is bimodal, there is still a
strong dichotomy between RL and RQ quasars. This is intriguing as the
SDSS quasar targeting and spectroscopic classification is such that we
might expect our samples to be predominantly comprised of massive,
bulge-dominated systems \citep{Floyd04}. Consequently, it is worth
noting that our quasar sample still shows some of the radio dichotomy
of the samples that cover a wider range of luminosity space such as
\citet{Sikora07}.

We note that \citet{Floyd04} found both the RL and RQ quasars (with
$M_V<-24$ and $z\sim0.4$) in their investigation to be bulge-dominated
systems and that \citet{Sikora07} found luminous RQ quasars hosted by
ellipticals. Specifically, while it has been long known that RL
quasars live in ellipticals, \citet{Floyd04} demonstrates that
ellipticals can host RQ quasars too. If luminous, high-redshift
quasars are predominantly fueled by major mergers, it must be the case
that the average quasar hosted by an elliptical galaxy is RQ. As such,
the RL/RQ dichotomy among luminous quasars is unlikely to be entirely
due to morphology.  While we cannot say with certainty that both
extrema of the RQ quasar distribution have the same hosts, there is no
compelling reason (other than the differences in radio properties) to
think that RL quasars and the RQ quasars that occupy the RL parameter
space have different hosts. Although it has been found that RL and RQ
quasars have different environments \citep{Malkan84, Smith86,
  Sikora07}, with RL quasars living in denser regions, such work has
only been done in the context of RL vs.\ RQ as a whole. In that case,
we would indeed expect there to be differences. What would be of more
interest is to redo environmental and clustering studies considering
RL quasars in comparison with different subsets of RQ quasars,
specifically dividing up RQ quasars into hard-spectrum RQs (HSRQs;
which may have similar properties as RL quasars) and soft-spectrum RQs
(SSRQs; which would be expected to have the largest environmental and
clustering differences in comparison with RL quasars).

If massive, low \lledd\ systems can be both RL and RQ and if
morphology and/or environment are not clearly different between RL
quasars and their (putative) parent RQ quasars, then we are left to
consider spin as the defining parameter. Major mergers (that are
hypothesized to be the dominant cause of the luminous quasars in our
investigation) are expected to produce high-spin systems
\citep{Wilson95, Volonteri07, Volonteri13}. In part, this is because
if enough matter is accreted in one feeding (and even if the accretion
is initially retrograde) the BH is spun down more efficiently than it
is spun up; this tends to drive quasars that have experienced a
recent major merger to high (prograde) spin \citep{Volonteri13}. This
leads to a paradox: if RL quasars have high spin and RQ quasars are not
spinning (e.g., \citealt{Wilson95}), then the accretion efficiency
suggested by the \citet{Soltan82} argument means that RL quasars
should not be as rare as they are. This further argues against RL
quasars having retrograde spin as in \citet{Garofalo09} and
\citet{Garofalo10}.

One way to reconcile all of this is if {\em both} the RL quasars and
those RQ quasars with otherwise similar properties (i.e., HSRQs) are
dominated by major mergers, resulting in elliptical hosts with high
spins. These are systems with high mass for their luminosity and low
\lledd. The quasars among these that become RL may be those that have
undergone a rare ``2nd generation merger'' and have been spun up to a
value above some threshold \citep{Sikora09}.  However, we note that
the spins are not likely to be all that different as significant
differences in the spin would lead to significant changes in the
accretion disk properties (in particular the inner radius) that would
be expected to produce continuum (and broad emission line region)
changes inconsistent with the results of \citet{Richards11}.

This suggests that RL quasars and HSRQ quasars both have high spin,
with the RL quasars being somewhat more extreme. The rest of the RQ
population (SSRQs) could be similarly high spin objects, may have
decreasing spin with increasing \lledd, or could have no spin. Again,
however, the \citet{Soltan82} argument would suggest that having some
spin is more likely than having no spin.

In terms of making connections to low-$z$ quasars, it is particularly
important to realize that low-$z$ quasars (that are the frequent focus
of detailed observing campaigns and reverberation mapping analysis)
and the luminous high-$z$ quasars that dominate our sample might be
rather different creatures, especially if low-$z$ quasars are primarily
accreting molecular clouds \citep{Volonteri13}. This might
explain the somewhat different results seen here and by
\citet{Boroson02} in terms of whether or not RL quasars are an extrema
of the quasar population in ways other than in the radio. While we
can confirm the finding of  \citet{Boroson02} that RL quasars tend to be high mass, low
\lledd\ sources, we do not find them to be unique. Indeed most high
mass, low \lledd\ quasars are RQ.

\subsection{Accretion Disk Winds}

Our analyses in Eigenvector 1 and \civ\ parameter spaces (Section
\ref{sec:CIV}) and even color space (Section \ref{sec:StacksColor})
paint a consistent picture in that we find different quasar properties
tracking together.  For example, \citet{Sulentic07} find a correlation
between the \civ\ blueshift and R(\ion{Fe}{2}), \citet{Reichard03}
find a correlation between \civ\ blueshift and quasar color, we find a
correlation between radio loudness and color (Fig.~\ref{fig:f28}),
while both \citet{Gallagher05} and \citet{Kruczek11} consider the
connection between X-ray properties and \civ\ emission.  The mean and
extreme radio properties in these parameter spaces track together such
that higher RLF and higher \emph{mean} radio-loudness are biased to
low blueshifts (Figure \ref{fig:f22}).  More specifically, radio-loud
quasars
exhibit behaviors (emission line and continuum properties) that are
consistent with one extreme (large H$\beta$ FWHM, low R(\ion{Fe}{2}),
small \civ\ blueshift, red color) being much more likely to host
quasars with stronger radio emission than the opposite. This result is
in agreement with the investigation by \citet{Zamfir2008}.

If the \civ\ blueshift is related to the strength of a radiation
line-driven wind, this finding is very interesting in terms of the
long-observed anti-correlation between radio-loud quasars and BALQSOs
(Broad Absorption Line QSOs; \citealt{Stocke92}). While sources with
strong radio lobes tend to avoid BALQSOs (or vice versa)
\citep{Stocke92, Reichard03, Richards11}, they are not mutually
exclusive \citep{Becker00, Welling14}\footnote{Though we note that RL
  BALQSO are often either not RL (due to dust obscuration in the
  optical) or have relatively narrow absorption troughs that may not
  be consistent with a strong radiation line driven wind.}.  In the
\citet{Richards11} picture, all quasars have some sort of wind; it is
just that objects displaying absorption troughs that meet the
traditional BALQSO definition will have stronger radiation line-driven
winds than quasars that do not. Further, \citet{Richards11} argue that
{\em emission} line properties can be used to determine the strength of
radiation line-driving and, thus, of seeing BAL troughs along other
lines of sight. As high \lledd\ might be most expected to lead to a
strong radiation line-driven wind, the general anti-correlation of
BALQSOs and RL quasars would be expected.  

Second generation quasars in the model of \citet{Sikora09} could
explain the existence of the rare RL BALQSOs.  RL BALQSOs could be
those BALQSOs undergoing a ``2nd major merger'' \citep{Sikora09} and
getting spun up enough to produce a radio jet or, alternatively, those
RL quasars which undergo a significant increase in accretion rate that
generates a radiation line-driven wind. Another possibility is that RL
BALQSOs could be related to radio emission resulting from interactions
between their outflows and the ISM \citep{Jiang10, Zakamska14}.

\subsection{Evolution in $L$ and $z$}

While the trends in the RLF as a function of mass, accretion rate, and
wind-dominance seem clear, it is frustrating that we are not able to
better constrain the evolution in optical luminosity and redshift.  It
is curious that, unlike in the $L$-$z$ parameter space (where the mean
and extreme radio properties of quasars run in opposite directions),
in the EV1 (Figure~\ref{fig:f21}), \civ\ (Figure~\ref{fig:f22}), and
BH (Figures~\ref{fig:f24} and \ref{fig:f25}) parameter spaces the RLF and mean
radio-loudness increase in the {\em same} direction. Specifically, we
see reasonable agreement between the directions of evolution of the
mean radio-loudness and the RLF when we are considering parameters
that are not a strong function of the apparent magnitude.  As a
result, an explanation for the differences seen in the $L$-$z$
evolution could be the incompleteness of the FIRST survey (see
Section~\ref{sec:firstlim}) assuming that the incompleteness is
(indirectly) a function of optical magnitude.  In this case, the
relative shallowness of the radio data could be masking the true $L-z$
evolution of the RLF.

While it would seem that the mean radio loudness from stacking is more
robust, Figure~\ref{fig:f13} showed that optically fainter quasars
have \aro\ values that are more radio loud, which could indicate a
bias in the stacking results instead.
Such a correlation could come about if bright quasars that are
extincted by dust are moving to larger (fainter) magnitudes and, thus,
appear to be more RL than they should be. However, we have excluded
quasars that are most heavily dust reddened/extincted,
$\Delta(g-i)>0.5$, and we have further argued using
Figure~\ref{fig:f19} that dust is unlikely to be dominating the trend.

In short, modulo any corrections for the $L$- and $z$-dependences of
\aro, 
we are left to conclude that the mean radio-loudness does indeed
evolve with {\em both} redshift and luminosity in a way that mimics a
trend in apparent magnitude.  The opposite trend of the RLF with
apparent magnitude is either also real or is an artifact of
incompleteness to RL objects with fainter magnitude.  We note that
there is no reason that the RLF and the mean radio-loudness have to
evolve together in either $L$ or $z$, but the similarity of the trends
in the other parameter spaces that we have considered may suggest that
the observed RLF evolution is less robust than evolution of the mean
radio loudness.  Both deeper radio observations within the SDSS/FIRST
footprint and observations targeted at objects at the extremes of
$L-z$ parameter space (e.g., high-$z$ quasars) would help to answer
this important question.

\subsection{On the Meaning of Radio-Quiet}

Our goal was to identify non-radio properties of quasars that could be
used to predict whether an {\em individual} quasar is likely to be a
strong radio source or not. In that sense we have failed: RL quasars
and at least some RQ quasars do not appear to be significantly
different.  That said, we have expanded the parameter space over which
the RL/RQ dichotomy has been thoroughly investigated and have
identified properties that suggest when an optical quasar is very {\em
  unlikely} to be a strong radio source.  To make further progress, it
would help to be able to estimate BH spins for large samples of
(distant) quasars.

Given the relative similarity of RL and (some) RQ quasars, we
emphasize that there is no such thing as a ``radio-quiet" quasar. We
mean this literally in that all quasars appear to have some minimum
level of radio flux based on direct detections from deep observations
\citep{Kimball11} and hinted at by stacking (Figure~\ref{fig:f14} and
\citealt{White07}) and demographic analyses \citep{Condon13}.
However, we also mean it figuratively in that the RQ population spans
a large range of continuum and emission line properties (e.g.,
\citealt{Sulentic00b} and \citealt{Richards11}) such that it cannot be
considered a single monolithic object class.  For example,
\citet{Richards11} compared the average RL quasar spectrum to the
average spectrum of RQs with similar \civ\ EQWs and low blueshifts and
found little difference, whereas composite spectra from the other extreme in the
RQ population have quite different emission line properties.

\citet{Sulentic00b} and collaborators have emphasized this finding by
dividing quasars into ``Population A'' and ``Population B''.  In
\citet{Kruczek11} we give these classes more physical meaning by
refering to them as, respecively, ``soft-spectrum'' and
``hard-spectrum'' sources---especially when referring to RQ quasars
where the extrema (within the continuum) can be denoted as
hard-spectrum radio quiet (HSRQ) and soft-spectrum radio quiet (SSRQ).


We argue that investigations that have naively split the quasar
population into two (RL/RQ) should be reconsidered. If a RL sample is
compared to a truly representative RQ sample, one would expect to see
differences since the RQ population spans a larger range of parameter
space than the RL. Comparisons of RL quasars separately with what we
have called HSRQ and SSRQs quasars would be extremely interesting.


\section{Conclusions}
\label{sec:Conclusions}

In Section~\ref{sec:Intro} we argued that the seemingly discordant
literature on the possible bimodality of the detected quasar
population is actually in good agreement. All investigations find that
the distribution of radio-loudness is poorly fit by a single
component: there is a minority population of radio-loud objects. As
noted by \citet{Laor03}, evolution of this population may cloud our
analysis of it through the use of a single dividing line at all
redshifts and luminosities.

Section \ref{sec:aro} explains why we adopt \aro\ as our measure of
radio-loudness instead of $\log R$. Although $\log R$ has been
commonly used among radio astronomers, we have decided to implement
\aro\ in our analyses to make our results accessible to those who do
not primarily work within the radio regime. \aro\ is universal in that
it directly describes the shape of a quasar's spectral energy
distribution, specifically between the radio and the optical.

In Section~\ref{sec:Lz} (Figure~\ref{fig:f20}) we showed that the
radio-loud fraction appears to evolve in both $L$ and $z$, in
agreement with \citet{Jiang07} and \citet{Balokovic}. This evolution
is such that the RLF most closely tracks the optical apparent
magnitude, which suggests a possible bias. A radio sample covering the
area of the FIRST survey to 3$\times$ its depth or deeper is needed to
resolve this issue. We further found that the {\em mean}
radio-loudness evolves in the exact opposite sense.  Thus, it appears
that the mean and extrema of the radio-loudness distribution do not
track each other.  
This difference could offer insight into the nature of radio emission
in quasars, perhaps suggesting different tracks for the RL and
radio-intermediate sources.  Alternatively, it could be an indication
that FIRST is indeed incomplete in a manner that clouds our
understanding of the RLF evolution.

We explored the evolution of radio properties in Eigenvector 1 and
\civ\ parameter spaces in Section~\ref{sec:CIV}. These properties may
trace the relative power of radiation line-driven accretion disk winds
\citep{Richards11}. The RLF is much higher in quasars without emission
properties that point to strong radiation line-driven winds. Indeed,
the RLF is essentially zero for quasars with the highest \civ\
blueshifts (Figure \ref{fig:f22}). The mean radio-loudness shows a
similar, albeit somewhat weaker, trend. The trends in RLF and mean
radio-loudnessin EV1 parameter space (Figure \ref{fig:f21}) are
broadly consistent with the \civ\ results. We further find that the
mean radio-loudness increases with increasing reddening of the optical
continuum (Figure \ref{fig:f28}), which is consistent with these other
findings. Contrary to \citet{Boroson02}, we find that while RL quasars
tend occupy only a fraction of the quasar parameter space, they not
occupy a {\em unique} parameter space; thus, it appears that RL and RQ
quasars are parallel sequences. Some additional parameter (spin?) must
contribute to an object being RL, where that parameter is strongly
biased to those objects without strong radiation line-driven winds.

Section~\ref{sec:BH} considers the radio properties of quasars as a
function of mass, accretion rate, and \lledd. We argue that BH mass
estimates from survey-quality spectral measurements of \civ\ have
catastrophic errors in luminous quasars.  This finding is relevant to
other investigation of BH masses in high-redshift quasars.  These
errors are identified by a radical change in the radio properties of
quasars as a function of BH mass with redshift and have led some
previous work to questionable conclusions regarding the redshift
evolution of \lledd\ for RL quasars. Ignoring the biased \civ\ BH mass
results (or assuming that the actual \civ\ BH masses are inverted from
their apparent values), we find that the radio-loud fraction is
highest for the largest BH masses (at a given luminosity) (Figure
\ref{fig:f24}). This means that the RLF is a function of \lledd, in
agreement with past results. The mean radio-loudness shows a similar,
but somewhat weaker trend.

Further progress must come in the context of the realization that
there is no typical RQ quasar with which to contrast the RL
population. Rather RL quasars should be compared to RQ quasars that
have similar (non-radio) properties (Population B in
\citealt{Sulentic00a,Sulentic00b} and HSRQ in \citealt{Kruczek11}) and
contrasted with those RQ quasars that exhibit dissimilar (non-radio)
properties (Population A in \citealt{Sulentic00a,Sulentic00b} and SSRQ
in \citealt{Kruczek11}).  As we find little to differentiate RL
quasars and HSRQs, the suggestion by \citet{Sikora07} of RL quasars
being spun-up by second generation mergers is an intriguing one.

\acknowledgements We thank our colleagues for comments on the
manuscript, particularly Rick White, Bob Becker and Amy Kimball.  We
thank Rick White further for providing the FIRST completeness
calculations and his object stacking code, Adam Myers for help
constructing the Master quasar catalog, Coleman Krawczyk for help with
the BH mass analysis, and Sarah Gallagher for discussions regarding
the interpretation of differential spectral indices.  This work was
supported by NSF AAG grant 1108798.  GTR acknowledges the generous
support of a research fellowship from the Alexander von Humboldt
Foundation at the Max-Planck-Institut f\"{u}r Astronomie and is
grateful for the hospitality of the Astronomisches Rechen-Institut.
We thank the referee for suggestions that significantly improved the
presentation of the paper.  Funding for the SDSS and SDSS-II was
provided by the Alfred P. Sloan Foundation, the Participating
Institutions, the National Science Foundation, the U.S. Department of
Energy, the National Aeronautics and Space Administration, the
Japanese Monbukagakusho, the Max Planck Society, and the Higher
Education Funding Council for England. The SDSS was managed by the
Astrophysical Research Consortium for the Participating Institutions.
Funding for SDSS-III has been provided by the Alfred P. Sloan
Foundation, the Participating Institutions, the National Science
Foundation, and the U.S. Department of Energy Office of Science. The
SDSS-III web site is http://www.sdss3.org/.  SDSS-III is managed by
the Astrophysical Research Consortium for the Participating
Institutions of the SDSS-III Collaboration including the University of
Arizona, the Brazilian Participation Group, Brookhaven National
Laboratory, Carnegie Mellon University, University of Florida, the
French Participation Group, the German Participation Group, Harvard
University, the Instituto de Astrofisica de Canarias, the Michigan
State/Notre Dame/JINA Participation Group, Johns Hopkins University,
Lawrence Berkeley National Laboratory, Max Planck Institute for
Astrophysics, Max Planck Institute for Extraterrestrial Physics, New
Mexico State University, New York University, Ohio State University,
Pennsylvania State University, University of Portsmouth, Princeton
University, the Spanish Participation Group, University of Tokyo,
University of Utah, Vanderbilt University, University of Virginia,
University of Washington, and Yale University.

\newpage
\bibliographystyle{apj3}
\bibliography{Thesis}

\begin{thebibliography}{144}
\expandafter\ifx\csname natexlab\endcsname\relax\def\natexlab#1{#1}\fi
\expandafter\ifx\csname url\endcsname\relax
  \def\url#1{{\tt #1}}\fi
\expandafter\ifx\csname urlprefix\endcsname\relax\def\urlprefix{URL }\fi
\providecommand{\eprint}[2][]{\url{#2}}

\bibitem[\protect\astroncite{{Abazajian} et~al.}{2009}]{DR7}
{Abazajian}, K.~N., et~al. 2009, \apjs, 182, 543, \eprint{0812.0649}

\bibitem[\protect\astroncite{{Ahn} et~al.}{2012}]{DR9}
{Ahn}, C.~P., et~al. 2012, \apjs, 203, 21, \eprint{1207.7137}

\bibitem[\protect\astroncite{{Assef} et~al.}{2011}]{Assef11}
{Assef}, R.~J., et~al. 2011, \apj, 742, 93, \eprint{1009.1145}

\bibitem[\protect\astroncite{{Avni} \& {Tananbaum}}{1982}]{Avni82}
{Avni}, Y. \& {Tananbaum}, H. 1982, \apjl, 262, L17

\bibitem[\protect\astroncite{{Balokovi{\'c}} et~al.}{2012}]{Balokovic}
{Balokovi{\'c}}, M., {Smol{\v c}i{\'c}}, V., {Ivezi{\'c}}, {\v Z}., {Zamorani},
  G., {Schinnerer}, E., \& {Kelly}, B.~C. 2012, \apj, 759, 30,
  \eprint{1209.1099}

\bibitem[\protect\astroncite{{Barthel}}{1989}]{Barthel89}
{Barthel}, P.~D. 1989, \apj, 336, 606

\bibitem[\protect\astroncite{{Barvainis} et~al.}{2005}]{Barvainis05}
{Barvainis}, R., {Leh{\'a}r}, J., {Birkinshaw}, M., {Falcke}, H., \&
  {Blundell}, K.~M. 2005, \apj, 618, 108, \eprint{astro-ph/0409554}

\bibitem[\protect\astroncite{{Becker} et~al.}{2000}]{Becker00}
{Becker}, R.~H., {White}, R.~L., {Gregg}, M.~D., {Brotherton}, M.~S.,
  {Laurent-Muehleisen}, S.~A., \& {Arav}, N. 2000, \apj, 538, 72,
  \eprint{astro-ph/0002470}

\bibitem[\protect\astroncite{{Becker} et~al.}{1995}]{FIRST95}
{Becker}, R.~H., {White}, R.~L., \& {Helfand}, D.~J. 1995, \apj, 450, 559

\bibitem[\protect\astroncite{{Blandford}}{1990}]{Blandford90}
{Blandford}, R.~D. 1990, in Active Galactic Nuclei, eds. R.~D. {Blandford},
  H.~{Netzer}, L.~{Woltjer}, T.~J.-L. {Courvoisier}, \& M.~{Mayor}, 161--275

\bibitem[\protect\astroncite{{Blandford} \& {Payne}}{1982}]{BP82}
{Blandford}, R.~D. \& {Payne}, D.~G. 1982, \mnras, 199, 883

\bibitem[\protect\astroncite{{Blandford} \& {Znajek}}{1977}]{BZ77}
{Blandford}, R.~D. \& {Znajek}, R.~L. 1977, \mnras, 179, 433

\bibitem[\protect\astroncite{{Blanton} et~al.}{2003}]{Blanton03}
{Blanton}, M.~R., et~al. 2003, \apj, 592, 819, \eprint{arXiv:astro-ph/0210215}

\bibitem[\protect\astroncite{{Blundell} \& {Beasley}}{1998}]{Blundell98}
{Blundell}, K.~M. \& {Beasley}, A.~J. 1998, \mnras, 299, 165,
  \eprint{astro-ph/9805169}

\bibitem[\protect\astroncite{{Blundell} \& {Kuncic}}{2007}]{Blundell07}
{Blundell}, K.~M. \& {Kuncic}, Z. 2007, \apjl, 668, L103, \eprint{0708.2929}

\bibitem[\protect\astroncite{{Bondi} et~al.}{2008}]{Bondi2008}
{Bondi}, M., {Ciliegi}, P., {Schinnerer}, E., {Smol{\v c}i{\'c}}, V., {Jahnke},
  K., {Carilli}, C., \& {Zamorani}, G. 2008, \apj, 681, 1129,
  \eprint{0804.1706}

\bibitem[\protect\astroncite{{Boroson}}{2002}]{Boroson02}
{Boroson}, T.~A. 2002, \apj, 565, 78, \eprint{astro-ph/0109317}

\bibitem[\protect\astroncite{{Boroson} \& {Green}}{1992}]{BG92}
{Boroson}, T.~A. \& {Green}, R.~F. 1992, \apjs, 80, 109

\bibitem[\protect\astroncite{{Bovy} et~al.}{2011}]{Bovy11}
{Bovy}, J., et~al. 2011, \apj, 729, 141, \eprint{1011.6392}

\bibitem[\protect\astroncite{{Brotherton} \& {Francis}}{1999}]{BF99}
{Brotherton}, M.~S. \& {Francis}, P.~J. 1999, in Quasars and Cosmology, eds.
  G.~{Ferland} \& J.~{Baldwin}, vol. 162 of {\em Astronomical Society of the
  Pacific Conference Series\/}, 395, \eprint{astro-ph/9811088}

\bibitem[\protect\astroncite{{Casebeer} et~al.}{2006}]{Casebeer06}
{Casebeer}, D.~A., {Leighly}, K.~M., \& {Baron}, E. 2006, \apj, 637, 157,
  \eprint{arXiv:astro-ph/0508503}

\bibitem[\protect\astroncite{{Cirasuolo} et~al.}{2003{\natexlab{a}}}]{Cira03a}
{Cirasuolo}, M., {Celotti}, A., {Magliocchetti}, M., \& {Danese}, L.
  2003{\natexlab{a}}, \mnras, 346, 447, \eprint{astro-ph/0306415}

\bibitem[\protect\astroncite{{Cirasuolo} et~al.}{2003{\natexlab{b}}}]{Cira03b}
{Cirasuolo}, M., {Magliocchetti}, M., {Celotti}, A., \& {Danese}, L.
  2003{\natexlab{b}}, \mnras, 341, 993, \eprint{astro-ph/0301526}

\bibitem[\protect\astroncite{{Condon} et~al.}{2002}]{Condon02}
{Condon}, J.~J., {Cotton}, W.~D., \& {Broderick}, J.~J. 2002, \aj, 124, 675

\bibitem[\protect\astroncite{{Condon} et~al.}{1998}]{NVSS}
{Condon}, J.~J., {Cotton}, W.~D., {Greisen}, E.~W., {Yin}, Q.~F., {Perley},
  R.~A., {Taylor}, G.~B., \& {Broderick}, J.~J. 1998, \aj, 115, 1693

\bibitem[\protect\astroncite{{Condon} et~al.}{2013}]{Condon13}
{Condon}, J.~J., {Kellermann}, K.~I., {Kimball}, A.~E., {Ivezi{\'c}}, {\v Z}.,
  \& {Perley}, R.~A. 2013, \apj, 768, 37

\bibitem[\protect\astroncite{{Croom} et~al.}{2004}]{2QZ}
{Croom}, S.~M., {Smith}, R.~J., {Boyle}, B.~J., {Shanks}, T., {Miller}, L.,
  {Outram}, P.~J., \& {Loaring}, N.~S. 2004, \mnras, 349, 1397,
  \eprint{arXiv:astro-ph/0403040}

\bibitem[\protect\astroncite{{Croom} et~al.}{2009}]{2SLAQ}
{Croom}, S.~M., et~al. 2009, \mnras, 392, 19, \eprint{0810.4955}

\bibitem[\protect\astroncite{{Denney}}{2012}]{Denney12}
{Denney}, K.~D. 2012, \apj, 759, 44, \eprint{1208.3465}

\bibitem[\protect\astroncite{{Denney} et~al.}{2013}]{Denney13}
{Denney}, K.~D., {Pogge}, R.~W., {Assef}, R.~J., {Kochanek}, C.~S., {Peterson},
  B.~M., \& {Vestergaard}, M. 2013, \apj, 775, 60, \eprint{1303.3889}

\bibitem[\protect\astroncite{{Dunlop} et~al.}{2003}]{Dunlop03}
{Dunlop}, J.~S., {McLure}, R.~J., {Kukula}, M.~J., {Baum}, S.~A., {O'Dea},
  C.~P., \& {Hughes}, D.~H. 2003, \mnras, 340, 1095, \eprint{astro-ph/0108397}

\bibitem[\protect\astroncite{{Edge} et~al.}{1959}]{Edge59}
{Edge}, D.~O., {Shakeshaft}, J.~R., {McAdam}, W.~B., {Baldwin}, J.~E., \&
  {Archer}, S. 1959, \memras, 68, 37

\bibitem[\protect\astroncite{{Elvis}}{2000}]{Elvis00}
{Elvis}, M. 2000, \apj, 545, 63, \eprint{arXiv:astro-ph/0008064}

\bibitem[\protect\astroncite{{Elvis} et~al.}{1994}]{Elvis94}
{Elvis}, M., et~al. 1994, \apjs, 95, 1

\bibitem[\protect\astroncite{{Fine} et~al.}{2010}]{Fine10}
{Fine}, S., {Croom}, S.~M., {Bland-Hawthorn}, J., {Pimbblet}, K.~A., {Ross},
  N.~P., {Schneider}, D.~P., \& {Shanks}, T. 2010, \mnras, 409, 591,
  \eprint{1005.5287}

\bibitem[\protect\astroncite{{Floyd} et~al.}{2004}]{Floyd04}
{Floyd}, D.~J.~E., {Kukula}, M.~J., {Dunlop}, J.~S., {McLure}, R.~J., {Miller},
  L., {Percival}, W.~J., {Baum}, S.~A., \& {O'Dea}, C.~P. 2004, \mnras, 355,
  196, \eprint{astro-ph/0308436}

\bibitem[\protect\astroncite{{Fukugita} et~al.}{1996}]{Fuku96}
{Fukugita}, M., {Ichikawa}, T., {Gunn}, J.~E., {Doi}, M., {Shimasaku}, K., \&
  {Schneider}, D.~P. 1996, \aj, 111, 1748

\bibitem[\protect\astroncite{{Gallagher} et~al.}{2005}]{Gallagher05}
{Gallagher}, S.~C., {Richards}, G.~T., {Hall}, P.~B., {Brandt}, W.~N.,
  {Schneider}, D.~P., \& {Vanden Berk}, D.~E. 2005, \aj, 129, 567,
  \eprint{astro-ph/0410641}

\bibitem[\protect\astroncite{{Garofalo}}{2009}]{Garofalo09}
{Garofalo}, D. 2009, \apjl, 699, L52, \eprint{0905.4782}

\bibitem[\protect\astroncite{{Garofalo} et~al.}{2010}]{Garofalo10}
{Garofalo}, D., {Evans}, D.~A., \& {Sambruna}, R.~M. 2010, \mnras, 406, 975,
  \eprint{1004.1166}

\bibitem[\protect\astroncite{{Gaskell}}{1982}]{Gaskell82}
{Gaskell}, C.~M. 1982, \apj, 263, 79

\bibitem[\protect\astroncite{{Glikman} et~al.}{2006}]{Glikman06}
{Glikman}, E., {Helfand}, D.~J., \& {White}, R.~L. 2006, \apj, 640, 579,
  \eprint{arXiv:astro-ph/0511640}

\bibitem[\protect\astroncite{{Goldschmidt} et~al.}{1999}]{Goldschmidt99}
{Goldschmidt}, P., {Kukula}, M.~J., {Miller}, L., \& {Dunlop}, J.~S. 1999,
  \apj, 511, 612, \eprint{arXiv:astro-ph/9810392}

\bibitem[\protect\astroncite{{Gunn} et~al.}{1998}]{Gunn98}
{Gunn}, J.~E., et~al. 1998, \aj, 116, 3040, \eprint{arXiv:astro-ph/9809085}

\bibitem[\protect\astroncite{{Gunn} et~al.}{2006}]{Gunn06}
--- 2006, \aj, 131, 2332, \eprint{arXiv:astro-ph/0602326}

\bibitem[\protect\astroncite{{Hewett} \& {Wild}}{2010}]{HewettWild}
{Hewett}, P.~C. \& {Wild}, V. 2010, \mnras, 405, 2302, \eprint{1003.3017}

\bibitem[\protect\astroncite{{Ho}}{2002}]{Ho02}
{Ho}, L.~C. 2002, \apj, 564, 120, \eprint{astro-ph/0110440}

\bibitem[\protect\astroncite{{Hodge} et~al.}{2011}]{Hodge2011}
{Hodge}, J.~A., {Becker}, R.~H., {White}, R.~L., {Richards}, G.~T., \&
  {Zeimann}, G.~R. 2011, \aj, 142, 3, \eprint{1103.5749}

\bibitem[\protect\astroncite{{Hogg} et~al.}{2001}]{Hogg01}
{Hogg}, D.~W., {Finkbeiner}, D.~P., {Schlegel}, D.~J., \& {Gunn}, J.~E. 2001,
  \aj, 122, 2129, \eprint{astro-ph/0106511}

\bibitem[\protect\astroncite{{Ivezi{\'c}} et~al.}{2002}]{Ivezic}
{Ivezi{\'c}}, {\v Z}., et~al. 2002, \aj, 124, 2364,
  \eprint{arXiv:astro-ph/0202408}

\bibitem[\protect\astroncite{{Ivezi{\'c}} et~al.}{2004}]{Ivezic04}
--- 2004, Astronomische Nachrichten, 325, 583, \eprint{astro-ph/0410195}

\bibitem[\protect\astroncite{{Jarvis} \& {McLure}}{2002}]{JarvisMcLure02}
{Jarvis}, M.~J. \& {McLure}, R.~J. 2002, \mnras, 336, L38,
  \eprint{astro-ph/0208390}

\bibitem[\protect\astroncite{{Jester}}{2005}]{Jester05}
{Jester}, S. 2005, \apj, 625, 667, \eprint{astro-ph/0502394}

\bibitem[\protect\astroncite{{Jiang} et~al.}{2007}]{Jiang07}
{Jiang}, L., {Fan}, X., {Ivezi{\'c}}, {\v Z}., {Richards}, G.~T., {Schneider},
  D.~P., {Strauss}, M.~A., \& {Kelly}, B.~C. 2007, \apj, 656, 680,
  \eprint{arXiv:astro-ph/0611453}

\bibitem[\protect\astroncite{{Jiang} et~al.}{2010}]{Jiang10}
{Jiang}, Y.-F., {Ciotti}, L., {Ostriker}, J.~P., \& {Spitkovsky}, A. 2010,
  \apj, 711, 125, \eprint{0904.4918}

\bibitem[\protect\astroncite{{Just} et~al.}{2007}]{Just07}
{Just}, D.~W., {Brandt}, W.~N., {Shemmer}, O., {Steffen}, A.~T., {Schneider},
  D.~P., {Chartas}, G., \& {Garmire}, G.~P. 2007, \apj, 665, 1004,
  \eprint{0705.3059}

\bibitem[\protect\astroncite{{Kellermann} et~al.}{1989}]{Kellermann89}
{Kellermann}, K.~I., {Sramek}, R., {Schmidt}, M., {Shaffer}, D.~B., \& {Green},
  R. 1989, \aj, 98, 1195

\bibitem[\protect\astroncite{{Kimball} \& {Ivezi{\'c}}}{2008}]{Kimball08}
{Kimball}, A.~E. \& {Ivezi{\'c}}, {\v Z}. 2008, \aj, 136, 684,
  \eprint{0806.0650}

\bibitem[\protect\astroncite{{Kimball} et~al.}{2011}]{Kimball11}
{Kimball}, A.~E., {Kellermann}, K.~I., {Condon}, J.~J., {Ivezi{\'c}}, {\v Z}.,
  \& {Perley}, R.~A. 2011, \apjl, 739, L29, \eprint{1107.3551}

\bibitem[\protect\astroncite{{Kochanek} et~al.}{2012}]{AGES}
{Kochanek}, C.~S., et~al. 2012, \apjs, 200, 8, \eprint{1110.4371}

\bibitem[\protect\astroncite{{Krawczyk} et~al.}{2013}]{Coleman13}
{Krawczyk}, C.~M., {Richards}, G.~T., {Mehta}, S.~S., {Vogeley}, M.~S.,
  {Gallagher}, S.~C., {Leighly}, K.~M., {Ross}, N.~P., \& {Schneider}, D.~P.
  2013, \apjs, 206, 4, \eprint{1304.5573}

\bibitem[\protect\astroncite{{Kruczek} et~al.}{2011}]{Kruczek11}
{Kruczek}, N.~E., et~al. 2011, \aj, 142, 130, \eprint{1109.1515}

\bibitem[\protect\astroncite{{Lacy} et~al.}{2001}]{Lacy01}
{Lacy}, M., {Laurent-Muehleisen}, S.~A., {Ridgway}, S.~E., {Becker}, R.~H., \&
  {White}, R.~L. 2001, \apjl, 551, L17, \eprint{astro-ph/0103087}

\bibitem[\protect\astroncite{{Laor}}{2000}]{Laor00}
{Laor}, A. 2000, \apjl, 543, L111, \eprint{astro-ph/0009192}

\bibitem[\protect\astroncite{{Laor}}{2003}]{Laor03}
--- 2003, ArXiv Astrophysics e-prints, \eprint{astro-ph/0312417}

\bibitem[\protect\astroncite{{Leighly}}{2004}]{Leighly04}
{Leighly}, K.~M. 2004, \apj, 611, 125, \eprint{arXiv:astro-ph/0402452}

\bibitem[\protect\astroncite{{Leighly} et~al.}{2007}]{Leighly07}
{Leighly}, K.~M., {Halpern}, J.~P., {Jenkins}, E.~B., {Grupe}, D., {Choi}, J.,
  \& {Prescott}, K.~B. 2007, \apj, 663, 103, \eprint{arXiv:astro-ph/0611349}

\bibitem[\protect\astroncite{{Lilly} et~al.}{2007}]{COSMOS07}
{Lilly}, S.~J., et~al. 2007, \apjs, 172, 70, \eprint{arXiv:astro-ph/0612291}

\bibitem[\protect\astroncite{{Lupton} et~al.}{2001}]{Lupton01}
{Lupton}, R., {Gunn}, J.~E., {Ivezi{\'c}}, Z., {Knapp}, G.~R., \& {Kent}, S.
  2001, in Astronomical Data Analysis Software and Systems X, eds. F.~R.
  {Harnden}, Jr., F.~A. {Primini}, \& H.~E. {Payne}, vol. 238 of {\em
  Astronomical Society of the Pacific Conference Series\/}, 269,
  \eprint{astro-ph/0101420}

\bibitem[\protect\astroncite{{Lusso} et~al.}{2010}]{Lusso10}
{Lusso}, E., et~al. 2010, \aap, 512, A34, \eprint{0912.4166}

\bibitem[\protect\astroncite{{Lynden-Bell}}{1969}]{LB69}
{Lynden-Bell}, D. 1969, \nat, 223, 690

\bibitem[\protect\astroncite{{Maddox} et~al.}{2012}]{Maddox12}
{Maddox}, N., {Hewett}, P.~C., {P{\'e}roux}, C., {Nestor}, D.~B., \&
  {Wisotzki}, L. 2012, \mnras, 424, 2876, \eprint{1206.1434}

\bibitem[\protect\astroncite{{Mahony} et~al.}{2012}]{Mahony12}
{Mahony}, E.~K., {Sadler}, E.~M., {Croom}, S.~M., {Ekers}, R.~D., {Feain},
  I.~J., \& {Murphy}, T. 2012, \apj, 754, 12, \eprint{1205.2233}

\bibitem[\protect\astroncite{{Malkan}}{1984}]{Malkan84}
{Malkan}, M.~A. 1984, \apj, 287, 555

\bibitem[\protect\astroncite{{McLure} \& {Jarvis}}{2002}]{McLureJarvis02}
{McLure}, R.~J. \& {Jarvis}, M.~J. 2002, \mnras, 337, 109,
  \eprint{astro-ph/0204473}

\bibitem[\protect\astroncite{{Miller} et~al.}{2011}]{Miller11}
{Miller}, B.~P., {Brandt}, W.~N., {Schneider}, D.~P., {Gibson}, R.~R.,
  {Steffen}, A.~T., \& {Wu}, J. 2011, \apj, 726, 20, \eprint{1010.4804}

\bibitem[\protect\astroncite{{Miller} et~al.}{1990}]{Miller90}
{Miller}, L., {Peacock}, J.~A., \& {Mead}, A.~R.~G. 1990, \mnras, 244, 207

\bibitem[\protect\astroncite{{Moore} \& {Stockman}}{1984}]{Moore84}
{Moore}, R.~L. \& {Stockman}, H.~S. 1984, \apj, 279, 465

\bibitem[\protect\astroncite{{Murray} et~al.}{1995}]{Murray95}
{Murray}, N., {Chiang}, J., {Grossman}, S.~A., \& {Voit}, G.~M. 1995, \apj,
  451, 498

\bibitem[\protect\astroncite{{Neugebauer} et~al.}{1986}]{Neugebauer86}
{Neugebauer}, G., {Miley}, G.~K., {Soifer}, B.~T., \& {Clegg}, P.~E. 1986,
  \apj, 308, 815

\bibitem[\protect\astroncite{{Oke} \& {Gunn}}{1983}]{OkeGunn83}
{Oke}, J.~B. \& {Gunn}, J.~E. 1983, \apj, 266, 713

\bibitem[\protect\astroncite{{Palanque-Delabrouille} et~al.}{2013}]{SDSSIII13}
{Palanque-Delabrouille}, N., et~al. 2013, \aap, 551, A29, \eprint{1209.3968}

\bibitem[\protect\astroncite{{Papovich} et~al.}{2006}]{Papovich06}
{Papovich}, C., et~al. 2006, \aj, 132, 231, \eprint{arXiv:astro-ph/0512623}

\bibitem[\protect\astroncite{{P{\^a}ris} et~al.}{2012}]{SDSSIII}
{P{\^a}ris}, I., et~al. 2012, \aap, 548, A66, \eprint{1210.5166}

\bibitem[\protect\astroncite{{Park} et~al.}{2013}]{Park13}
{Park}, D., {Woo}, J.-H., {Denney}, K.~D., \& {Shin}, J. 2013, \apj, 770, 87,
  \eprint{1304.7281}

\bibitem[\protect\astroncite{{Peacock} et~al.}{1986}]{Peacock86}
{Peacock}, J.~A., {Miller}, L., \& {Longair}, M.~S. 1986, \mnras, 218, 265

\bibitem[\protect\astroncite{{Pier} et~al.}{2003}]{Pier03}
{Pier}, J.~R., {Munn}, J.~A., {Hindsley}, R.~B., {Hennessy}, G.~S., {Kent},
  S.~M., {Lupton}, R.~H., \& {Ivezi{\'c}}, {\v Z}. 2003, \aj, 125, 1559,
  \eprint{astro-ph/0211375}

\bibitem[\protect\astroncite{{Proga} et~al.}{2000}]{Proga00}
{Proga}, D., {Stone}, J.~M., \& {Kallman}, T.~R. 2000, \apj, 543, 686,
  \eprint{arXiv:astro-ph/0005315}

\bibitem[\protect\astroncite{{Rafiee} \& {Hall}}{2011}]{Rafiee11}
{Rafiee}, A. \& {Hall}, P.~B. 2011, \mnras, 415, 2932, \eprint{1011.1268}

\bibitem[\protect\astroncite{{Rafter} et~al.}{2009}]{Rafter09}
{Rafter}, S.~E., {Crenshaw}, D.~M., \& {Wiita}, P.~J. 2009, \aj, 137, 42,
  \eprint{0809.3977}

\bibitem[\protect\astroncite{{Reichard} et~al.}{2003}]{Reichard03}
{Reichard}, T.~A., et~al. 2003, \aj, 126, 2594, \eprint{astro-ph/0308508}

\bibitem[\protect\astroncite{{Richards}
  et~al.}{2002{\natexlab{a}}}]{Richards02b}
{Richards}, G.~T., {Vanden Berk}, D.~E., {Reichard}, T.~A., {Hall}, P.~B.,
  {Schneider}, D.~P., {SubbaRao}, M., {Thakar}, A.~R., \& {York}, D.~G.
  2002{\natexlab{a}}, \aj, 124, 1, \eprint{arXiv:astro-ph/0204162}

\bibitem[\protect\astroncite{{Richards}
  et~al.}{2002{\natexlab{b}}}]{Richards02a}
{Richards}, G.~T., et~al. 2002{\natexlab{b}}, \aj, 123, 2945,
  \eprint{arXiv:astro-ph/0202251}

\bibitem[\protect\astroncite{{Richards} et~al.}{2003}]{Richards03}
--- 2003, \aj, 126, 1131, \eprint{arXiv:astro-ph/0305305}

\bibitem[\protect\astroncite{{Richards} et~al.}{2006}]{Richards06}
--- 2006, \aj, 131, 2766, \eprint{arXiv:astro-ph/0601434}

\bibitem[\protect\astroncite{{Richards} et~al.}{2009}]{Richards09}
--- 2009, \apjs, 180, 67, \eprint{0809.3952}

\bibitem[\protect\astroncite{{Richards} et~al.}{2011}]{Richards11}
--- 2011, \aj, 141, 167, \eprint{1011.2282}

\bibitem[\protect\astroncite{{Runnoe} et~al.}{2013}]{Runnoe13}
{Runnoe}, J.~C., {Brotherton}, M.~S., {Shang}, Z., \& {DiPompeo}, M.~A. 2013,
  \mnras, 434, 848, \eprint{1306.3521}

\bibitem[\protect\astroncite{{Sanders} et~al.}{1989}]{Sanders89}
{Sanders}, D.~B., {Phinney}, E.~S., {Neugebauer}, G., {Soifer}, B.~T., \&
  {Matthews}, K. 1989, \apj, 347, 29

\bibitem[\protect\astroncite{{Schlegel} et~al.}{1998}]{Schlegel98}
{Schlegel}, D.~J., {Finkbeiner}, D.~P., \& {Davis}, M. 1998, \apj, 500, 525,
  \eprint{astro-ph/9710327}

\bibitem[\protect\astroncite{{Schmidt}}{1963}]{Schmidt63}
{Schmidt}, M. 1963, \nat, 197, 1040

\bibitem[\protect\astroncite{{Schmidt}}{1970}]{Schmidt70}
--- 1970, \apj, 162, 371

\bibitem[\protect\astroncite{{Schmidt} \& {Green}}{1983}]{SG83}
{Schmidt}, M. \& {Green}, R.~F. 1983, \apj, 269, 352

\bibitem[\protect\astroncite{{Schneider} et~al.}{2010}]{DR7QC}
{Schneider}, D.~P., et~al. 2010, \aj, 139, 2360, \eprint{1004.1167}

\bibitem[\protect\astroncite{{Shankar} et~al.}{2010}]{Shankar10}
{Shankar}, F., {Sivakoff}, G.~R., {Vestergaard}, M., \& {Dai}, X. 2010, \mnras,
  401, 1869, \eprint{0909.4092}

\bibitem[\protect\astroncite{{Shen} et~al.}{2011}]{Shen11}
{Shen}, Y., et~al. 2011, \apjs, 194, 45, \eprint{1006.5178}

\bibitem[\protect\astroncite{{Sikora}}{2009}]{Sikora09}
{Sikora}, M. 2009, Astronomische Nachrichten, 330, 291, \eprint{0811.3105}

\bibitem[\protect\astroncite{{Sikora} et~al.}{2007}]{Sikora07}
{Sikora}, M., {Stawarz}, {\L}., \& {Lasota}, J.-P. 2007, \apj, 658, 815,
  \eprint{astro-ph/0604095}

\bibitem[\protect\astroncite{{Singal} et~al.}{2011}]{Singal11}
{Singal}, J., {Petrosian}, V., {Lawrence}, A., \& {Stawarz}, {\L}. 2011, \apj,
  743, 104, \eprint{1101.2930}

\bibitem[\protect\astroncite{{Singal} et~al.}{2013}]{Singal13}
{Singal}, J., {Petrosian}, V., {Stawarz}, {\L}., \& {Lawrence}, A. 2013, \apj,
  764, 43, \eprint{1207.3396}

\bibitem[\protect\astroncite{{Smith} et~al.}{1986}]{Smith86}
{Smith}, E.~P., {Heckman}, T.~M., {Bothun}, G.~D., {Romanishin}, W., \&
  {Balick}, B. 1986, \apj, 306, 64

\bibitem[\protect\astroncite{{Soltan}}{1982}]{Soltan82}
{Soltan}, A. 1982, \mnras, 200, 115

\bibitem[\protect\astroncite{{Spergel} et~al.}{2007}]{Spergel07}
{Spergel}, D.~N., et~al. 2007, \apjs, 170, 377, \eprint{astro-ph/0603449}

\bibitem[\protect\astroncite{{Steffen} et~al.}{2006}]{Steffen06}
{Steffen}, A.~T., {Strateva}, I., {Brandt}, W.~N., {Alexander}, D.~M.,
  {Koekemoer}, A.~M., {Lehmer}, B.~D., {Schneider}, D.~P., \& {Vignali}, C.
  2006, \aj, 131, 2826, \eprint{arXiv:astro-ph/0602407}

\bibitem[\protect\astroncite{{Steidel} \& {Sargent}}{1991}]{Steidel91}
{Steidel}, C.~C. \& {Sargent}, W.~L.~W. 1991, \apj, 382, 433

\bibitem[\protect\astroncite{{Stocke} et~al.}{1992}]{Stocke92}
{Stocke}, J.~T., {Morris}, S.~L., {Weymann}, R.~J., \& {Foltz}, C.~B. 1992,
  \apj, 396, 487

\bibitem[\protect\astroncite{{Stoughton} et~al.}{2002}]{Stoughton02}
{Stoughton}, C., et~al. 2002, \aj, 123, 485

\bibitem[\protect\astroncite{{Strittmatter} et~al.}{1980}]{Strittmatter80}
{Strittmatter}, P.~A., {Hill}, P., {Pauliny-Toth}, I.~I.~K., {Steppe}, H., \&
  {Witzel}, A. 1980, \aap, 88, L12

\bibitem[\protect\astroncite{{Sulentic} et~al.}{2007}]{Sulentic07}
{Sulentic}, J.~W., {Bachev}, R., {Marziani}, P., {Negrete}, C.~A., \&
  {Dultzin}, D. 2007, \apj, 666, 757, \eprint{0705.1895}

\bibitem[\protect\astroncite{{Sulentic}
  et~al.}{2000{\natexlab{a}}}]{Sulentic00a}
{Sulentic}, J.~W., {Marziani}, P., \& {Dultzin-Hacyan}, D. 2000{\natexlab{a}},
  \araa, 38, 521

\bibitem[\protect\astroncite{{Sulentic}
  et~al.}{2000{\natexlab{b}}}]{Sulentic00b}
{Sulentic}, J.~W., {Zwitter}, T., {Marziani}, P., \& {Dultzin-Hacyan}, D.
  2000{\natexlab{b}}, \apjl, 536, L5, \eprint{astro-ph/0005177}

\bibitem[\protect\astroncite{{Trump} et~al.}{2009}]{COSMOS09}
{Trump}, J.~R., et~al. 2009, \apj, 696, 1195, \eprint{0811.3977}

\bibitem[\protect\astroncite{{Ulvestad} et~al.}{2005}]{Ulvestad05}
{Ulvestad}, J.~S., {Antonucci}, R.~R.~J., \& {Barvainis}, R. 2005, \apj, 621,
  123, \eprint{astro-ph/0411678}

\bibitem[\protect\astroncite{{Urry} \& {Padovani}}{1995}]{UrryandP95}
{Urry}, C.~M. \& {Padovani}, P. 1995, \pasp, 107, 803,
  \eprint{arXiv:astro-ph/9506063}

\bibitem[\protect\astroncite{{Vanden Berk} et~al.}{2001}]{VDBerk01}
{Vanden Berk}, D.~E., et~al. 2001, \aj, 122, 549,
  \eprint{arXiv:astro-ph/0105231}

\bibitem[\protect\astroncite{{Vanden Berk} et~al.}{2005}]{VB05}
--- 2005, \aj, 129, 2047, \eprint{astro-ph/0501113}

\bibitem[\protect\astroncite{{Vestergaard} \& {Peterson}}{2006}]{VP06}
{Vestergaard}, M. \& {Peterson}, B.~M. 2006, \apj, 641, 689,
  \eprint{astro-ph/0601303}

\bibitem[\protect\astroncite{{Visnovsky} et~al.}{1992}]{Visnovsky92}
{Visnovsky}, K.~L., {Impey}, C.~D., {Foltz}, C.~B., {Hewett}, P.~C., {Weymann},
  R.~J., \& {Morris}, S.~L. 1992, \apj, 391, 560

\bibitem[\protect\astroncite{{Volonteri} et~al.}{2007}]{Volonteri07}
{Volonteri}, M., {Sikora}, M., \& {Lasota}, J.-P. 2007, \apj, 667, 704,
  \eprint{0706.3900}

\bibitem[\protect\astroncite{{Volonteri} et~al.}{2013}]{Volonteri13}
{Volonteri}, M., {Sikora}, M., {Lasota}, J.-P., \& {Merloni}, A. 2013, \apj,
  775, 94, \eprint{1210.1025}

\bibitem[\protect\astroncite{{Wals} et~al.}{2005}]{Wals05}
{Wals}, M., {Boyle}, B.~J., {Croom}, S.~M., {Miller}, L., {Smith}, R.,
  {Shanks}, T., \& {Outram}, P. 2005, \mnras, 360, 453,
  \eprint{arXiv:astro-ph/0502401}

\bibitem[\protect\astroncite{{Welling} et~al.}{2014}]{Welling14}
{Welling}, C.~A., {Miller}, B.~P., {Brandt}, W.~N., {Capellupo}, D.~M., \&
  {Gibson}, R.~R. 2014, ArXiv e-prints, \eprint{1403.0958}

\bibitem[\protect\astroncite{{White} et~al.}{2007}]{White07}
{White}, R.~L., {Helfand}, D.~J., {Becker}, R.~H., {Glikman}, E., \& {de
  Vries}, W. 2007, \apj, 654, 99, \eprint{arXiv:astro-ph/0607335}

\bibitem[\protect\astroncite{{White} et~al.}{2000}]{White00}
{White}, R.~L., et~al. 2000, \apjs, 126, 133, \eprint{astro-ph/9912215}

\bibitem[\protect\astroncite{{Wilkes}}{1984}]{Wilkes84}
{Wilkes}, B.~J. 1984, \mnras, 207, 73

\bibitem[\protect\astroncite{{Wilson} \& {Colbert}}{1995}]{Wilson95}
{Wilson}, A.~S. \& {Colbert}, E.~J.~M. 1995, \apj, 438, 62,
  \eprint{astro-ph/9408005}

\bibitem[\protect\astroncite{{Wisotzki}}{2000}]{Wisotzki2000}
{Wisotzki}, L. 2000, \aap, 353, 861

\bibitem[\protect\astroncite{{Woo} \& {Urry}}{2002}]{WooUrry02}
{Woo}, J.-H. \& {Urry}, C.~M. 2002, \apjl, 581, L5, \eprint{astro-ph/0211118}

\bibitem[\protect\astroncite{{Worseck} \& {Prochaska}}{2011}]{Worseck11}
{Worseck}, G. \& {Prochaska}, J.~X. 2011, \apj, 728, 23, \eprint{1004.3347}

\bibitem[\protect\astroncite{{Wu} et~al.}{2012}]{Wu12}
{Wu}, J., {Brandt}, W.~N., {Anderson}, S.~F., {Diamond-Stanic}, A.~M., {Hall},
  P.~B., {Plotkin}, R.~M., {Schneider}, D.~P., \& {Shemmer}, O. 2012, \apj,
  747, 10, \eprint{1112.2228}

\bibitem[\protect\astroncite{{Xu} et~al.}{1999}]{Xu99}
{Xu}, C., {Livio}, M., \& {Baum}, S. 1999, \aj, 118, 1169,
  \eprint{astro-ph/9905322}

\bibitem[\protect\astroncite{{York} et~al.}{2000}]{York00}
{York}, D.~G., et~al. 2000, \aj, 120, 1579, \eprint{arXiv:astro-ph/0006396}

\bibitem[\protect\astroncite{{Zakamska} \& {Greene}}{2014}]{Zakamska14}
{Zakamska}, N.~L. \& {Greene}, J.~E. 2014, ArXiv e-prints, \eprint{1402.6736}

\bibitem[\protect\astroncite{{Zamfir} et~al.}{2008}]{Zamfir2008}
{Zamfir}, S., {Sulentic}, J.~W., \& {Marziani}, P. 2008, \mnras, 387, 856,
  \eprint{0804.0788}

\end{thebibliography}
\end{document}